\documentclass[12pt,oneside,reqno]{amsart}
\usepackage{graphicx}
\usepackage{cancel}
\pdfoutput=1
\usepackage[margin=1in]{geometry}
\usepackage{abstract}
\usepackage{amssymb, amsmath, amsthm, mathrsfs, amscd, amsaddr}
\usepackage{subcaption, color, graphicx}
\usepackage[dvipsnames]{xcolor}
\usepackage{microtype, siunitx, fancyhdr, polynom}
\usepackage{tikz, tikz-cd, physics, multicol, multirow}
\usepackage{hyperref, makecell}
\usepackage[all, cmtip]{xy}
\usepackage{comment}
\usepackage{float}
\usepackage{placeins}
\usepackage{cancel}

\newcommand\qt{\widetilde{q}}
\def\ff#1#2{{\textstyle\frac{#1}{#2}}}

 \usepackage{stackrel,nicefrac}
 \usepackage{mathtools}

\hypersetup{colorlinks,linkcolor=blue,citecolor=cyan}

\usetikzlibrary{decorations.markings}

\allowdisplaybreaks

\tikzset{->-/.style={decoration={
  markings,
  mark=at position #1 with {\arrow{>}}},postaction={decorate}}}

\numberwithin{equation}{section}

\def\tilde{\widetilde}

\newcommand{\conjref}[1]{Conjecture~\ref{#1}}
\newtheorem{conjecture}{Conjecture}

\begin{document}

    \vspace*{-1cm}%
    \begin{minipage}[t]{16cm}
      \begin{flushright}
      BONN-TH-2025-13\\
      \end{flushright}
    \end{minipage}%
  \vspace*{1cm}

\title{Orientation Reversal and the Chern-Simons Natural Boundary}
\author{Griffen Adams$^1$, Ovidiu Costin$^2$, Gerald V. Dunne$^1$, Sergei Gukov$^{3}$, Oğuz \"Oner$^4$  }

\address{$^1$Physics Department, University of Connecticut, 196 Auditorium Road,
Storrs, CT 06269, USA}
\address{$^2$Mathematics Department, The Ohio State University, 231 W. 18th Avenue,  Columbus, Ohio 43210, USA}
\address{$^3$Division of Physics, Mathematics and Astronomy, California Institute of Technology, 1200~E.~California Blvd., Pasadena, CA 91125, USA}
\address{$^4$Bethe Center for Theoretical Physics, Universit\"at Bonn, D-53115, Germany}

\setcounter{tocdepth}{2}
\email{griffen.adams@uconn.edu}
\email{costin@math.osu.edu}
\email{gerald.dunne@uconn.edu}
\email{gukov@math.caltech.edu}
\email{oguzoner@uni-bonn.de}


\maketitle

\begin{abstract}

We show that the fundamental property of preservation of relations, underlying resurgent analysis, provides a new perspective on crossing a natural boundary, an important general problem in theoretical and mathematical physics. This reveals a deeper rigidity aspect of resurgence in a quantum field theory path integral. The physical context here is the non-perturbative completion of complex Chern-Simons theory that associates to a 3-manifold a collection of $q$-series invariants labeled by Spin$^c$ structures, for which crossing the natural boundary corresponds to orientation reversal of the 3-manifold. Our new resurgent perspective leads to a practical numerical algorithm that generates $q$-series which are dual to unary $q$-series composed of false theta functions. Until recently, these duals were only known in a limited number of cases, essentially based on Ramanujan's mock theta functions, and the common belief was that the duals might not even exist in the general case. Resurgence analysis identifies as primary objects Mordell integrals: up to changes of variables, they are Laplace transforms of resurgent functions. Their unique Borel summed transseries decomposition on either side of the Stokes line is simply the unique decomposition into real and imaginary parts. In turn, the latter are combinations of unary $q$-series in terms of $q$ and its modular counterpart $\tilde{q}$, and are resurgent by construction. The Mordell integral is analytic across the natural boundary of the $q$ and $\tilde{q}$ series, and uniqueness of a similar decomposition which preserves algebraic relations on the other side of the boundary defines the unique boundary crossing of the $q$ series. We demonstrate that this continuation can be efficiently implemented numerically.
In the cases where unique mock modular identities are known, they are found by this numerical procedure, but the procedure can go well beyond the known list of identities. A particularly interesting feature of the resurgent approach is that it reveals new aspects, and is very different from other known approaches based on indefinite theta series, Appell-Lerch sums, and representation theory of logarithmic vertex operator algebras.

\end{abstract}

\setcounter{tocdepth}{2}
\newpage
\tableofcontents

\section{Introduction and Motivation}

Crossing a natural boundary is a well-known important problem in theoretical and mathematical physics. It naturally arises in a variety of problems, including the hard hexagon model \cite{baxter,andrews}, the 2d Ising model susceptibility \cite{nickel,perk, mccoy}, and (super)minimal models \cite{BMO95}. It has recently been studied in quantum field theory (QFT), e.g. for non-perturbative completion of complex Chern-Simons theory \cite{GMP,CCFGH,CFG19,CCKPG,CDGG}, where reversal of the orientation has the effect of transforming the coupling constant $\hbar \to - \hbar$, or equivalently,
\begin{equation}
    q \; \to \; q^{-1}
\end{equation}
In the context of magnets and spin systems, this operation is the familiar sign change of the coupling constant
\begin{equation}
    J \; \to \; -J
\end{equation}
that relates ferromagnetic and anti-ferromagnetic regimes.

Much like ferromagnetic and anti-ferromagnetic regimes exhibit very different magnetization and other parameters, the partition function of non-perturbative complex Chern-Simons theory on manifolds related by orientation reversal are, in general, qualitatively different. This curious property makes such $q$-series invariants labeled by Spin$^c$ structures promising for applications to low-dimensional topology, in particular to the study of exotic smooth structures on 4-manifolds (see e.g. \cite{CDGG} for more details).
For certain manifolds crossing the boundary can be achieved in terms of explicit $q$-Pochhammer representations of the relevant $q$-series, intimately related to Ramanujan's mock modular theta functions \cite{Ram,Watson,GM12}.
However, this is only a limited class, based on special linear combinations of false theta functions, and relying on special $q$-series identities, pioneered by Ramanujan. The QFT perspective suggests that this interpretation of orientation reversal should apply in greater generality.

Here we present a new approach to this problem, based on resurgent asymptotics \cite{Co08,Marino:2012zq,Dorigoni:2014hea,ABS19,CD20b}, by invoking the fundamental property of preservation of relations under unique continuation of resurgent functions \cite{sauzin}. This is a much deeper feature of resurgence, which is relatively unexplored, and we show here that it can have significant consequences. This approach leads to a novel numerical algorithmic procedure to find dual $q$-series both inside and outside the unit disk. This method not only recovers the known mock theta cases, but also extends far beyond these examples.

\begin{figure}[h!]
\centering
  \centering
  \includegraphics[width=1\linewidth]{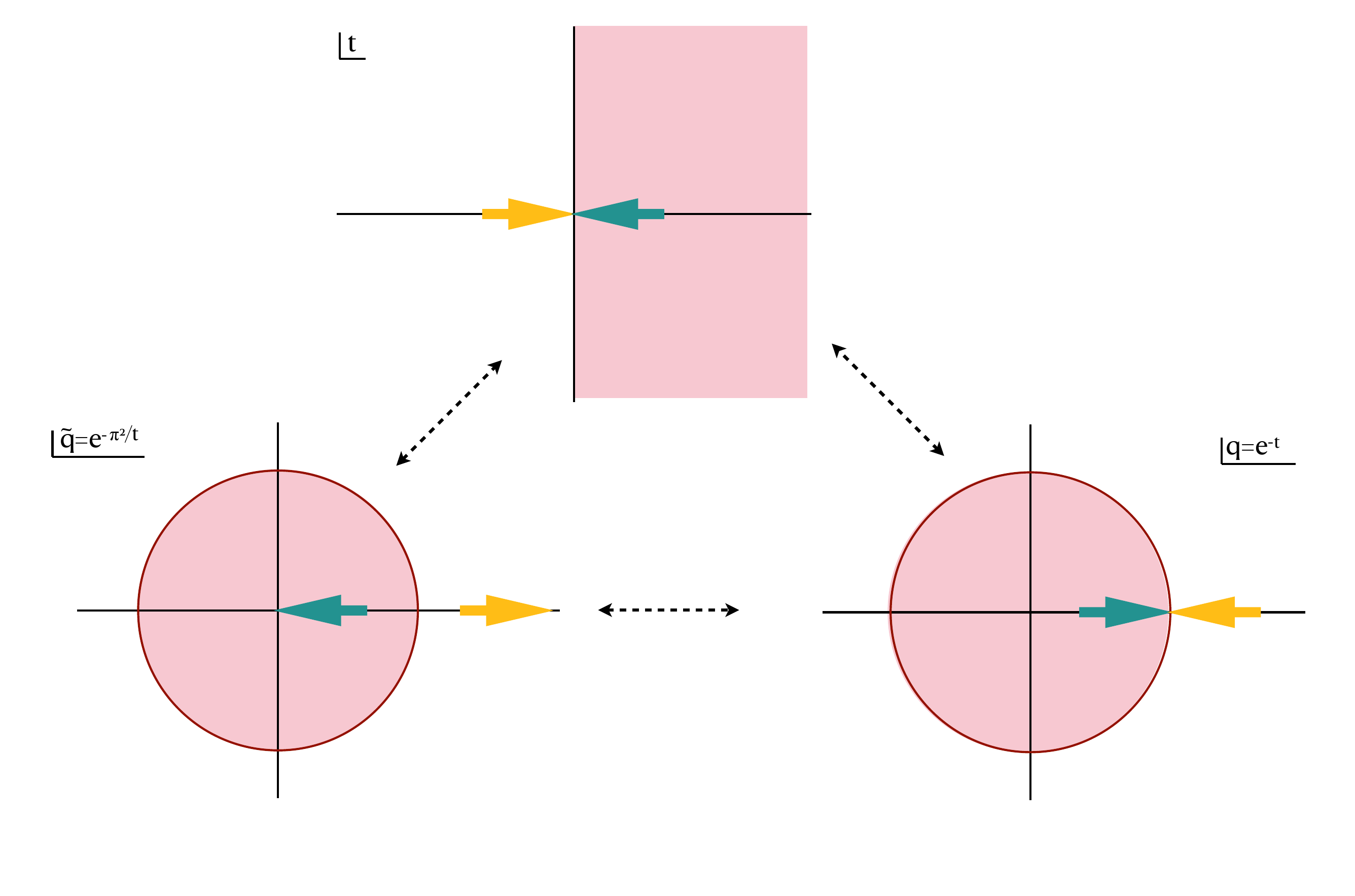}
\caption{Mappings between the $t$ plane and the $q=e^{-t}$ and $\qt=e^{-\pi^2/t}$ planes. The natural boundary is the imaginary axis in $t$ and the unit circle in $q$ and $\qt$.}
\label{fig:nb}
\end{figure}

The $q$-series of interest in this paper are expanded in terms of $q$ and $\qt$, defined as:
\begin{equation}
    q:=e^{-t}\qquad; \qquad \qt:=e^{-\pi^2/t}.
    \label{eq:qqt}
\end{equation}
See Figure \ref{fig:nb}. We use the symbol $t$ in place of $\hbar$ to conform to standard notation in the literature.
\graphicspath{ {./paper plots/} }
In \cite{CDGG} it was shown that both the $t\to 0^+$ and $t\to +\infty$ expansions of the relevant Mordell-Borel integrals (see \eqref{eq:js}-\eqref{eq:jc}) are given by a factorially divergent power series, and that there is a Stokes line along the negative real axis: $t\in (-\infty, 0]$. This translates to a Stokes line along the negative real axis in the Borel plane. In the analysis of resurgent functions, the richest source of information about a function is the vicinity of its Stokes line(s). When the Borel contour grazes the Stokes line above and below, the integrals have a {\it unique} decomposition into real and imaginary parts. This decomposition corresponds precisely to the transseries decomposition of the integral. Since the formal power series of the integral is real-valued, the associated Borel function is also real on the real axis. By the Schwarz reflection principle, the analytic continuations of this Borel function into the upper and lower half-planes are complex conjugates of each other. Therefore, the real part of the integral is given by the half-sum of the integrals above and below the cut. Since the singularities are poles, this half-sum is equal to the Cauchy principal value (PV) of the integral. For the same reason, any well-behaved \'Ecalle averages coincide with the PV. Therefore the PV integral produces the purely perturbative part of the transseries, and the imaginary part of the integral corresponds to the non-perturbative component of the transseries.

This has the important further consequence that on the Stokes line the transseries decomposition of the integral is in terms of {\it unary} $q$-series and $\qt$-series: i.e., expansions with coefficients  only $0$ or $\pm 1$. In fact, for the examples considered in this paper, the Stokes line transseries decomposition of the integral is expressed in terms of {\it special} unary $q$-series: Ramanujan's unary false theta functions. See Section \ref{sec:false}. The basic structure involves {\it vector-valued} Mordell-Borel integrals \cite{Zwe08,Zag09}, and we find that these decompose into closed orbits under the action of the $SL(2, \mathbb Z)$ generators which probe the natural boundary $|q|=1$. We refer to these basis blocks as \underline{\it $SL(2, \mathbb Z)$ resurgence orbits}, or \underline{\it resurgence orbits} for short.

Since there is no obstacle to unique continuation of the integrals under $t\to -t$, the property of resurgent functions of {\it preservation of relations} suggests that the same cyclic orbit structure is preserved on the other side of the natural boundary. Here we develop a numerical algorithm to impose this structure, which generates predictions for the dual $q$-series expansions. Remarkably, the numerical expansion coefficients are to high precision integer-valued. For known examples of false/mock theta function dualities, this algorithm generates the correct dual $q$-series expansions, but the numerical method extends far beyond these known cases.

In our resurgent approach, the Mordell-Borel integrals are, up to simple changes of variables, the Borel
sums of the asymptotic series and therefore contain all the information about the Borel plane. As such they are the fundamental objects, rather than the $q$-series. 
This change of perspective means that "mock modular" identities of the form
\begin{eqnarray}
  q^{-\alpha_1} \,Y_1(q)  = \sqrt{\frac{\pi}{t}} 
    \qt^{\, -\alpha_2}\, Y_2(\qt) + J(t),
    \label{eq:hyy1}
\end{eqnarray}
where the integral $J(t)$ is viewed as an "obstruction to modularity", are instead viewed as a "transseries decomposition" of the Borel integral $J(t)$:
\begin{eqnarray}
    J(t) = q^{-\alpha_1} \,Y_1(q)-\sqrt{\frac{\pi}{t}}\, 
    \qt^{\, -\alpha_2}\, Y_2(\qt).
    \label{eq:hyy2}
\end{eqnarray}
By construction, $Y_1(q)$ and $Y_2(q)$ are also resurgent. Therefore, since this decomposition is uniquely determined on the Stokes line ($t<0$), when they cross the natural boundary they transform into the corresponding transseries for $J(t)$. 
In particular, in \eqref{eq:hyy1}-\eqref{eq:hyy2} the exponents $\alpha_1$ and $\alpha_2$, and the $\sqrt{\frac{\pi}{t}}$ factor, must be preserved under the crossing of the boundary. This is indeed a feature of all known examples, and we argue here that it can be used as a guiding principle for crossing the boundary, and also implemented numerically.

\subsection{Summary of Main Results}
\label{sec:summary}

In this paper we consider two natural decompositions into {\it resurgence orbits} formed from vectors of Mordell-Borel integrals. This generalizes and significantly extends the results in \cite{CDGG}. These resurgence orbits are characterized by a positive integer $p$, and are defined as follows:
\begin{itemize}
    \item Those associated with {\it individual} false theta functions $\Psi_p^{(a)}(q)$, with $p$ an odd integer and $a=1, 2, \dots, (p-1)$: see Section \ref{sec:class1}.
    \item Those associated with {\it special linear combinations of} four false theta functions $\Psi_p^{(a)}(q)$, arising in non-perturbative completion of complex Chern-Simons theory on Brieskorn spheres: see Section \ref{sec:class2}.
\end{itemize}
The low degree examples of each orbit agree with known duality identities relating mock theta functions and false theta functions, and we present a rigorous systematic way to extend them further to higher values of $p$, with the only apparent limitation being the amount of computational resources.

These decompositions have the following properties within each orbit:
\begin{enumerate}
\item
Within each orbit, the Mordell-Borel integrals $J_j(t)$ close under the modular transformation: $t\to \frac{\pi^2}{t}$.  See Section \ref{sec:both-asymptotic}.
\item
The expansions of $J_j(t)$ at  small $t$ and large $t$ are {\bf both} asymptotic, with factorially divergent coefficients. Moreover, the expansion coefficients at small and large $t$ are explicitly related to one another in a simple way: each encodes the other. See Section \ref{sec:small-large}.
\item
When analytically continued from $t>0$ to the Stokes line $t<0$, these Mordell-Borel integrals $J_j(t)$ split {\bf  uniquely} into a real part  in terms of unary false theta functions in the variable $q=e^{-t}$, each with an overall prefactor (a rational power of $q$), together with an imaginary part that is a linear combination of unary false theta functions in the variable $\tilde{q}=e^{-\pi^2/t}$, each with an overall prefactor (a rational power of $\tilde{q}$). This algebraic decomposition is rigid, since it is precisely the unique transseries decomposition of the integral, and furthermore is completely determined by any one of the integrals via parameter transformations. 
\item
Since the Mordell-Borel integrals $J_j(t)$ are analytic in $t$ along the half-circle of rotation, there is no obstruction to analytically continuing them across the boundary ${\rm Re}(t)=0$ (i.e. $|q|=1$). Moreover, by the {\it preservation of relations} for resurgent functions \cite{sauzin}, when continued back to $t>0$ the algebraic decomposition structure from the unary side must be preserved.    
Therefore, the problem of continuing the unary $q$-series and $\tilde{q}$-series across the natural boundary,  $|q|=1$,  becomes that of finding on the other side {\it the same algebraic decomposition} of the Mordell-Borel integrals $J_j(t)$. See Sections \ref{sec:preservation1} and \ref{sec:preservation2}.

\item Unique continuation is a fundamental concept in analysis, governed by two principal mechanisms:

\textit{Structural Unique Continuation:} This mechanism relies on a governing equation, such as an underlying ODE or PDE, to enforce the unique continuation property.

\textit{Intrinsic Unique Continuation:} This mechanism arises from intrinsic properties of the function itself, without reliance on external equations. It can occur in two ways:
   \begin{enumerate}
      \item[(a)] The Carleman condition provides a necessary and sufficient criterion for quasi-analyticity \cite{rudin}. It constrains the growth of the Taylor coefficients of a $C^\infty$ function to a rate slightly faster than geometric, ensuring the function is uniquely determined throughout its domain.
      \item[(b)] {\it Resurgence at a point} combined with its property of the {\it preservation of  functional relations} can also ensure unique continuation. This applies even to functions whose coefficients grow factorially, as resurgence encodes structural consistency across different domains.
   \end{enumerate}
In this paper we are in the intrinsic case (b). We have  resurgence as well as functional relations. The latter are the transformations under the action of the generators of $SL(2,\mathbb{Z})$. In the unit disk, in Watson's notation, common in the literature \cite{Watson,GM12}, these are $q\to -q$ and $q\mapsto \tilde{q}$. 

We also note that equations based on these transformations are the hyperbolic analog of difference equations in $\mathbb{R}^n$, since they describe the evolution under the isometries of the ambient space. 

But we are not aware of any global structural equation or external condition that imposes, without the use of resurgence, the Mordell-Borel integral decomposition in $q$-series to have the same powers and coefficients on both sides of the boundary. 
\item 
The resurgence perspective reveals that the standard mock theta functions can be naturally grouped into two kinds of orbits, whose decomposition structure is completely fixed by the unique decompositions of the Mordell-Borel integrals into real and imaginary parts on the unary side $(t<0)$. A strong indication of the rigidity of this structure is the fact that the rational exponents of $q$ and $\tilde{q}$ appearing in the unary decompositions match exactly the rational exponents appearing in the decomposition identities on the non-unary side, and moreover the algebraic mixing matrices are the same on the unary and non-unary sides.
\item
We present a constructive algebraic fitting procedure that can be achieved efficiently in the vicinity of the self-dual point $t=\pi$, where $q=\tilde{q}=e^{-\pi}\approx 0.043214$, and where this duality behavior produces extreme sensitivity. See Section \ref{sec:algorithm}.
\item
In the cases where the vector-valued Mordell-Borel integrals $J_j(t)$ correspond to known classical results from the Ramanujan-Watson theory of mock theta functions \cite{Ram,Watson,GM12}, our numerical fitting procedure discovers the same identities. The first two examples of the first resurgence orbit correspond to what are conventionally referred to as order 3 and order 10 mock theta functions, and the first two examples of the second resurgence orbit correspond to the conventional order 5 and order 7 mock theta functions. See Sections \ref{sec:p3}-\ref{sec:p5}, and \ref{sec:mock5}-\ref{sec:mock7}. 
\item
Our resurgent continuation method can be systematically applied to higher order cases of each kind of resurgence orbit,  going well beyond established mock theta function identities. For the first resurgence orbits, the duals of pure false theta functions $\Psi_p^{(a)}(q)$, we present new results for the duals for odd integers $p$ up to $p=19$, and with all $a=1, 2,  \dots, (p-1)$. See Section \ref{sec:results-p-class}, and Appendices \ref{sec:appendix-results1}. For the second resurgence orbits, associated with Chern-Simons theory on Brieskorn sphere manifolds $\Sigma(2,3, 6k\pm 1)$, the duals are known for $\Sigma(2,3,5)$ and $\Sigma(2,3,7)$, but we present new results up to $\Sigma(2,3, 35)$. These constitute new results for Chern-Simons theory on the \textbf{orientation reversed} manifolds $\overline{\Sigma(2,3,6k\pm1)}$. See Section \ref{sec:results-mock-class}, and Appendix \ref{sec:appendix-results2}.
\item
We do not {\it impose} the condition that the fitting procedure produces {\bf integer valued} expansion coefficients for the resulting $q$-series. However, the fitting procedures result in integer coefficients to high precision. This is consistent with the physical Chern-Simons theory interpretation of these decompositions as partition functions for BPS states \cite{GMP,CCFGH,CDGG}. 
\item
In a related paper \cite{ceff2} we derive analytic estimates for the growth rate of the integer-valued coefficients of the dual $q$-series, including subleading corrections. These growth rate expressions follow from the algebraic structure of the Mordell-Borel integrals on the unary side, and serve as a consistency check of our numerical results.  The exponent $c$ in the basic growth rate $\exp[2\pi\sqrt{c\, n}]/\sqrt{n}$ has a direct physical interpretation in terms of the central charge in associated 3 dimensional $\mathcal N=2$ superconformal field theories \cite{CDGG,GJ,ceff2}.
\item
In \cite{CD25}, it is proved for the first example of each resurgence orbit that the decomposition into $q$-series and $\qt$-series on the non-unary side is unique when the complete $SL(2, \mathbb Z)$ algebraic structure from the unary side is enforced on the non-unary side. Physically this amounts to imposing the algebraic structure that arises from the physical operations probing the transverse and longitudinal shifts at the natural boundary $|q|=1$.
\item
The only apparent obstacles to our numerical procedure at even higher orders appear to stem from computational complexity.

\end{enumerate}

\section{Basis of Mordell-Borel Integrals}
\label{sec:mordell-borel}

We define the following notation for the Mordell-Borel integrals which form the building blocks for our analysis of mock theta functions \cite{GM12} which arise in the study of complex Chern-Simons theory \cite{GMP}:
\begin{eqnarray}
JS_{(p,a)}(t)&:=&\frac{1}{t} \int_0^\infty du\, e^{-p u^2/t} \, \frac{\sinh[(p-a)u]}{\sinh[p u]}, 
\label{eq:js}\\
JC_{(p,a)}(t)&:=&\frac{1}{t} \int_0^\infty du\, e^{-p u^2/t} \, \frac{\cosh[(p-a)u]}{\cosh[p u]}. 
\label{eq:jc}
\end{eqnarray}
At this point we only require $p>0$ and $a>0$, but we will later specialize to integer values of $p$, and integer and half-integer values of $a$, with $0<a\leq p$. In Section \ref{sec:class1} we further specialize to consider $p$ to be an odd integer, while in Section \ref{sec:class2} we consider $p$ odd, with $p=6k\pm 1$,  $k\in \mathbb Z^+$. 
In Appendix \ref{sec:other-brieskorn} we present results for some other Brieskorn sphere manifolds, but we defer a comprehensive analysis of irreducible structures that arise for other parametric choices for $p$ to a future publication.

Using the basic Fourier transform identity
\begin{eqnarray}
e^{-u^2/t}=\sqrt{\frac{t}{\pi}}\int_0^\infty dv\, e^{-v^2 t/4}\, \cos(v u), 
\label{eq:fourier}
\end{eqnarray}
we obtain the dual Borel integral representations \cite{GM12}:
\begin{eqnarray}
{JS}_{(p,a)}(t)&=& \frac{1}{\sqrt{4p\pi t}} 
   \int_0^\infty dv  \, e^{-p v^2 t /(4\pi^2)} \frac{\sin\left(\frac{a\pi}{p}\right) }{\cosh[v] -\cos\left(\frac{a \pi}{p}\right)},
   \label{eq:jsd}
   \\
   {JC}_{(p,a)}(t)&=& \frac{1}{\sqrt{4p\pi t}} 
   \int_0^\infty dv  \, e^{-p v^2 t /(4\pi^2)} \frac{2 \sin\left(\frac{a\pi}{2p}\right) \cosh[v/2]}{\cosh[v] -\cos\left(\frac{a \pi}{p}\right)}.
   \label{eq:jcd}
\end{eqnarray}
\noindent{\bf Remarks}
\begin{enumerate}
    \item In these defining expressions \eqref{eq:js}-\eqref{eq:jcd} we take $t>0$, but a key point of our analysis will be to analytically continue these integral representations to $t<0$. See Section \ref{sec:false} below.
    \item Notice the $t\to\frac{4\pi^2}{t}$ transformation in the exponent of the Gaussian factors in \eqref{eq:js}-\eqref{eq:jc} compared to \eqref{eq:jsd}-\eqref{eq:jcd}.  This will be important. See Sections \ref{sec:both-asymptotic} and \ref{sec:small-large}.
\end{enumerate}

\subsection{Both Small $t$ and Large $t$ Expansions are Divergent}
\label{sec:both-asymptotic}
An interesting feature of the Mordell-Borel integrals \eqref{eq:js}-\eqref{eq:jc} is that {\it both} the small $t$ and large $t$ expansions are factorially divergent. In particular, this shows that both limits are singular, in the sense that $t=0$ and $t=\infty$ are both irregular singular points. Moreover, the small $t$ and large $t$ expansions are directly related to one another, in the sense that the expansion coefficients of one encode those of the other.

The small $t$ expansions can be generated from the following small $u$ expansion identities of the Borel kernels in terms of Hurwitz zeta functions:
\begin{eqnarray}
\frac{\sinh[(p-a)u]}{\sinh[p u]}&=& \frac{1}{2}\sum_{n=0}^\infty \frac{\left(4 p u\right)^{2n+1}}{\Gamma(2n+1)} \left[ \zeta_{-2n}\left(\frac{a}{4p}\right) 
-\zeta_{-2n}\left(1-\frac{a}{4p}\right)
\right.
\nonumber\\
&& \hskip 2cm \left.
+\zeta_{-2n}\left(\frac{1}{2}+\frac{a}{4p}\right)
-\zeta_{-2n}\left(\frac{1}{2}-\frac{a}{4p}\right)
\right], 
\label{eq:sinh-bernoulli}\\
\frac{\cosh[(p-a)u]}{\cosh[p u]}&=& 
\frac{1}{2}\sum_{n=0}^\infty \frac{\left(4 p u\right)^{2n+1}}{\Gamma(2n+1)} \left[ \zeta_{-2n}\left(\frac{a}{4p}\right) 
+\zeta_{-2n}\left(1-\frac{a}{4p}\right) \right.
\nonumber\\
&& \hskip 2cm \left.
-\zeta_{-2n}\left(\frac{1}{2}+\frac{a}{4p}\right)
-\zeta_{-2n}\left(\frac{1}{2}-\frac{a}{4p}\right)
\right]. 
\label{eq:cosh-euler}
\end{eqnarray}
These expansions can alternatively be expressed in terms of Bernoulli and Euler polynomials, but we have written them as above in order to emphasize the duality between the small $t$ and large $t$ expansions, as explained below.

The large $t$ expansions can be generated from the following small $v$ expansion identities:
\begin{eqnarray}
\frac{\sin\left(\frac{a\pi}{p}\right)}{\cosh[v] -\cos\left(\frac{a \pi}{p}\right)}
&=&
\sum_{n=0}^\infty \frac{\left(v/2\right)^{2n}}{\Gamma(2n+1)}{\Im}\left[{\rm Li}_{-2n}\left(e^{\frac{i \pi a}{2p}}\right)+{\rm Li}_{-2n}\left(-e^{\frac{i \pi a}{2p}}\right)\right],   \label{eq:jsd2}
   \\
    \frac{2 \sin\left(\frac{a\pi}{2p}\right) \cosh[v/2]}{\cosh[v] -\cos\left(\frac{a \pi}{p}\right)}
    &=&
\sum_{n=0}^\infty \frac{\left(v/2\right)^{2n}}{\Gamma(2n+1)}{\Im}\left[{\rm Li}_{-2n}\left(e^{\frac{i \pi a}{2p}}\right)-{\rm Li}_{-2n}\left(-e^{\frac{i \pi a}{2p}}\right)\right].  
   \label{eq:jcd2}
\end{eqnarray}

Using these identities we find the small $t$ expansions ($t\to 0^+$):
\begin{eqnarray}
JS_{(p, a)}(t) &\sim&
\frac{1}{\pi} \frac{1}{\sqrt{p\, t}} 
 \sum _{n=0}^{\infty }(-1)^n \left(\frac{4p \, t}{\pi^2}\right)^n \,  \Gamma\left(n+\frac{1}{2}\right) 
 {\rm Im}\left[ {\rm Li}_{2n+1}\left(e^{\frac{i \pi a}{2p}}\right)
 +{\rm Li}_{2n+1}\left(-e^{\frac{i \pi a}{2p}}\right)\right],
\nonumber\\
   \label{eq:js-small-t}
   \\
   JC_{(p, a)}(t) &\sim&
\frac{1}{\pi} \frac{1}{\sqrt{p\, t}} 
 \sum _{n=0}^{\infty }(-1)^n \left(\frac{4 p \, t}{\pi^2}\right)^n \,  \Gamma\left(n+\frac{1}{2}\right) 
 {\rm Im} \left[{\rm Li}_{2n+1}\left(e^{\frac{i \pi a}{2p}}\right)- {\rm Li}_{2n+1}\left(-e^{\frac{i \pi a}{2p}}\right)\right].
\nonumber\\
\label{eq:jc-small-t}
\end{eqnarray}
At large $t$ we obtain the following asymptotic expansions ($t\to +\infty$):
   \begin{eqnarray}
JS_{(p, a)}(t) &\sim&
\frac{1}{4 p\, t \sqrt{\pi} } 
 \sum _{n=0}^{\infty }(-1)^n  \frac{1}{(4p \, t)^n} \,  \Gamma\left(n+\frac{1}{2}\right)
  \left[\zeta_{2n+1}\left(\frac{a}{4p}\right) - \zeta_{2n+1}\left(1-\frac{a}{4p}\right)    \right.
  \nonumber\\
  &&\hskip 3cm \left. -\zeta_{2n+1}\left(\frac{1}{2}- \frac{a}{4p}\right) + \zeta_{2n+1}\left(\frac{1}{2}+ \frac{a}{4p}\right)  \right],
   \label{eq:js-large-t}
   \\
   JC_{(p, a)}(t) &\sim&
\frac{1}{4 p\, t \sqrt{\pi} } 
 \sum _{n=0}^{\infty }(-1)^n \frac{1}{(4 p \, t)^n}  \,  \Gamma\left(n+\frac{1}{2}\right)
  \left[\zeta_{2n+1}\left(\frac{a}{4p}\right) - \zeta_{2n+1}\left(1-\frac{a}{4p}\right)    \right.
  \nonumber\\
  &&\hskip 3cm \left. +\zeta_{2n+1}\left(\frac{1}{2}- \frac{a}{4p}\right) - \zeta_{2n+1}\left(\frac{1}{2}+ \frac{a}{4p}\right)  \right].
   \label{eq:jc-large-t}
   \end{eqnarray}
These small $t$ and large $t$ expansions, \eqref{eq:js-small-t}-\eqref{eq:jc-small-t} and \eqref{eq:js-large-t}-\eqref{eq:jc-large-t}, follow directly from the expansions \eqref{eq:sinh-bernoulli}-\eqref{eq:cosh-euler}
and \eqref{eq:jsd2}-\eqref{eq:jcd2},
combined with the Joncqui\`ere identities:
   \begin{eqnarray}
       {\rm Im}\, {\rm Li}_{2n+1}\left(e^{i \pi \frac{a}{2p}}\right)&=& \frac{(-1)^n}{4} \frac{(2\pi)^{2n+1}}{\Gamma(2n+1)}\left[\zeta_{-2n}\left(\frac{a}{4p}\right)-\zeta_{-2n}\left(1-\frac{a}{4p}\right)\right],
       \label{eq:jonc1}
   \end{eqnarray}
   \begin{eqnarray}
 \left[\zeta_{2n+1}\left(\frac{a}{4p}\right)-\zeta_{2n+1}\left(1-\frac{a}{4p}\right)\right]
 &=& (-1)^n \frac{(2\pi)^{2n+1}}{\Gamma(2n+1)}
       {\rm Im}\, {\rm Li}_{-2n}\left(e^{i \pi \frac{a}{2p}}\right).
       \label{eq:jonc2}
   \end{eqnarray}

\vspace{5mm}

\noindent{\bf Remarks:}
\begin{enumerate}
   \item 
   We have expressed these small and large $t$ expansions, \eqref{eq:js-small-t}-\eqref{eq:jc-small-t} and \eqref{eq:js-large-t}-\eqref{eq:jc-large-t},  in this form to emphasize that in each case the leading growth of the coefficients at large order, $n\to +\infty$,  has the characteristic form of factorial-over-power, $(-1)^n A^n \Gamma\left(n+\frac{1}{2}\right)$, and with alternating sign, owing to the large order expressions, for $n\to \infty$:
   \begin{eqnarray}
   {\rm Im}\, {\rm Li}_{2n+1}\left(\pm e^{\frac{i \pi a}{2p}}\right)&\sim& \pm \sin \left(\frac{a\pi }{2p}\right)+
   \frac{1}{2^{2n+1}} \sin \left(\frac{a\pi }{p}\right)+\dots,
    \\
     \left[\zeta_{2n+1}\left(\frac{a}{4p}\right)-\zeta_{2n+1}\left(1-\frac{a}{4p}\right)\right] &\sim& \left(\frac{4p}{a}\right)^{2n+1}\left(1-\left(\frac{\frac{a}{4p}}{1-\frac{a}{4p}}\right)^{2n+1} +\dots \right). 
\end{eqnarray}

\item 
The alternating-sign factorial growth of the asymptotic expansions in \eqref{eq:js-small-t}-\eqref{eq:jc-small-t} and \eqref{eq:js-large-t}-\eqref{eq:jc-large-t} tells us that the Stokes line for the Mordell-Borel integrals is the line $t<0$, and therefore this is the most natural and most important place to begin analysis of the asymptotics of the Mordell-Borel integrals. See Section \ref{sec:false}.

\item 
The  small $t$ and large $t$ asymptotic expansions \eqref{eq:js-small-t}-\eqref{eq:jc-large-t} have also been expressed in a form that illustrates the duality between the small $t$ and large $t$ asymptotic expansions. This implies that the small $t$ expansion encodes the large $t$ expansion and {\it vice versa}. This duality between the small $t$ and large $t$ expansions is reflected in the interchange of polylogarithm and zeta functions, mediated by the Joncqui\`ere identities \eqref{eq:jonc1}-\eqref{eq:jonc2}.

\item 
When $p$ and $a$ take integer values, there is an even more explicit realization of this duality between the small and large $t$ expansions. See Section \ref{sec:small-large} below.

\item 
The fact that both the small and large $t$ expansions are asymptotic implies that the transseries of the Mordell-Borel integrals should contain expansions in exponential terms of the form 
\begin{eqnarray}
    q:=e^{-t} \qquad {\rm and }\qquad \tilde{q}:=e^{-\pi^2/t}
    \label{eq:qs}
\end{eqnarray}
And since the small $t$ and large $t$ asymptotic expansions are related, we expect that in the transseries expansions the $q$-series and $\tilde{q}$-series should also be closely related. This will be important in our analysis.

\end{enumerate}

\subsection{Small $t$ to Large $t$ Duality of Mordell-Borel Integrals}
\label{sec:small-large}

So far, all identities and expansions have been derived without restricting $p$ or $a$ to integer or half-integer values.
However, for integer-valued $p$, and $a$ integer or half-integer, the Mordell-Borel integrals have additional symmetry properties, arising from the action of the generators of $SL(2, \mathbb Z)$, which probe the natural boundary $|q|=1$.

For example, when $p$ and $a$ are integers (with $0< a\leq p$),  under the duality transformation $t\to 4\pi^2/t$ the Mordell-Borel integrals in \eqref{eq:js}-\eqref{eq:jc} [and equivalently in \eqref{eq:jsd}-\eqref{eq:jcd}] yield a {\bf finite} linear combination of exactly the same Mordell-Borel building block integrals:
\begin{eqnarray}
 JS_{(p,a)}(t)&=& \left(\frac{2\pi}{t}\right)^{3/2}\, \sum_{b=1}^{p}   \sqrt{\frac{2}{p}}\, \sin\left(\frac{a b \pi}{p}\right)JS_{(p,b)}\left(\frac{4\pi^2}{t}\right),
\label{eq:js-4pi}
\\
 JC_{(p,a)}(t)&=& \left(\frac{2\pi}{t}\right)^{3/2}\, \sum_{b=1}^{p}   \sqrt{\frac{2}{p}}\, \sin\left(\frac{a \left(b-\frac{1}{2}\right) \pi}{p}\right)JS_{\left(p,b-\frac{1}{2}\right)}\left(\frac{4\pi^2}{t}\right).
\label{eq:jcs-4pi}
\end{eqnarray}
These duality transformations involve a discrete Fourier transform with an ``$S$-matrix'' kernel:
\begin{eqnarray}
    \mathcal S_{ab}=\sqrt{\frac{2}{p}} \sin\left(\frac{a b \pi}{p}\right).
    \label{eq:mixing}
\end{eqnarray}
The matrix $\mathcal S$ has eigenvalues $\pm 1$ (and also 0 if we include $a=p$ or $b=p$). $S$ is equal to the modular $S$-matrix in a rational CFT (WZW model) and, as argued in \cite{GPV}, determines the transformation of flat connections under the action of the $SL(2,\mathbb{Z})$ modular group. It also has an interpretation in the 3d-3d correspondence, namely in terms of the category of line operators MTC$[M_3]$ in 3d $\mathcal{N}=2$ theory \cite{CCFGH}.

There are similar identities when the a index is half-integer:
\begin{eqnarray}
 JS_{(p,a-\frac{1}{2})}(t)&=& \left(\frac{2\pi}{t}\right)^{3/2}\left\{ \frac{\cos(a \pi)}{\sqrt{2p}} JC_{(p,p)}\left(\frac{4\pi^2}{t}\right)
 \right.
 \nonumber\\
 && \left. 
 +\sum_{b=1}^{p}  \sqrt{\frac{2}{p}}\, \sin\left(\frac{\left(a-\frac{1}{2}\right) b \pi}{p}\right)JC_{(p, b)}\left(\frac{4\pi^2}{t}\right)   \right\},
\label{eq:jsc-4pi}
\\
JC_{(p,a-\frac{1}{2})}(t)&=& \left(\frac{2\pi}{t}\right)^{3/2}\, \sum_{b=1}^{p}   \sqrt{\frac{2}{p}}\, \sin\left(\frac{\left(a-\frac{1}{2}\right) \left(b-\frac{1}{2}\right) \pi}{p}\right)JC_{(p, b-\frac{1}{2})}\left(\frac{4\pi^2}{t}\right).
\label{eq:jc-4pi}
\end{eqnarray}
These duality identities \eqref{eq:js-4pi}-\eqref{eq:jc-4pi} follow from the following basic trigonometric identities, valid for $p$ and $a$ positive integers with $0<a\leq p$: 
\begin{eqnarray}
\frac{\sinh[(p-a)u]}{\sinh[p u]}&=&\frac{1}{p}\sum_{b=1}^{p} \sin\left( \frac{a b \pi}{p}\right)\,
 \frac{\sin\left(\frac{b\pi}{p}\right)}{\cosh[u]-\cos\left(\frac{b\pi}{p}\right)},
 \label{eq:js-identity1}
\\
 \frac{\sin\left(\frac{a\pi}{p}\right)}{\cosh[u]-\cos\left(\frac{a\pi}{p}\right)}
 &=& 2\sum_{b=1}^{p} \sin\left( \frac{a b \pi}{p}\right)\,\frac{\sinh[(p-b)u]}{\sinh[p u]},
 \label{eq:js-identity2}
 \\
 \frac{\cosh[(p-a)u]}{\cosh[p u]}&=&\frac{1}{p}\sum_{b=1}^{p} \sin\left( \frac{a \left(b-\frac{1}{2}\right) \pi}{p}\right)\,
 \frac{\sin\left(\left(b-\frac{1}{2}\right)\frac{\pi}{p}\right)}{\cosh[u]-\cos\left(\frac{\left(b-\frac{1}{2}\right)\pi}{p}\right)},
 \label{eq:jc-identity1}
\\
 \frac{\sin\left(\frac{\left(a-\frac{1}{2}\right)\pi}{p}\right)}{\cosh[u]-\cos\left(\frac{\left(a-\frac{1}{2}\right)\pi}{p}\right)}
 &=&\frac{\cos(a\, \pi)} {\cosh[p\, u]}+ 2\sum_{b=1}^{p} \sin\left( \frac{\left(a-\frac{1}{2}\right) b \pi}{p}\right)\,\frac{\cosh[(p-b)u]}{\cosh[p u]}.
 \label{eq:jc-identity2}
 \end{eqnarray}

 \noindent{\bf Remarks:}
\begin{enumerate}

\item
The small $t$ to large $t$ duality identities in \eqref{eq:js-4pi}-\eqref{eq:jc-4pi} are fundamental properties of the Mordell-Borel integrals, and this places rigid constraints on the structure of the transseries expansions of the integrals.

\item 
Since the Mordell-Borel integrals involve meromorphic functions, there are no fluctuations about any of the Borel poles. However, because of the duality between small and large $t$, there are explicit relations between the associated Borel pole structures. 
\item 
The small to large duality means that the small $t$ expansion encodes the large $t$ expansion, and {\it vice versa}. Ultimately this follows simply from the Fourier transform duality in \eqref{eq:js}-\eqref{eq:jcd}, and is particularly explicit when $p$ and $a$ are integers. To illustrate this, consider the integrals $JS_{(p,a)}(t)$ with $p$ odd and $a$ even.\footnote{The details depend on the parity of $p$ and $a$, but the other cases work similarly.}\\

Define the coefficients multiplying the universal $(-1)^n \Gamma\left(n+\frac{1}{2}\right)$ behavior in the small $t$ and large $t$ expansions \eqref{eq:js-small-t} and \eqref{eq:js-large-t} as
\begin{eqnarray}
c_n^{\rm small}(p, a)&:=&  \frac{1}{\pi} \frac{1}{\sqrt{p}} \left(\frac{4p}{\pi^2}\right)^n
 {\rm Im}\left[ {\rm Li}_{2n+1}\left(e^{\frac{i \pi a}{2p}}\right)
 +{\rm Li}_{2n+1}\left(-e^{\frac{i \pi a}{2p}}\right)\right],
\end{eqnarray}
\begin{eqnarray}
    c_n^{\rm large}(p, a)&:=& \frac{1}{\sqrt{\pi} } 
  \frac{1}{(4p)^{n+1}} \left[\zeta_{2n+1}\left(\frac{a}{4p}\right) - \zeta_{2n+1}\left(1-\frac{a}{4p}\right)    \right.
  \nonumber\\
  &&\hskip 3cm \left. -\zeta_{2n+1}\left(\frac{1}{2}- \frac{a}{4p}\right) + \zeta_{2n+1}\left(\frac{1}{2}+ \frac{a}{4p}\right)  \right].
\end{eqnarray}
Then $c_n^{\rm small}(p, a)$ can be expressed in terms of Bernoulli polynomials:
\begin{eqnarray}
c_n^{\rm small}(p, a)&=& \frac{1}{\pi} \frac{1}{\sqrt{p}} \left(\frac{p}{\pi^2}\right)^n \sum_{k=1}^\infty \frac{\sin\left(\frac{k \pi a}{p}\right)}{k^{2n+1}}.
\end{eqnarray}
This can further be expressed as a finite Fourier expansion.
When $p$ is an integer and $a$ is an even integer, we obtain the following finite Fourier sum of Hurwitz zeta functions:
\begin{eqnarray}
c_n^{\rm small}(p, a)=\frac{1}{\pi} \frac{1}{\sqrt{p}} \left(\frac{p}{\pi^2}\right)^n \frac{1}{p^{2n+1}}\sum_{r=1}^p \sin\left(\frac{ a r \pi }{p}\right) \zeta_{2n+1}\left(\frac{r}{p}\right). 
\end{eqnarray}
For $p$ odd (and $a$ even) this finite sum can be rewritten as
\begin{eqnarray}
c_n^{\rm small}(p, a)&=& \frac{2}{\pi \sqrt{p}} \frac{1}{(4p)^{n+1}} \frac{1}{\pi^{2n}} 
\sum_{b=2, 4, 6, \dots}^{p-1} \sin\left(\frac{ a b \pi }{2p}\right) 
\left[\zeta_{2n+1}\left(\frac{b}{4p}\right) - \zeta_{2n+1}\left(1-\frac{b}{4p}\right)    \right.
  \nonumber\\
  &&\hskip 3cm \left. -\zeta_{2n+1}\left(\frac{1}{2}- \frac{b}{4p}\right) + \zeta_{2n+1}\left(\frac{1}{2}+ \frac{b}{4p}\right)\right]
  \\
  &=& \frac{2}{\sqrt{p\, \pi}} \frac{1}{\pi^{2n}} 
\sum_{b=2, 4, 6, \dots}^{p-1} \sin\left(\frac{ a b \pi }{2p}\right)\, c_n^{\rm large}(p, b).
\end{eqnarray}
Thus, the small $t$ coefficients (with $a$ even) are expressed as a finite linear combination of the large $t$ coefficients (again with $b$ even). This argument can be inverted, so we see that the small $t$ coefficients determine the large $t$ coefficients, and {\it vice versa}. Note that this is an identity for all $n$, not just an asymptotic large order $n$ relation.

\end{enumerate}

\section{False Theta Function Decomposition at the Stokes Line}
\label{sec:false}

The Stokes line $t<0$ is the richest source of information about the Mordell-Borel integrals $JS_{(p, a)}(t)$ and $JC_{(p, a)}(t)$.
These Mordell-Borel integrals are initially defined in (\ref{eq:js})-(\ref{eq:jc})  and  (\ref{eq:jsd})-(\ref{eq:jcd}) for $t>0$, 
but can be analytically continued by rotating the contour of integration. The following expressions are valid for $t<0$ and $\epsilon\to 0^+$:
\begin{eqnarray}
JS_{(p,a)}^{(\pm)}(t)&:=&\mp i\, e^{\pm i\epsilon} \int_0^\infty du\, e^{p t (e^{\pm i\epsilon} u)^2} \, \frac{\sin[(p - a) (e^{\pm i\epsilon} u)\, t]}{\sin[p \, (e^{\pm i\epsilon} u)\, t]}, 
\label{eq:jspm}
\\
JC_{(p,a)}^{(\pm)}(t)&:=&\mp i\, e^{\pm i\epsilon} \int_0^\infty du\, e^{p t (e^{\pm i\epsilon} u)^2} \, \frac{\cos[(p - a) (e^{\pm i\epsilon} u)\, t]}{\cos[p \, (e^{\pm i\epsilon} u)\, t]}. 
\label{eq:jcpm}
\end{eqnarray}
Similarly for the dual integrals in \eqref{eq:jsd}-\eqref{eq:jcd} we have for $t<0$ and $\epsilon\to 0^+$:
\begin{eqnarray}
JS_{(p,a)}^{(\pm)}(t)&:=&   \frac{e^{\mp i \epsilon}\sin{\left(\frac{a\pi}{p}\right)}}{2\sqrt{-p \pi t}} \int_0^\infty dv\,e^{p (e^{\mp i \epsilon}v)^2 t/(4\pi^2)}\frac{1}{\cos(e^{\mp i \epsilon}v)-\cos{\left(a\pi/p\right)}},
\label{eq:jsdpm}
\\
JC_{(p,a)}^{(\pm)}(t)&:=& \frac{e^{\mp i \epsilon}\sin{\left(\frac{a\pi}{2p}\right)}}{\sqrt{-p \pi t}} \int_0^\infty dv\,e^{p (e^{\mp i \epsilon}v)^2 t/(4\pi^2)}\frac{\cos(e^{\mp i \epsilon}v/2)}{\cos(e^{\mp i \epsilon}v)-\cos{\left(a\pi/p\right)}}.
\label{eq:jcdpm}
\end{eqnarray}
Note that the sinh (cosh) functions are replaced by sin (cos) functions. Therefore, we find that under analytic continuation from $t>0$ to  $t<0$ the Mordell-Borel  integrals (\ref{eq:js})-\eqref{eq:jc}, and (\ref{eq:jsd})-(\ref{eq:jcd}), acquire both a real and imaginary part.
This decomposition is unique and is, in fact, their resummed transseries decomposition. Recall item (5) in the Summary of Main Results in Section \ref{sec:summary}. 

For $t<0$, the real part follows from a straightforward residue analysis of the dual integrals \eqref{eq:jsd}-\eqref{eq:jcd}, while the imaginary part follows from a straightforward residue analysis of the original integrals \eqref{eq:js}-\eqref{eq:jc} (see \cite{CDGG}). We then have the following expressions for the real and imaginary parts, obtained by averaging the rotated integrals on both sides of the Stokes line:
\begin{equation}
\operatorname{Re}\left[\sqrt{\frac{4 p t}{\pi}} JS_{(p, a)}(t)\right]=\sqrt{-\frac{4 p t}{\pi}} \frac{i}{2}\left(JS_{(p, a)}^{(+)}(t)-JS_{(p, a)}^{(-)}(t)\right),
\label{eq:real-js}
\end{equation}

\begin{equation}
\operatorname{Im}\left[\sqrt{\frac{4 p t}{\pi}} JS_{(p, a)}(t)\right]=\sqrt{-\frac{4 p t}{\pi}} \frac{1}{2}\left(JS_{(p, a)}^{(+)}(t)+JS_{(p, a)}^{(-)}(t)\right),
\label{eq:imag-js}
\end{equation}

\begin{equation}
\operatorname{Re}\left[\sqrt{\frac{4 p t}{\pi}} JC_{(p, a)}(t)\right]= \sqrt{-\frac{4 p t}{\pi}} \frac{i}{2}\left(JC_{(p, a)}^{(+)}(t)-JC_{(p, a)}^{(-)}(t)\right),  
\label{eq:real-jc}
\end{equation}

\begin{equation}
\operatorname{Im}\left[\sqrt{\frac{4 p t}{\pi}} JC_{(p, a)}(t)\right]=\sqrt{-\frac{4 p t}{\pi}} \frac{1}{2}\left(JC_{(p, a)}^{(+)}(t)+JC_{(p, a)}^{(-)}(t)\right). 
\label{eq:imag-jc}
\end{equation}

We specialize here to $p\in \mathbb Z$, $p\geq 2$, and  $a\in \mathbb Z$, $0<a\leq p$. Then we discover that the expansions on the Stokes line naturally arise as unary series (i.e. series with coefficients being only $\pm 1$ or $0$) in both $1/q$ and $1/\tilde{q}$, as defined in \eqref{eq:qs}. Note that these unary expansions arise as expansions in $1/q$ and $1/\tilde{q}$, which are both $<1$ in the $t<0$ region:
\begin{eqnarray}
    {\rm non-unary}:\qquad t>0 \quad &\Rightarrow& \quad q=e^{-t}<1 \quad, \quad \qt=e^{-\pi^2/t}<1 \hspace{2mm} ,
    \label{eq:tpos}
    \\
    {\rm unary}:\qquad t<0 \quad &\Rightarrow& \quad q=e^{-t}>1 \quad, \quad \qt=e^{-\pi^2/t}>1 \hspace{2mm}.
    \label{eq:tneg}
\end{eqnarray}
The expansions for $t<0$ are unary because they come from residue expansions of Borel-type integrals involving Borel kernels that are rational functions of exponentials \cite{Watson}. For this reason we will refer to the $t<0$ region as the ``unary side'', in contrast with the ``non-unary side'' where $t>0$. 

Furthermore, the decompositions on the unary side of the Mordell-Borel integrals (\ref{eq:js})-\eqref{eq:jc}, and (\ref{eq:jsd})-(\ref{eq:jcd}), into real and imaginary parts are expressed in terms of linear combinations of false theta functions. We adopt the following notation\footnote{Note that we write $\Psi^{(a)}_p (q)$ rather than $\widetilde{\Psi}^{(a)}_p (q)$, dropping the tilde over $\Psi$, to simplify the notation and to avoid potential confusion with $q$ and $\qt$.} for false theta functions \cite{Bring15,GMP,CCFGH,HLSS22}, for $p, a \in \mathbb Z$:
\begin{equation}
\Psi^{(a)}_p (q) \;  :=  \; \sum_{n=0}^\infty \psi^{(a)}_{2p}(n) q^{\frac{n^2}{4p}} \qquad \in q^\frac{a^2}{4p}\,\mathbb{Z}[[q]]
\label{eq:false}
\end{equation}
where
\begin{eqnarray}
\psi^{(a)}_{2p}(n)  =  \left\{
\begin{array}{cl}
\pm 1, & n\equiv \pm a \quad ({\rm mod}\,\, 2p)\,, \\
0, & \text{otherwise}.
\end{array} \right.
\end{eqnarray}
It is convenient to define a related notation, in which we extract the fractional power of $q$,
\begin{equation}
\Psi^{(a)}_p (q) \;  :=  q^\frac{a^2}{4p}\, \Phi^{(a)}_p (q)
\label{eq:false-phi},
\end{equation}
thereby isolating the unary $q$-series part $\Phi^{(a)}_p (q)$. 
Equivalently, we can write the unary expansions in terms of the Kronecker symbol $\left(\dfrac{*}{*}\right)$:
\begin{equation}
\Psi^{(a)}_p (q) \;  :=  \;  \sum_{n=0}^\infty \left(\frac{-4 p}{n}\right) q^{\frac{n^2}{4 p}}  
\qquad, \quad n \equiv \pm a (\bmod 2 p).
\label{eq:false2}
\end{equation}

\begin{figure}[H]
\centering
\begin{subfigure}{.5\textwidth}
  \centering
  \includegraphics[width=.8\linewidth]{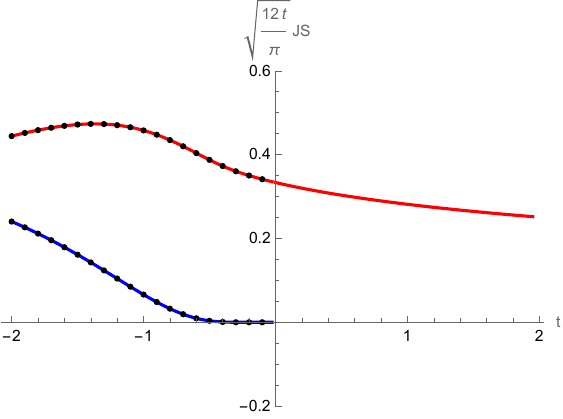}
\end{subfigure}%
\begin{subfigure}{.5\textwidth}
  \centering
  \includegraphics[width=.8\linewidth]{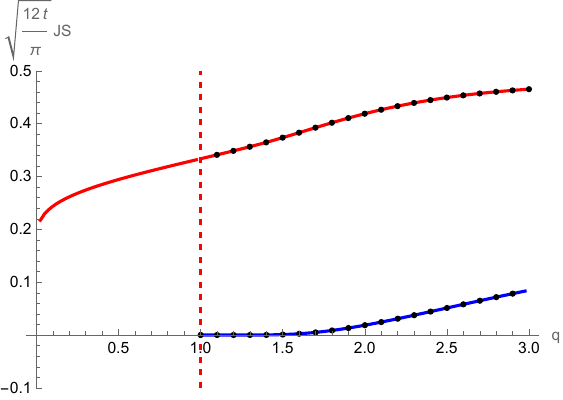}
\end{subfigure}
\caption{Plots of the real and imaginary parts of $\sqrt{12t/\pi}\, JS_{(3,2)}(t)$, as functions of $t$ [left], and as functions of $q=e^{-t}$ [right]. On the non-unary side ($t>0$, $q<1$) the expression is real and shown by a solid red line.  On the unary side ($t<0$, $q>1$) the real and imaginary parts are shown in red and blue lines respectively. The unary side $q$-series on the RHS of expressions \eqref{eq:js-false} for the real and imaginary parts are shown as black dots.}
\label{fig:js32}
\end{figure}

\begin{figure}[H]
\centering
\begin{subfigure}{.5\textwidth}
  \centering
  \includegraphics[width=.8\linewidth]{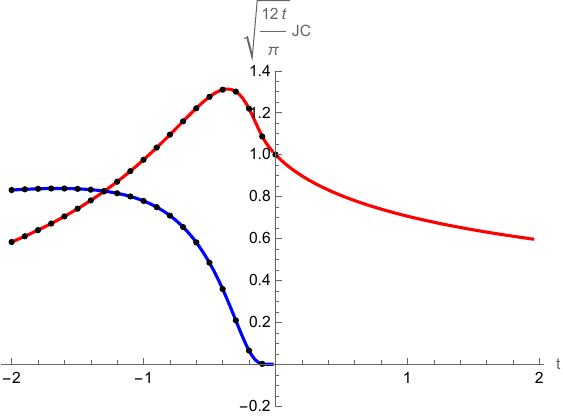}
\end{subfigure}%
\begin{subfigure}{.5\textwidth}
  \centering
  \includegraphics[width=.8\linewidth]{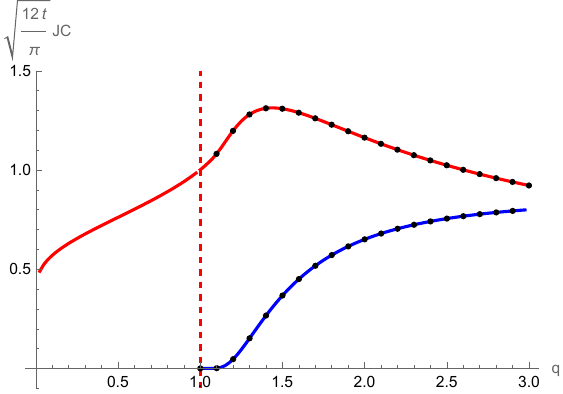}
\end{subfigure}
\caption{
Plots of the real and imaginary parts of $\sqrt{12t/\pi}\, JC_{(3,2)}(t)$, as functions of $t$ (left), and as functions of $q=e^{-t}$ (right). On the non-unary side ($t>0$, $q<1$) the expression is real and shown by a solid red line.  On the unary side ($t<0$, $q>1$) the real and imaginary parts are shown in red and blue lines respectively. The unary side $q$-series on the RHS of expressions \eqref{eq:jc-false} for the real and imaginary parts are shown as black dots.}
\label{fig:jc32}
\end{figure}

These false theta functions are important because the Mordell-Borel integrals decompose on the Stokes line $t<0$ into various combinations (depending on the parity of $p$ and $a$) of false theta functions, involving unary series in terms of $1/q$ and $1/\tilde{q}$. We concentrate here primarily on {\it odd} $p$ values.
For odd $p$, we find the following decompositions of the integrals on the left-hand-side into real and imaginary parts:
\begin{eqnarray}
i \sqrt{\frac{4p|t|}{\pi}}\, JS_{(p, a)}(t) \!\!\! &=& \!\!\! q^{-\frac{a^2}{4p}}\, \Phi_{p}^{(a)}\left(\frac{1}{q}\right) +i \sqrt{\frac{\pi}{|t|}}\sum_{b=1}^{p} \sqrt{\frac{4}{p}} \sin\left(\frac{a b \pi}{p}\right)\,\qt^{\,-\frac{b^2}{p}}\Phi^{(2b)}_{2p}\left(\frac{1}{\qt^{\, 2}}\right),
\label{eq:js-false}
\end{eqnarray}
\begin{eqnarray}
i \sqrt{\frac{4p|t|}{\pi}}\, JC_{(p, a)}(t) \!\!\! &=& \!\!\!
q^{-\frac{a^2}{4p}}\,  \Phi_{2p}^{(a)}\left(\frac{1}{q^2}\right)+q^{-\frac{(2p-a)^2}{4p}}\,\Phi_{2p}^{(2p-a)}\left(\frac{1}{q^2}\right) 
\nonumber\\ 
\hskip -4cm && + i\sqrt{\frac{\pi}{|t|}}\sum_{b=1}^{p} \sqrt{\frac{4}{p}} \sin\left(\frac{a (b-1/2) \pi}{p}\right)\, \qt^{\,-\frac{(2b-1)^2}{4p}}\Phi_{2p}^{(2b-1)}\left(\frac{1}{\qt^{\, 2}}\right).
\label{eq:jc-false}
\end{eqnarray}
These decompositions into real and imaginary parts, in terms of false theta functions, are illustrated in Figures \ref{fig:js32} and \ref{fig:jc32}, plotted as functions of $t$ and as functions of $q$. The natural boundary at $q=1$ is shown as a vertical dashed line in the right hand plots.
\graphicspath{ {./paper plots/} }

\noindent{\bf Remarks:} 
\begin{enumerate}
\item The decompositions \eqref{eq:js-false}-\eqref{eq:jc-false} into real and imaginary parts are uniquely defined on the Stokes line.    

\item In these expressions \eqref{eq:js-false}-\eqref{eq:jc-false}, since $t<0$, it follows that $q\equiv e^{-t}>1$ and $\qt \equiv e^{-\pi^2/t}>1$. The expressions \eqref{eq:js-false}-\eqref{eq:jc-false} should be understood accordingly.

\item 
The identities \eqref{eq:js-false}-\eqref{eq:jc-false} are consistent with the modular properties of the Mordell-Borel integrals in \eqref{eq:js-4pi}-\eqref{eq:jc-4pi}.

\end{enumerate}

\section{Duals of False Theta Functions}

\label{sec:class1}

The first resurgence orbit that we study is a set of Mordell-Borel integrals defined by the condition that the real part on the Stokes line $t<0$ is a pure false theta function $\Psi_p^{(a)}$. 
We restrict attention here to $p$ odd, with $p\geq 3$ and $0<a\leq p$.
\subsection{Closed Resurgence Orbit of Vectors of Mordell-Borel Integrals}
\label{sec:vectors1}

With this choice of $(p, a)$, it follows that the $p\times p$ algebraic structure of the non-unary-side identities \eqref{eq:js-4pi}-\eqref{eq:jc-4pi}, and also of the unary-side identities \eqref{eq:js-false}-\eqref{eq:jc-false}, can be reduced to a $\frac{(p-1)}{2}\times \frac{(p-1)}{2}$ sub-orbit structure, such that the two structures match across the natural boundary.

This algebraic reduction is achieved by defining the following four $\frac{(p-1)}{2}$-component vector-valued Mordell-Borel integrals, characterized by whether or not the $a$ index of the Mordell-Borel integral $JS_{(p, a)}$ or $JC_{(p, a)}$ is even or odd: 
\begin{eqnarray}
J_j^{(A, p)}(t)&:=& JS_{(p, 2j)}(t), \label{eq:a}
\\
J_j^{(B, p)}(t)&:=& \sqrt{2}\left(JC_{(2p, 2j-1)}(t)+(-1)^{(p-1)/2+j} JC_{(2p, 2p-(2j-1))}(t)\right),
\label{eq:b}
\\
J_j^{(C, p)}(t)&:=& JS_{(p, 2j-1)}(t),
\label{eq:c}
\\
J_j^{(D, p)}(t)&:=& JC_{(p, 2j)}(t).
\label{eq:d}
\end{eqnarray}
The $\sqrt{2}$ factor in the definition of $J_j^{(B, p)}(t)$ is simply for notational convenience, but the plus/minus signs between the two terms are crucial. Each vector $J^{(*, p)}(t)$ has $\frac{(p-1)}{2}$ elements: $j=1, 2, \dots , \frac{(p-1)}{2}$. The superscripts $*=A, B, C, D$ refer to the mixing matrices [defined below in \eqref{eq:reduced-mixing}] which enter the following modular transformation properties of these vectors:
\begin{eqnarray}
J_j^{(A, p)}(t)&=&\left(\frac{\pi}{t}\right)^{3/2}  \sum_{k=1}^{\frac{(p-1)}{2}} A_{jk} \, J_k^{(A, p)}\left(\frac{\pi^2}{t}\right), 
\label{eq:at}
\\
J_j^{(B, p)}(t)&=&\left(\frac{\pi}{t}\right)^{3/2}  \sum_{k=1}^{\frac{(p-1)}{2}} B_{jk} \, J_k^{(B, p)}\left(\frac{\pi^2}{t}\right), 
\label{eq:bt}
\\
J_j^{(C, p)}(t)&=&\left(\frac{\pi}{t}\right)^{3/2}  \sum_{k=1}^{\frac{(p-1)}{2}} C_{jk} \, J_k^{(D, p)}\left(\frac{\pi^2}{t}\right), 
\label{eq:ct}
\\
J_j^{(D, p)}(t)&=&\left(\frac{\pi}{t}\right)^{3/2}  \sum_{k=1}^{\frac{(p-1)}{2}} D_{jk} \, J_k^{(C, p)}\left(\frac{\pi^2}{t}\right). 
\label{eq:dt}
\end{eqnarray}
The mixing matrices for the vectors labeled by A, B, C and D  are defined as:
\begin{align}
A_{jk}&= \sqrt{\frac{4}{p}}\sin\left(\frac{2jk\pi}{p}\right)\,;
\qquad 
B _{jk}= \sqrt{\frac{4}{p}}\sin\left( \frac{(j-1/2)(k-1/2)\pi}{p} + (-1)^{j+k+\frac{p+1}{2}} \frac{\pi}{4} \right) 
\nonumber\\
C_{jk}&= \sqrt{\frac{4}{p}}\sin\left(\frac{(2j-1)k\pi}{p}\right) \qquad; \qquad D_{jk}=C_{kj} \hspace{2mm}.
\label{eq:reduced-mixing}
\end{align}
The modular mixing matrices $A$ and $B$ have eigenvalues $\pm 1$, while those of $C$ and $D$ have eigenvalues with modulus $1$.
And note that these are $\frac{(p-1)}{2}\times \frac{(p-1)}{2}$ matrices, in contrast to the larger $p \times p$ mixing matrices in \eqref{eq:mixing}.

Using the general unary-side decompositions \eqref{eq:js-false}-\eqref{eq:jc-false} of the Mordell-Borel integrals into false theta functions for $t<0$ ($q>1$), the vector-valued Mordell-Borel integrals defined in \eqref{eq:a}-\eqref{eq:d} uniquely decompose into real and imaginary parts in a way that involves the same mixing matrices as in \eqref{eq:reduced-mixing}:
\begin{eqnarray}
\hskip -15 pt i\sqrt{\frac{4p|t|}{\pi}}\,  J_j^{(A, p)}(t)&\hskip -35pt  = \hskip -25pt & \hskip -10pt q^{-\frac{j^2}{p}} \, \Phi_{p}^{(2j)}\left(
\frac{1}{q}\right) +i  \sqrt{\frac{\pi}{|t|}}  \sum_{k=1}^{\frac{(p-1)}{2}} A_{jk} \, \qt^{\, \,-\frac{k^2}{p}} \, \Phi_{p}^{(2k)}\left(\frac{1}{\qt}\right),  
\label{eq:aq}
\\
\hskip -15 pt i\sqrt{\frac{4p|t|}{\pi}}\,  J_j^{(B, p)}\!(t)  & \hskip -35pt \! = \! \hskip -25pt & \hskip -10pt q^{-\frac{(2j-1)^2}{8p}} \, \Phi_{p}^{(2j-1)}\!\!\left(\frac{i}{\sqrt{q}}\right) \!+\!i  \sqrt{\frac{\pi}{|t|}}  \sum_{k=1}^{\frac{(p-1)}{2}}\! B_{jk} \, \qt^{\, \,-\frac{(2k-1)^2}{8p}} \, \Phi_{p}^{(2k-1)}\!\!\left(\frac{i}{\sqrt{\qt}}\right),  
\label{eq:bq}
\\
i\sqrt{\frac{4p|t|}{\pi}}\,  J_j^{(C, p)}(t)&=&q^{-\frac{(2j-1)^2}{4p}} \, \Phi_{p}^{(2j-1)}\left(\frac{1}{q}\right) +i  \sqrt{\frac{\pi}{|t|}}  \sum_{k=1}^{\frac{(p-1)}{2}} C_{jk} \, \qt^{\, \,-\frac{k^2}{p}} \, \Phi_{p}^{(2k)}\left(-\frac{1}{\qt}\right),  
\label{eq:cq}
\\
i\sqrt{\frac{4p|t|}{\pi}}\,  J_j^{(D, p)}(t)&=&q^{-\frac{j^2}{p}} \, \Phi_{p}^{(2j)}\left(-\frac{1}{q}\right) +i  \sqrt{\frac{\pi}{|t|}}  \sum_{k=1}^{\frac{(p-1)}{2}} D_{jk} \, \qt^{\, \,-\frac{(2k-1)^2}{4p}} \, \Phi_{p}^{(2k-1)}\left(\frac{1}{\qt}\right).  
\label{eq:dq}
\end{eqnarray}
\noindent{\bf Remarks:}
\begin{enumerate}
\item These identities are reductions of the more general identities \eqref{eq:js-false}-\eqref{eq:jc-false}, under the condition that $p$ is odd, and by distinguishing between even and odd indices $a$.
\item 
If $p$ and $a$ are both odd, then $\Phi_p^{(a)}(q)$ has a small $q$ unary expansion that is alternating in sign and in {\bf even} powers of $q$. Therefore the large $q$ expansions for  the $q$-series appearing in the decomposition of $J_j^{(B, p)}(t)$  are in fact expansions in integer powers of $\frac{1}{q}$, and non-alternating in sign. 
\item Furthermore, this means that the unary expansions of {\bf all} the $\Phi_p^{(a)}$ appearing in the decomposition identities \eqref{eq:aq}-\eqref{eq:dq} are real. Therefore, the 
decomposition identities \eqref{eq:aq}-\eqref{eq:dq} are {\it unique} decompositions of the LHS into real and imaginary parts, expressed in terms of $1/q$ and $1/\qt$, respectively.
\item Hence, the defining property of the integrals $J_j^{(A, p)}(t)$, $J_j^{(B, p)}(t)$, $J_j^{(C, p)}(t)$, and $J_j^{(D, p)}(t)$ is that for each $j$ the real part on the RHS, when $t<0$, is a pure false theta function.
\item We stress that even though there are {\it four} different $\frac{(p-1)}{2}$-dimensional vectors of Mordell-Borel integrals defined in \eqref{eq:a}-\eqref{eq:d}, their algebraic decomposition relations \eqref{eq:aq}-\eqref{eq:dq} on the unary side involve only {\it two} $\frac{(p-1)}{2}$-dimensional vectors of unary $q$-series, $\Phi_{p}^{(2j)}$ and $\Phi_{p}^{(2j-1)}$, having even and odd $a$ parameters, respectively.
So, for example, the sets A and C contain all the $JS$ integrals for $a<p$, and so we cover all the false theta functions $\Psi_{p}^{(a)}$, for $a=1, 2, \dots, (p-1)$.
\item
The first two of the unary-side decompositions, \eqref{eq:aq}-\eqref{eq:bq}, are {\bf self-dual} under $t\to\frac{\pi^2}{t}$, as a result of the duality properties of $J_j^{(A, p)}(t)$ and $J_j^{(B, p)}(t)$ listed in \eqref{eq:at}-\eqref{eq:bt}. In our discussion of this class of Mordell-Borel integrals, and associated $q$-series and $\tilde q$-series, we will refer to \eqref{eq:aq}-\eqref{eq:bq} as the {\it self-dual} identities.
\item The latter  two of the unary-side decompositions, \eqref{eq:cq}-\eqref{eq:dq}, {\bf interchange} with one another under $t\to\frac{\pi^2}{t}$, as a result of the duality properties of $J_j^{(C, p)}(t)$ and $J_j^{(D, p)}(t)$ listed in \eqref{eq:ct}-\eqref{eq:dt}. Thus, they actually contain exactly the same information. We will refer to \eqref{eq:cq}-\eqref{eq:dq} as the {\it mixed} identities.
\end{enumerate}
\subsection{Resurgence and the Property of Preservation of Relations}
\label{sec:preservation1}

The algebraic decomposition relations \eqref{eq:aq}-\eqref{eq:dq} on the unary side constitute a fundamental property of the original Mordell-Borel integrals $JS_{(p, a)}(t)$ and $JC_{(p, a)}(t)$, when $p$ is an odd integer. Since they are decompositions into real and imaginary parts, they are manifestly unique on the unary side $t<0$.
Therefore,  these relations must be preserved under unique continuation from $t<0$ back to the $t>0$ region in which the integrals were originally defined. Hence we are led to the following conjecture for the vectors of Mordell-Borel integral defined in \eqref{eq:a}-\eqref{eq:d}:

\begin{conjecture}\label{conj:conjecture1}
For the class of Mordell-Borel integrals defined in \eqref{eq:a}-\eqref{eq:d}, for all odd $p$ values,  the Mordell-Borel integrals for $t>0$ decompose uniquely into $q$-series and $\qt$-series as
\begin{align}
\sqrt{\ff{4pt}{\pi}}J^{(A,p)}_j(t) &= q^{\,-\frac{j^2}{p}} \Phi_{p}^{(2j)}(q)^{\vee} + \sqrt{\ff{\pi}{t}} \sum\limits_{k=1}^{\frac{p-1}{2}} A_{jk} \, \qt^{\,-\frac{k^2}{p}} \Phi_{p}^{(2k)}(\qt)^{\vee}, 
\label{eq:aq-dual}\\
\sqrt{\ff{4pt}{\pi}}J^{(B,p)}_j(t) &= q^{-\frac{(2j-1)^2}{8p}} \Phi_{p}^{(2j-1)}(i \sqrt{q})^{\vee} + \sqrt{\ff{\pi}{t}} \sum\limits_{k=1}^{\frac{p-1}{2}} B_{jk} \, \qt^{\,-\frac{(2k-1)^2}{8p}} \Phi_{p}^{(2k-1)}(i \sqrt{\qt})^{\vee}, 
\label{eq:bq-dual}\\
\sqrt{\ff{4pt}{\pi}}J^{(C,p)}_j(t) &= q^{-\frac{(2j-1)^2}{4p}} \Phi_{p}^{(2j-1)}(q)^{\vee} + \sqrt{\ff{\pi}{t}} \sum\limits_{k=1}^{\frac{p-1}{2}} C_{jk} \, \qt^{\,-\frac{k^2}{p}} \Phi_{p}^{(2k)}(-\qt)^{\vee}, \label{eq:cq-dual}\\
\sqrt{\ff{4pt}{\pi}}J^{(D,p)}_j(t) &= q^{-
\frac{j^2}{p}} \Phi_{p}^{(2j)}(-q)^{\vee} + \sqrt{\ff{\pi}{t}} \sum\limits_{k=1}^{\frac{p-1}{2}} D_{jk} \, \qt^{\,-\frac{(2k-1)^2}{4p}} \Phi_{p}^{(2k-1)}(\qt)^{\vee},
\label{eq:dq-dual}
\end{align}
where $\Phi^{(a)}_p(q)^{\vee}$, for $a=1, 2, ..., (p-1)$, are  $q$-series  with integer coefficients, up to an overall multiplicative factor (which turns out to be an inverse power of $2$), and $A$, $B$, $C$, $D$ are the same mixing matrices defined in \eqref{eq:reduced-mixing} and which appear in the unary-side decompositions \eqref{eq:js-false}-\eqref{eq:jc-false}.

Including the prefactor rational powers of $q^{-\frac{a^2}{4p}}$, we therefore identify the duals to the false theta functions $\Psi^{(a)}_p(q)$ under the operation $q \to \dfrac{1}{q}$ as:
\begin{eqnarray}
t \to -t\,\left(q \to \dfrac{1}{q}\right) \quad \Rightarrow\quad
\Psi^{(a)}_p(q) \longleftrightarrow  
\Psi^{(a)}_p(q)^{\vee} := q^{-\frac{a^2}{4p}}\Phi^{(a)}_{p}(q)^{\vee}
\quad,  a=1, \dots, p-1 \hspace{2mm}.
\label{eq:false-theta-duals}
\end{eqnarray}

\end{conjecture}

\noindent{\bf Remarks:}
\begin{enumerate}
\item The lowest examples of this orbit are $p=3$ and $p=5$, in which case the decompositions \eqref{eq:aq-dual}-\eqref{eq:dq-dual} agree with the known identities for order 3 and order 10 mock theta functions \cite{GM12}. See Sections \ref{sec:p3} and \ref{sec:p5} below.

\item In Section \ref{sec:algorithm} we present a numerical algorithm that goes well beyond these known $p=3$ and $p=5$ cases, determining the expansions of the dual $q$-series $\Phi^{(a)}_p(q)^{\vee}$ in \eqref{eq:false-theta-duals}, which appear on the RHS of the decompositions
\eqref{eq:aq-dual}-\eqref{eq:dq-dual}. The resulting explicit $q$-series expressions for $\Psi^{(a)}_p(q)^{\vee}$ appear in Section \ref{sec:results-p-class} for $p=7$ and $p=9$,  and in Appendix \ref{sec:appendix-results1} for $p=11$, $p=13$, $p=15$, $p=17$, and $p=19$. 
All these results for $p\geq 7$ are new. 
\item 

The mixing matrices $A$, $B$, $C$, and $D$ which appear in Conjecture \ref{conj:conjecture1} in the $q$-series decompositions on the non-unary side, \eqref{eq:aq-dual}-\eqref{eq:dq-dual}, are precisely the same mixing matrices that appear in the $q$-series decompositions on the unary side, \eqref{eq:aq}-\eqref{eq:dq}, and also that appear in the modular relations \eqref{eq:at}-\eqref{eq:dt} for the Mordell-Borel integrals. This shows that the mixing matrices in Conjecture \ref{conj:conjecture1} are uniquely determined by the Mordell-Borel integrals themselves, ultimately coming from linear combinations of the elementary identities \eqref{eq:js-identity1}-\eqref{eq:jc-identity2}.

\item The rational powers of $q$ and $\tilde q$ which appear in Conjecture \ref{conj:conjecture1} in the $q$-series decompositions on the non-unary side, \eqref{eq:aq-dual}-\eqref{eq:dq-dual}, are precisely the same rational powers of $q$ and $\tilde q$  appearing in the $q$-series decompositions on the unary side, \eqref{eq:aq}-\eqref{eq:dq}. Thus, the full algebraic structure of Conjecture \ref{conj:conjecture1} is actually encoded in the vector-valued Mordell-Borel integrals \eqref{eq:a}-\eqref{eq:d} themselves.

\item
The property of preservation of relations for resurgent functions suggests that the dual $q$-series in \eqref{eq:false-theta-duals} should be unique. We discuss the question of uniqueness further in Section \ref{sec:p-uniqueness} and in \cite{CDGG,CD25}.
\end{enumerate}

Given \textbf{\conjref{conj:conjecture1}},  the question of ``crossing the natural boundary'', $|q|=1$, for the false theta functions \eqref{eq:false}-\eqref{eq:false-phi} becomes a different set of questions:

\begin{itemize}

\item Is it possible to {\bf uniquely} decompose, in the $t>0$ region where $q<1$, the vector-valued Mordell-Borel integrals defined in \eqref{eq:a}-\eqref{eq:d}, into $q$-series and $\tilde{q}$-series such that the structure on the right-hand-sides in \eqref{eq:aq}-\eqref{eq:dq} is preserved, namely in the form in \eqref{eq:aq-dual}-\eqref{eq:dq-dual}? 

    \item 
If the answer is yes, then there is a unique continuation across the natural boundary that identifies the dual $\Psi^{(a)}_p(q)^\vee$ of each false theta function $\Psi^{(a)}_p(q)$, as in \eqref{eq:false-theta-duals}.

\item We then ask: are the expansion coefficients of the $q$-series part, $\Phi^{(a)}_p(q)^\vee$, integer-valued?  This is expected on physical grounds from the connection with Chern-Simons theory, since these dual $q$-series have the interpretation of partition function decompositions, with coefficients that count states.

\item And: is there an efficient and constructive way to generate these duals?

\end{itemize}

We argue below that the answer is "yes" to each of  these questions. We first  show that Conjecture \ref{conj:conjecture1} agrees with known results when they are available: see Sections \ref{sec:p3} and \ref{sec:p5}, for $p=3$ and $p=5$, respectively. Then, in Section \ref{sec:algorithm} we introduce a numerical algebraic algorithm that generates the dual $q$-series appearing in Conjecture \ref{conj:conjecture1}. We confirm that this method reproduces the known results for $p=3$ and $p=5$, and we then present new results that go well beyond the known ones, for higher odd $p$ values: For $p=7$ and $p=9$ see  Section \ref{sec:results-p-class}, and the results for $p=11$, $p=13$, $p=15$, $p=17$, and $p=19$ are presented in Appendix \ref{sec:appendix-results1}. 

In \cite{ceff2} it is shown how the algebraic structure of Conjecture \ref{conj:conjecture1} places strong constraints on the growth rate of the coefficients of the $q$-series appearing in the decompositions \eqref{eq:aq-dual}-\eqref{eq:dq-dual}, providing  a consistency check on our numerical algorithm.

\subsection{Example: Order 3 Mock Theta Functions}
\label{sec:p3}

Here we show that for $p=3$, the known results for order 3 mock theta functions satisfy \textbf{\conjref{conj:conjecture1}}. Uniqueness is proven in \cite{CDGG,CD25}.

Choosing $p=3$, so that $\frac{(p-1)}{2}=1$, there is in fact no vector or matrix structure. The four corresponding integrals in \eqref{eq:a}-\eqref{eq:d} are
\begin{eqnarray}
J^{(A, 3)}(t)&=&JS_{(3,2)}(t) = \frac{1}{t}\int_0^\infty du\, e^{-3u^2/t} \frac{\sinh(u)}{\sinh(3u)}, 
\\
J^{(B, 3)}(t)&=& \sqrt{2}\left(JC_{(6,1)}(t)+JC_{(6,5)}(t)\right) = \frac{ \sqrt{2}}{t}\int_0^\infty du\, e^{-6u^2/t} \frac{\cosh(5u)+\cosh(u)}{\cosh(6u)}, 
\\
J^{(C, 3)}(t)&=&JS_{(3,1)}(t) = \frac{1}{t}\int_0^\infty du\, e^{-3u^2/t} \frac{\sinh(2u)}{\sinh(3u)}, 
\\
J^{(D, 3)}(t)&=&JC_{(3,2)}(t) = \frac{1}{t}\int_0^\infty du\, e^{-3u^2/t} \frac{\cosh(u)}{\cosh(3u)}. 
\label{eq:p3-mordell}
\end{eqnarray}
These are identified with the integrals $W_3(t)$, $\frac{1}{\sqrt{2}} W(t)$, $W_1(2t)$ and $\frac{1}{2}W_2\left(\frac{t}{2}\right)$, respectively, in the notation of the Gordon-McIntosh review article \cite{GM12}: see page 115.

The unary-side  decompositions \eqref{eq:aq-dual}-\eqref{eq:dq-dual} into real and imaginary parts for $t<0$ read:
\begin{eqnarray}
i\sqrt{\frac{12|t|}{\pi}}\,  J^{(A, 3)}(t)&=&q^{-\frac{1}{3}} \, \Phi_{3}^{(2)}\left(\frac{1}{q}\right) +i  \sqrt{\frac{\pi}{|t|}} \, \qt^{\, -\frac{1}{3}} \, \Phi_{3}^{(2)}\left(\frac{1}{\qt}\right),  
\label{eq:aq3}
\\
i\sqrt{\frac{12|t|}{\pi}}\,  J^{(B, 3)}(t)&=&q^{-\frac{1}{24}} \, \Phi_{3}^{(1)}\left(\frac{i}{\sqrt{q}}\right) +i  \sqrt{\frac{\pi}{|t|}} \, \qt^{\, -\frac{1}{24}} \, \Phi_{3}^{(1)}\left(\frac{i}{\sqrt{\qt}}\right),  
\label{eq:bq3}
\\
i\sqrt{\frac{12|t|}{\pi}}\,  J^{(C, 3)}(t)&=&q^{-\frac{1}{12}} \, \Phi_{3}^{(1)}\left(\frac{1}{q}\right) +i  \sqrt{\frac{\pi}{|t|}}  \,  \qt^{\, -\frac{1}{3}} \, \Phi_{3}^{(2)}\left(-\frac{1}{\qt}\right),  
\label{eq:cq3}
\\
i\sqrt{\frac{12|t|}{\pi}}\,  J^{(D, 3)}(t)&=&q^{-\frac{1}{3}} \, \Phi_{3}^{(2)}\left(-\frac{1}{q}\right) +i  \sqrt{\frac{\pi}{|t|}}   \, \qt^{\, -\frac{1}{12}} \, \Phi_{3}^{(1)}\left(\frac{1}{\qt}\right).  
\label{eq:dq3}
\end{eqnarray}
The mixing matrices all reduce to $1$ because $\sin\left(\frac{\pi}{3}\right)=\sin\left(\frac{2\pi}{3}\right)=\frac{\sqrt{3}}{2}$.
We compare these with the identities \cite{GM12} for the same integrals on the non-unary side ($t>0$, $q<1$):
\begin{eqnarray}
\sqrt{\frac{12 t}{\pi}}\,  J^{(A, 3)}(t)&=& q^{-\frac{1}{3}} \, q \, \omega(-q) + \sqrt{\frac{\pi}{t}}  \, \qt^{\, -\frac{1}{3}} \, \qt\, \omega(-\qt),  
\label{eq:aq3-plus}
\\
\sqrt{\frac{12 t}{\pi}}\,  J^{(B, 3)}(t)&=&q^{-\frac{1}{24}} \,\frac{1}{2}\, f(-q) + \sqrt{\frac{\pi}{t}} \, \qt^{\, -\frac{1}{24}} \,\frac{1}{2}\, f(-\qt),  
\label{eq:bq3-plus}
\\
\sqrt{\frac{12 t}{\pi}}\,  J^{(C, 3)}(t)&=&q^{-\frac{1}{12}} \,\frac{1}{2}\, f(q^2) - \sqrt{\frac{\pi}{t}} \,  \qt^{\, -\frac{1}{3}} \, \qt \, \omega(\qt), 
\label{eq:cq3-plus}
\\
\sqrt{\frac{12 t}{\pi}}\,  J^{(D, 3)}(t)&=& - q^{-\frac{1}{3}} \, q\, \omega(q) +  \sqrt{\frac{\pi}{t}}    \, \qt^{\, -\frac{1}{12}} \, \frac{1}{2}\,  f(\qt^2).  
\label{eq:dq3-plus}
\end{eqnarray}
This leads to the following identifications under the orientation reversal operation $t \to -t$, or $q \to 1/q$:
\begin{eqnarray}
\Psi^{(1)}_3(q) &\longleftrightarrow&
\Psi^{(1)}_3(q)^{\vee} := q^{-\frac{1}{12}}\Phi^{(1)}_{3}(q)^{\vee},\\
\Psi^{(2)}_3(q) &\longleftrightarrow&
\Psi^{(2)}_3(q)^{\vee} := q^{-\frac{1}{3}}\Phi^{(2)}_{3}(q)^{\vee},
\label{eq:p3-duals}
\end{eqnarray}
where the dual $q$-series are
\begin{eqnarray}
 \Phi_{3}^{(1)}(q)^\vee &=&   \frac{1}{2}f(q^2),\\
  \Phi_{3}^{(2)}(q)^\vee &=&   q\,\omega(-q).
\end{eqnarray}
This agrees with known results \cite{GM12}. We observe that $\Phi_{3}^{(1)}(q)^\vee$ is indeed an expansion in $q^2$, in agreement with the general result for odd $a$, as noted in the second remark in Section \ref{sec:vectors1}.

Recall the following expressions for the small $q$ and large $q$ expansions
\begin{eqnarray}
f(q)&:=& 1-\sum_{n=1}^\infty \frac{(-1)^n q^{n}}{(-q; q)_{n}}
\label{eq:fq}
\\
&=&1+q-2q^2+3q^3-3q^4+3q^5-5q^6+\dots\quad,\quad |q|\to 0
\label{eq:fq-small}
\\
&=&2\left(1-\frac{1}{q}+ \frac{1}{q^2}-\frac{1}{q^5}+\frac{1}{q^7}-\frac{1}{q^{12}}+\frac{1}{q^{15}}+
\dots \right) \quad,\quad |q|\to\infty 
\label{eq:fq-large}
\end{eqnarray}
\begin{eqnarray}
\omega(q)&:=& \sum_{n=0}^\infty \frac{q^{n}}{(q; q^2)_{n+1}}
\label{eq:wq}
\\
&=&1+2 q+3 q^2+4 q^3+6 q^4+8 q^5+10 q^6+\dots\quad,\quad |q|\to 0
\label{eq:wq-small}
\\
&=&-\frac{1}{q}-\frac{1}{q^2}+\frac{1}{q^6}+
\frac{1}{q^9}-\frac{1}{q^{17}}+
\dots \quad,\quad |q|\to\infty 
\label{eq:wq-large}
\end{eqnarray}
Notice that the large $q$ expansions are unary, up to a factor of $\frac{1}{2}$ for $f(q)$, while the small $q$ expansions are non-unary but have integer-valued coefficients.
\graphicspath{ {./paper plots/} }
\begin{figure}[h!]
\centering
\begin{subfigure}[t]{.75\textwidth}
  \centering
  \includegraphics[width=.8\linewidth]{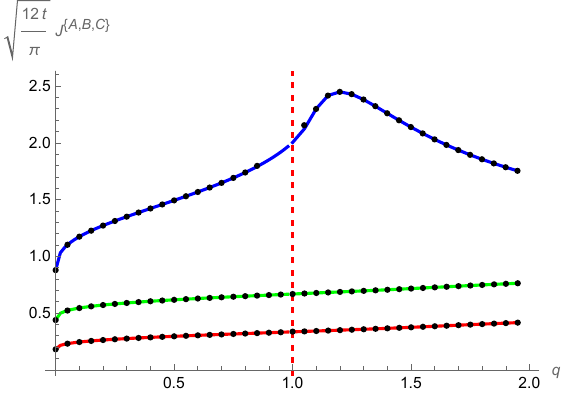}
\end{subfigure}
\caption{Crossing of the natural boundary for $p=3$. The solid lines represent the real parts of the integrals on the left hand side of the relations (\ref{eq:aq3})-(\ref{eq:cq3}) on the unary side, with $q>1$, and of the relations (\ref{eq:aq3-plus})-(\ref{eq:cq3-plus}) on the non-unary side, with $q<1$, for sets A (red), B (blue), and C (green). The black dots represent the corresponding $q$-series expressions on the right hand sides of these relations. For $q>1$, the large $q$ unary expansions of $f(q)$ and $\omega(q)$ are used, while for $q<1$ the small $q$ non-unary expansions are used.}
\label{fig:p3}
\end{figure}
Figure \ref{fig:p3} shows the crossing of the natural boundary for $p=3$, for the sets A (red), B (blue), and C (green), as in (\ref{eq:aq3})-(\ref{eq:cq3}) and in (\ref{eq:aq3-plus})-(\ref{eq:cq3-plus}). (Note that we do not need to include plots of the decompositions for set D, because these are equivalent to those for set C, by taking $t\rightarrow \frac{\pi^2}{t}$, or $q\to\qt$.) These plots show how the real parts of the decompositions (\ref{eq:aq3})-(\ref{eq:cq3}) on the unary side, with $q>1$, map smoothly to the decompositions (\ref{eq:aq3-plus})-(\ref{eq:cq3-plus}) on the non-unary side, where $q<1$ and the decompositions are purely real.

The solid lines in Figure \ref{fig:p3} show the real parts of the integrals on the LHS of the decomposition identities (\ref{eq:aq3})-(\ref{eq:cq3}) on the unary side, and on the LHS of the decomposition identities  (\ref{eq:aq3-plus})-(\ref{eq:cq3-plus}) on the non-unary side. The black points show the real part of the $q$-series expressions on the RHS of the decomposition identities (\ref{eq:aq3})-(\ref{eq:cq3}) on the unary side, and on the RHS of the decomposition identities  (\ref{eq:aq3-plus})-(\ref{eq:cq3-plus}) on the non-unary side, where both the $q$-series and $\qt$-series parts are real.

\subsection{The Importance of Uniqueness}
\label{sec:p-uniqueness}

\begin{figure}[h!]
\begin{subfigure}[t]{.5\textwidth}
  \centering
  \includegraphics[width=.8\linewidth]{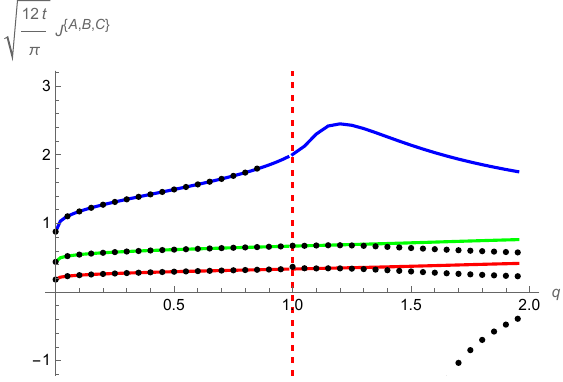}
\end{subfigure}
~
\begin{subfigure}[t]{.5\textwidth}
  \centering
  \includegraphics[width=.8\linewidth]{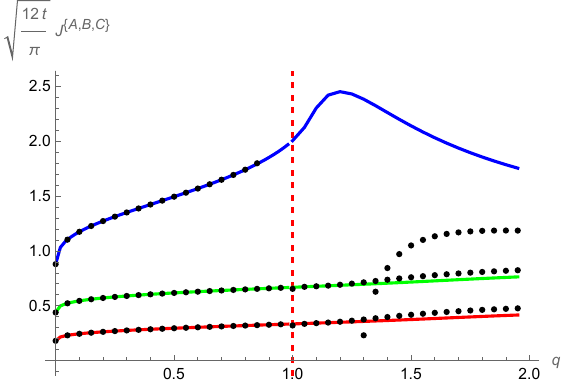}
\end{subfigure}
  \caption{Left: The failure of $\omega_2(q)$ and $f_2(q)$ to cross the natural boundary at $q=1$. Right: The failure of $\omega_3(q)$ and $f_3(q)$ to cross the natural boundary at $q=1$. In both plots, the solid lines represent the real parts of the integrals in the decompositions (\ref{eq:aq3})-(\ref{eq:cq3}) and (\ref{eq:aq3-plus})-(\ref{eq:cq3-plus}) for sets A (red), B (blue), and C (green) on both sides of the boundary, as described in the text. These lines are exactly the same as in Figure \ref{fig:p3}. The black dots represent the $q$ series expansions on the right hand sides of the relations (\ref{eq:aq3})-(\ref{eq:cq3}) and (\ref{eq:aq3-plus})-(\ref{eq:cq3-plus}).  We use the $q$-series expressions \eqref{eq:f2q} and \eqref{eq:w2q} [left], and \eqref{eq:f3q} and \eqref{eq:w3q} [right]. For the non-unary side, $q<1$,
  these dots are also the same as in Figure \ref{fig:p3}, but on the unary side, $q>1$, they do not match those in Figure \ref{fig:p3} and hence do not match the integrals.}
  \label{fig:p3allq123}
\end{figure}

It is important to note that there exist alternative $q$-Pochhammer $q$-series expressions for $f(q)$ and $\omega(q)$ in the $q<1$ region, which are different from \eqref{eq:fq} and \eqref{eq:wq} \cite{Bringmann}:
\begin{eqnarray}
    f_2(q)&=&  \sum_{n=0}^\infty \frac{q^{n^2}}{(-q; q)_n^2}  
    \label{eq:f2q}\\
    &\sim&  1+q-2 q^2+3 q^3-3 q^4+3 q^5-5q^6+\dots \quad,\quad |q|\to 0\\
    &\sim& 1+\frac{1}{q}-\frac{1}{q^2}+\frac{2}{q^3}-\frac{
   4}{q^4}+\frac{5}{q^5}-\frac{6
   }{q^6}+ \dots\quad,\quad |q|\to \infty
\end{eqnarray}
\begin{eqnarray}
    f_3(q)&=& 2-2 \sum_{n=1}^\infty \frac{q^{n(n-1)}}{(-1; q)_n (-q; q)_n}  
    \label{eq:f3q}\\
    &\sim&  1+q-2 q^2+3 q^3-3 q^4+3 q^5-5 q^6+\dots \quad,\quad |q|\to 0\\
    &\sim& 2-\frac{1}{q}+\frac{1}{q^5}-\frac{1}{q^6}+\frac{1}{q^7}-\frac{1}{q^8}+\frac{2}{q^9}-\frac{3}{q^{
   10}}+ \dots\quad,\quad |q|\to \infty
\end{eqnarray}
\begin{eqnarray}
    \omega_2(q)&=&  \sum_{n=0}^\infty \frac{q^{2n(n+1)}}{(q; q^2)_{n+1}^2}  
    \label{eq:w2q}\\
    &\sim&  1+2 q+3 q^2+4 q^3+6 q^4+8 q^5+10 q^6+\dots \quad,\quad |q|\to 0\\
    &\sim& \frac{1}{q^2}+\frac
   {2}{q^3}+\frac{4}{q^4}+\frac{
   6}{q^5}+\frac{9}{q^6}+\frac{1
   4}{q^7}+\frac{20}{q^8}+\frac{
   28}{q^9}+\frac{40}{q^{10}}+ \dots\quad,\quad |q|\to \infty
\end{eqnarray}
\begin{eqnarray}
    \omega_3(q)&=& -\frac{1}{q}+\frac{1}{q(1-q)} \sum_{n=0}^\infty \frac{q^{2n^2}}{(q; q^2)_n (q^3; q^2)_n}  
    \label{eq:w3q}\\
    &\sim&   1+2 q+3 q^2+4 q^3+6 q^4+8 q^5+10 q^6+\dots \quad,\quad |q|\to 0\\
    &\sim& -\frac{1}{q}-\frac{1}{q^2}-\frac{1}{q^3}-\frac{2}{q^4}-\frac{3}{
   q^5}-\frac{5}{q^6}-\frac{8}{q
   ^7}-\frac{12}{q^8}-\frac{18}{
   q^9}+\dots\quad,\quad |q|\to \infty \hspace{2mm} .
\end{eqnarray}
These alternative expressions for $f(q)$ and $\omega(q)$ have the same small $q$ expansions as \eqref{eq:fq-small} and \eqref{eq:wq-small}, but they {\it do not have the same large $q$ expansions} under the term-by-term inversion $q\to 1/q$: contrast the large $q$ expansions here with \eqref{eq:fq-large} and \eqref{eq:wq-large}.  Furthermore, it is not only that these large $q$ expansions are different from \eqref{eq:fq-large} and \eqref{eq:wq-large}. More importantly, none of $f_2(q)$, $f_3(q)$, 
$\omega_2(q)$ or $\omega_3(q)$ produces a {\bf unary} series in the large $q$ regime. Therefore, they do not satisfy the (unique!) decompositions into real and imaginary parts of the Mordell-Borel integrals \eqref{eq:p3-mordell} on the unary side where $t<0$ ($q>1$). In Figure \ref{fig:p3allq123} we show plots that graphically display the failure of these expressions for $\omega_2(q)$, $\omega_3(q)$, $f_2(q)$, and $f_3(q)$ to cross the natural boundary at $|q|=1$. The large $q$ expansions of these functions are not unary, and thus do not match the corresponding Mordell-Borel integrals when rotated to $t<0$ ($q>1$). Contrast Figure \ref{fig:p3allq123} with Figure \ref{fig:p3}: everything matches on the non-unary side where $q<1$, but the $q$-series expressions do not match the uniquely continued integrals on the unary side where $q>1$.
\graphicspath{ {./paper plots/} }
In addition, it can be proven rigorously that given any solution $f(q)$ in \eqref{eq:fq} to the self-dual relation \eqref{eq:bq3-plus}, and any solution $\omega(q)$ in \eqref{eq:wq} to the self-dual relation \eqref{eq:aq3-plus}, there is a {\it unique} solution for $f(q)$ and $\omega(q)$ that also satisfies the mixed identities \eqref{eq:cq3-plus}-\eqref{eq:dq3-plus} \cite{CDGG,CD25}. This unique pair is the pair $f(q)$ and $\omega(q)$ in \eqref{eq:fq} and \eqref{eq:wq}.

\subsection{Example: Order 10 Mock Theta Functions}
\label{sec:p5}

The next example is $p=5$, which now illustrates the $\frac{(p-1)}{2}\times \frac{(p-1)}{2}$ algebraic structure of Conjecture \ref{conj:conjecture1}.
Choosing $p=5$, so that $\frac{(p-1)}{2}=2$, we have $2$-component vectors and $2\times 2$ matrix structures. The four corresponding vector-valued Mordell-Borel integrals are:
\begin{eqnarray}
J^{(A, 5)}(t)=\begin{pmatrix} JS_{(5,2)}(t)\\ JS_{(5,4)}(t)\end{pmatrix}
\quad &;& 
J^{(B, 5)}(t)=\sqrt{2} \begin{pmatrix} JC_{(10,1)}(t)-JC_{(10,9)}(t)\\ JC_{(10,3)}(t)+JC_{(10,7)}(t)\end{pmatrix}
\label{eq:jab5}
\\
J^{(C, 5)}(t)=\begin{pmatrix} JS_{(5,1)}(t)\\ JS_{(5,3)}(t)\end{pmatrix}
\quad &;&
J^{(D, 5)}(t)=\begin{pmatrix} JC_{(5,2)}(t)\\ JC_{(5,4)}(t)\end{pmatrix}.
\label{eq:jcd5}
\end{eqnarray}
Notice the different relative signs between the two terms in the $B$ class of integrals, as in \eqref{eq:b}. The integrals in \eqref{eq:jab5}-\eqref{eq:jcd5} correspond to the eight integrals $J_n(t)$ and $K_n(t)$ ($n=4, 5,6,7$) appearing in the transformation identities for the order 10 mock theta functions in the notation of the Gordon-McIntosh review \cite{GM12}. The $J_n(t)$ integrals in \cite{GM12} should be rescaled, $t\to 2t$, in order to make the $t\to \frac{\pi^2}{t}$ duality manifest. Continuing these integrals to $t<0$, we obtain the unique decompositions \eqref{eq:aq}-\eqref{eq:dq} into real and imaginary parts on the unary side ($t<0$):
\begin{align}
    i\sqrt{\frac{20|t|}{\pi}}
    J^{(A,5)}(t)
    =
    \begin{pmatrix}
        q^{-\frac{1}{5}}\Phi^{(2)}_5\left(\frac{1}{q}\right) 
        \\
        q^{-\frac{4}{5}}\Phi^{(4)}_5\left(\frac{1}{q}\right)
    \end{pmatrix}
    + 
    i \sqrt{\frac{\pi}{|t|}} \frac{2}{\sqrt{5}}
    \begin{pmatrix}
        \sin(\frac{2\pi}{5}) & \sin(\frac{4\pi}{5}) \\
        \sin(\frac{4\pi}{5}) & \sin(\frac{8\pi}{5})
    \end{pmatrix}
    \begin{pmatrix}
        \qt^{\,-\frac{1}{5}}\Phi^{(2)}_5\left(\frac{1}{\qt} \right) 
        \\
        \qt^{\,-\frac{4}{5}}\Phi^{(4)}_5\left(\frac{1}{\qt} \right)
    \end{pmatrix},
    \label{eq:ja5qunary}
\end{align}
\begin{align}
    i\sqrt{\frac{20|t|}{\pi}}
    J^{(B,5)}(t)
    &=
    \begin{pmatrix}
        q^{-\frac{1}{40}}\Phi^{(1)}_5\left(\frac{i}{\sqrt{q}} \right) 
        \\
        q^{-\frac{9}{40}}\Phi^{(3)}_5\left(\frac{i}{\sqrt{q}} \right)
    \end{pmatrix}
    + 
    i \sqrt{\frac{\pi}{|t|}} \frac{2}{\sqrt{5}}
    \begin{pmatrix}
        \sin(\frac{6\pi}{5}) & \sin(\frac{2\pi}{5}) \\
        \sin(\frac{2\pi}{5}) & \sin(\frac{1\pi}{5})
    \end{pmatrix}
    \begin{pmatrix}
        \qt^{\,-\frac{1}{40}}\Phi^{(1)}_5\left(\frac{i}{\sqrt{\qt}} \right) 
        \\
        \qt^{\,-\frac{9}{40}}\Phi^{(3)}_5\left(\frac{i}{\sqrt{\qt}} \right)
    \end{pmatrix}, 
     \label{eq:jb5qunary}
\end{align}
\begin{align}
    i\sqrt{\frac{20|t|}{\pi}}
    J^{(C,5)}(t)
    =
    \begin{pmatrix}
        q^{-\frac{1}{20}}\Phi^{(1)}_5\left(\frac{1}{q} \right) 
        \\
        q^{-\frac{9}{20}}\Phi^{(3)}_5\left(\frac{1}{q} \right)
    \end{pmatrix}
    + 
    i \sqrt{\frac{\pi}{|t|}} \frac{2}{\sqrt{5}}
    \begin{pmatrix}
        \sin(\frac{\pi}{5}) & \sin(\frac{2\pi}{5}) \\
        \sin(\frac{3\pi}{5}) & \sin(\frac{6\pi}{5})
    \end{pmatrix}
    \begin{pmatrix}    
    \qt^{\,-\frac{1}{5}}\Phi^{(2)}_5\left(-\frac{1}{\qt}\right) 
    \\
    \qt^{\,-\frac{4}{5}}\Phi^{(4)}_5\left(-\frac{1}{\qt}\right) 
    \end{pmatrix}, 
     \label{eq:jc5qunary}
\end{align}
\begin{align}
    i\sqrt{\frac{20|t|}{\pi}}
    J^{(D,5)}(t)
    &=
    \begin{pmatrix}
        q^{-\frac{1}{5}}\Phi^{(2)}_5\left( -\frac{1}{q} \right) 
        \\
        q^{-\frac{4}{5}}\Phi^{(4)}_5\left( -\frac{1}{q} \right)
    \end{pmatrix}
    + 
    i \sqrt{\frac{\pi}{|t|}} \frac{2}{\sqrt{5}}
    \begin{pmatrix}
        \sin(\frac{\pi}{5}) & \sin(\frac{3\pi}{5}) \\
        \sin(\frac{2\pi}{5}) & \sin(\frac{6\pi}{5})
    \end{pmatrix}
    \begin{pmatrix}
        \qt^{\,-\frac{1}{20}}\Phi^{(1)}_5\left( \frac{1}{\qt} \right) 
        \\
        \qt^{\,-\frac{9}{20}}\Phi^{(3)}_5\left( \frac{1}{\qt}\right)
    \end{pmatrix}. 
     \label{eq:jd5qunary}
\end{align}
Now compare these unary-side identities with the non-unary-side ($t>0$) decompositions of the same Mordell-Borel integrals (see pages 129-130 in \cite{GM12}), organized in the natural $2\times 2$ algebraic structure of Conjecture \ref{conj:conjecture1}:
\begin{align}
    \sqrt{\frac{20t}{\pi}}
    J^{(A,5)}(t)
    =
    \begin{pmatrix}
        -q^{-\frac{1}{5}}\, \psi(-q) 
        \\
        q^{-\frac{4}{5}}\, q\, \phi(-q)
    \end{pmatrix}
    + 
     \sqrt{\frac{\pi}{t}}\frac{2}{\sqrt{5}}
    \begin{pmatrix}
        \sin(\frac{2\pi}{5}) & \sin(\frac{4\pi}{5}) \\
        \sin(\frac{4\pi}{5}) & \sin(\frac{8\pi}{5})
    \end{pmatrix}
    \begin{pmatrix}
        -\qt^{\,-\frac{1}{5}}\, \psi(-\qt) 
        \\
        \qt^{\,-\frac{4}{5}}\,\qt \, \phi(-\qt)
    \end{pmatrix},
    \label{eq:ja5q}
\end{align}
\begin{align}
    \sqrt{\frac{20t}{\pi}}
    J^{(B,5)}(t)
    &=
    \begin{pmatrix}
        q^{-\frac{1}{40}}X(-q)
        \\
        q^{-\frac{9}{40}}\chi(-q)
    \end{pmatrix}
    + 
   \sqrt{\frac{\pi}{t}}\frac{2}{\sqrt{5}}
    \begin{pmatrix}
        \sin(\frac{6\pi}{5}) & \sin(\frac{2\pi}{5}) \\
        \sin(\frac{2\pi}{5}) & \sin(\frac{\pi}{5})
    \end{pmatrix}
    \begin{pmatrix}
        \qt^{\, -\frac{1}{40}}\, X(-\qt) 
        \\
        \qt^{-\frac{9}{40}}\, \chi(-\qt)
    \end{pmatrix}, 
     \label{eq:jb5q}
\end{align}
\begin{align}
   \sqrt{\frac{20t}{\pi}}
    J^{(C,5)}(t)
    =
    \begin{pmatrix}
        q^{-\frac{1}{20}}\, X(q^2)
        \\
        q^{-\frac{9}{20}}\, \chi(q^2)
    \end{pmatrix}
    + 
   \sqrt{\frac{\pi}{t}}\frac{2}{\sqrt{5}}
    \begin{pmatrix}
        \sin(\frac{\pi}{5}) & \sin(\frac{2\pi}{5}) \\
        \sin(\frac{3\pi}{5}) & \sin(\frac{6\pi}{5})
    \end{pmatrix}
    \begin{pmatrix}    
    -\qt^{\, -\frac{1}{5}}\, \psi(\qt)
    \\
    -\qt^{\, -\frac{4}{5}}\, \phi(\qt)
    \end{pmatrix}, 
     \label{eq:jc5q}
\end{align}
\begin{align}
    \sqrt{\frac{20t}{\pi}}
    J^{(D,5)}(t)
    &=
    \begin{pmatrix}
        -q^{-\frac{1}{5}}\, \psi(q)
        \\
       - q^{-\frac{4}{5}}\, q\, \phi(q)
    \end{pmatrix}
    + 
   \sqrt{\frac{\pi}{t}}\frac{2}{\sqrt{5}}
    \begin{pmatrix}
        \sin(\frac{\pi}{5}) & \sin(\frac{3\pi}{5}) \\
        \sin(\frac{2\pi}{5}) & \sin(\frac{6\pi}{5})
    \end{pmatrix}
    \begin{pmatrix}
        \qt^{\, -\frac{1}{20}}\, X(\qt^2)
        \\
        \qt^{\, -\frac{9}{20}}\, \chi(\qt^2)
    \end{pmatrix}. 
     \label{eq:jd5q}
\end{align}
Comparing \eqref{eq:ja5q}-\eqref{eq:jd5q} with \eqref{eq:ja5qunary}-\eqref{eq:jd5qunary}
we therefore identify the $\Phi^{(a)}_5(q)^\vee$ with the order 10 mock theta functions:     
\begin{align}
    \Phi^{(1)}_5(q)^{\vee} &=X(q^2) \, ,\qquad \qquad
    \Phi^{(2)}_5(q)^{\vee}= -\psi(-q) 
    \label{eq:p5-ids-x-chi}\\
    \Phi^{(3)}_5(q)^{\vee} &= \chi(q^2) \, , \qquad \qquad
    \Phi^{(4)}_5(q)^{\vee}= q\,\phi(-q) \hspace{2mm}.
    \label{eq:p5-ids-psi-phi}
\end{align}
where
\begin{eqnarray}
    X(q)= \sum_{n=0}^\infty \frac{(-1)^n q^{n^2}}{(-q; q)_{2n}} 
    \qquad &;& \qquad 
    \chi(q) = \sum_{n=0}^\infty \frac{(-1)^n q^{(n+1)^2}}{(-q; q)_{2n+1}} 
    \label{eq:p5-x-chi}
    \\
    \psi(q)= \sum_{n=0}^\infty \frac{q^{(n+1)(n+2)/2}}{(q; q^2)_{n+1}}  
    \qquad &;& 
    \qquad \phi(q) = \sum_{n=0}^\infty \frac{q^{n(n+1)/2}}{(q; q^2)_{n+1}} \hspace{2mm}.
    \label{eq:p5-psi-phi}
\end{eqnarray}
\noindent{\bf Remarks:}
\begin{enumerate}
\item 
As noted before, when $a$ is odd, both the false theta $q$-series $\Phi_{p}^{(a)}(q)$ and their duals $\Phi_{p}^{(a)}(q)^\vee$ have $q$-series expansions in integer powers of $q^2$. The $\Phi_{p}^{(a)}(q)$ are unary, while the duals $\Phi_{p}^{(a)}(q)^\vee$ have integer coefficients but are not unary.

\item Here we have confirmed that the order 3 and order 10 mock theta functions fit the structure of \textbf{\conjref{conj:conjecture1}}, for $p=3$ and $p=5$, respectively. In Section \ref{sec:algorithm} we introduce a {\bf numerical} method that determines the dual $q$-series, in agreement with the known for $p=3$ and $p=5$ closed-form results described above. But our numerical method extends to higher values of (odd) $p$, for which there are no known explicit expressions for the dual $q$-series. These new results for the duals of higher order false thetas are presented in Section \ref{sec:results-p-class} for $p=7$ and $p=9$, and  results for $p=11$, $p=13$, $p=15$, $p=17$, and $p=19$ are presented in Appendix \ref{sec:appendix-results1}. 

\item 
In \cite{ceff2} we verify that the large order growth of the coefficients of the new dual $q$-series matches the analytic form that follows from the algebraic structure of Conjecture \ref{conj:conjecture1}.

\end{enumerate}

\section{Mordell-Borel Integrals for Orientation Reversed Brieskorn Spheres}
\label{sec:class2}

The second resurgence orbit of Mordell-Borel integrals that we study in this paper are those that arise in the analysis of complex Chern-Simons theory on orientation reversed Brieskorn spheres $\overline{\Sigma(2, 3, 6k\pm 1)}$, for $k=1, 2, ...$, where the Brieskorn sphere $\Sigma(s_1, s_2, s_3)$ (for $s_1$, $s_2$, $s_3$ pairwise coprime positive integers) is defined as
\begin{equation}
\Sigma(s_1, s_2, s_3)=\left\{(x, y, z) \in \mathbb{C}^3 \mid x^{s_1}+y^{s_2}+z^{s_3}=0\right\} \cap S^5.
\label{eq:brieskorn}
\end{equation}
For $k=1$, these are the manifolds $\overline{\Sigma(2,3,5)}$ and $\overline{\Sigma(2,3,7)}$, which have been investigated with regards to the $\widehat{Z}$ quantum invariants proposed in \cite{GPPV}. For these manifolds, the $\widehat{Z}$ invariants are related to the order $5$ and $7$ mock theta functions $\chi_0(q)$ and $\mathcal{F}_0(q)$. Furthermore, in \cite{CCKPG}, it was shown that the other mock theta functions of orders $5$ and $7$ can be found as the $\widehat{Z}$ invariants with supersymmetric defects inserted into the corresponding $3d$ $\mathcal{N}=2$ theory $T[\overline{\Sigma(2,3,6\pm1)}]$. Other properties of Chern-Simons theory on Brieskorn spheres and Seifert manifolds have been studied in \cite{Chun17,Wu20,Chung20,AM22}.

In Section \ref{sec:vectors2} we propose special linear combinations of Mordell-Borel integrals that capture the entire vector-valued mock modular structure associated to the orientation reversed Brieskorn spheres. For the two lowest examples, $\Sigma(2,3,5)$ and $\Sigma(2,3,7)$, our numerical continuation method reproduces the dual $q$-series associated with the orientation reversed manifolds $\overline{\Sigma(2,3,5)}$ and $\overline{\Sigma(2,3,7)}$. 
In addition, we can go much further to higher $k$ values. We first confirm that our numerical method for crossing the natural boundary reproduces the results for $\Sigma(2, 3, 5)$ and $\Sigma(2, 3, 7)$, and then we present new results for the $q$-series duals for orientation reversed $\overline{\Sigma(2, 3, 11)}$ and $\overline{\Sigma(2, 3, 13)}$, in Section \ref{sec:2311}. In Section \ref{sec:comparison} we compare with other results in the literature. Results up to $\overline{\Sigma(2, 3, 19)}$ appear in Appendix \ref{sec:appendix-results2}. In addition, we also show results for $\overline{\Sigma(2, 5, 7)}$, $\overline{\Sigma(2, 5, 11)}$, $\overline{\Sigma(3,4,5)}$, and $\overline{\Sigma(3, 5, 7)}$, in Appendix \ref{sec:257}.

\subsection{Closed Resurgence Orbit of Vectors of Mordell-Borel Integrals}
\label{sec:vectors2}

The second resurgence orbit that we study is a set of  Mordell-Borel integrals defined as specially chosen linear combinations of four $JS_{(p, a)}(t)$ or $JC_{(p, a)}(t)$ integrals, in such a way that they form self-dual vectors under $t \rightarrow \pi^2/t$.
We define two such combinations, one of which is a linear combination of 4 of the $JS$ integrals in \eqref{eq:js}, while the other is a linear combination of 4 of the $JC$ integrals in \eqref{eq:jc}. We define, for $p$ odd and $n=1, 2, \dots, \frac{(p-1)}{2}$:
\begin{align}
L^{(p, n)}(t)&:= \text{sign}(6n-p)JS_{(12p,|12n-2p|)}(t) + JS_{(12p,14p-12n)}(t) 
\nonumber\\
    & \qquad + JS_{(12p,12n+2p)}(t) + JS_{(12p,10p-12n)}(t) ,
    \label{eq:ls}
\\
L_1^{(p, n)}(t)&:=  \text{sign}(12n-p) JC_{(6p,12n-p)}(t) + JC_{(6p,12n+p)}(t) 
\nonumber\\
    &\qquad + JC_{(6p,7p-12n)}(t) + \text{sign}(12n-5p)JC_{(6p,|12n-5p|)}(t).
    \label{eq:l1s}
\end{align}
These integrals are closely related to (but importantly different from, as explained below in Section \ref{sec:mock7}) the integrals $L(r, \alpha)$ and $L_1(r, \alpha)$ defined on page 118 of \cite{GM12}. 

We further specialize to $p=6k\pm1$, which is coprime to $2$ and $3$ for all $k$. With this choice, the integrals we define are related to the partition functions of Chern-Simons theory on the Brieskorn spheres $\overline{\Sigma(2,3,6k\pm1)}$. For $k=1$ these integrals appear in the mock theta relations of mock theta functions of orders $5$ and $7$ \cite{GM12}, providing a direct comparison for our analysis for the $k=1$ case. Using these relations as motivation, we define two $\ell=\frac{(p-1)}{2}$-component vectors of integrals:
\begin{eqnarray}
K_p(t):= 
\begin{pmatrix} L^{(p,1)}(t) \\ L^{(p,2)}(t) \\L^{(p,3)}(t) \\ \vdots\\ L^{(p,\ell)}(t)\end{pmatrix}
\qquad; \qquad
L_p(t):= 
\begin{pmatrix} L_1^{(p,\ell)}(t) \\ L_1^{(p,1)} (t) \\L_1^{(p,\ell-1)}(t) \\ \vdots\\ L_1^{p,\lfloor \frac{\ell+1}{2} \rfloor}(t) \end{pmatrix},
\label{eq:eq-kl}
\end{eqnarray}
where $\lfloor ... \rfloor$ denotes the floor function.

When analytically continued to the unary side ($t<0$) these vector-valued integrals have a unique decomposition into real and imaginary parts with the following structure, which follows from the earlier expressions for the analytic continuation of the individual $JS$ and $JC$ integrals (recall Section \ref{sec:false}):
\begin{align}
i\sqrt{\frac{48p|t|}{\pi}}\, K_p^{(j)}(t) &= q^{-2\Delta_{(p,j)}}\, X^{(j)}_p \left(\frac{1}{q^2}\right) +i \sqrt{\frac{\pi}{|t|}} \sum_{k=1}^{\frac{(p-1)}{2}}  M_{jk} \, q^{-2\Delta_{(p,k)}}\, X^{(k)}_p \left(\frac{1}{\qt^2}\right),
\label{eq:mock_q2}
\\
i\sqrt{\frac{24p |t|}{\pi}}\, L_p^{(j)}(t) &= (-1)^j q^{-\Delta_{(p,j)}}\, X^{(j)}_p\left(-\frac{1}{q}\right) +i \sqrt{\frac{\pi}{|t|}} \sum_{k=1}^{\frac{(p-1)}{2}}  (P_{\pi} N P^T_{\pi})_{jk}\, (-1)^k q^{-\Delta_{(p,k)}}\, X^{(k)}_p\left(-\frac{1}{\qt}\right).
\nonumber\\
\label{eq:mock_minusq}
\end{align}
Here $X_p$ is a $\frac{(p-1)}{2}$-component vector of $q$-series, with the $j^{th}$ entry multiplied by a rational power  $q^{-\Delta_{(p,j)}}$,  with
\begin{eqnarray}
    \Delta_{(p,j)} = \frac{(6j-p)^2}{24p}
    \qquad , \qquad j=1, 2, \dots, \frac{(p-1)}{2} \hspace{2mm}.
    \label{eq:delta-mock}
    \end{eqnarray}
    The $\frac{(p-1)}{2}\times \frac{(p-1)}{2}$ mixing matrices $M$ and $N$ in \eqref{eq:mock_q2}-\eqref{eq:mock_minusq} are  given by
\begin{align}
M_{jk} &= \frac{2}{\sqrt{p}}(-1)^{j+k+\lfloor \frac{p+3}{6} \rfloor}\sin{\left(\frac{6jk\pi}{p}\right)},
\label{eq:pm}
\\
N_{jk} &= \frac{2}{\sqrt{p}}(-1)^{j+k+\lfloor \frac{p+6}{12} \rfloor}\sin{\left(\frac{12jk\pi}{p}\right)},
\label{eq:pn}
\end{align}
and $P_{\pi}$ is the cyclic permutation matrix $(1,2,\ldots,\ell)\mapsto (\ell,1,\ell-1,2,\ell-2,\ldots,\lfloor \frac{\ell+1}{2} \rfloor)$.

We stress that there are two different vectors of Mordell-Borel integrals, $K_p(t)$ and $L_p(t)$, defined in \eqref{eq:eq-kl}, but their decompositions into real and imaginary parts on the unary side are expressed in terms of the very same vector of $q$-series in $X_p$, but evaluated at $1/q^2$ in one case, and at $(-1/q)$ in the other case.  

Moreover, on the unary side each entry in $X_p$ is proportional to a sum of four different false theta functions:
\begin{align}
    X^{(j)}_p(q) = q^{-\Delta_{(p,j)}}\Psi_{6p}^{(a_1)+(a_2)+(a_3)+(a_4)}(q) \, \in \mathbb{Z}[[q]]
    \label{eq:mock-x}
\end{align}
where
\begin{align}
    a_1=6j-p, \quad a_2=7p-6j, \quad a_3=6j+p, \quad a_4 = 5p-6j.
    \label{eq:as}
\end{align}
Here we have used the notation
\begin{align}
    \Psi^{n_{a_1}(a_1)+n_{a_2}(a_2)+\dots}_p := n_{a_1} \Psi^{(a_1)}_p + n_{a_2} \Psi^{(a_2)}_p + \dots
    \label{eq:ns}
\end{align}
together with the identities $\Psi^{(-a)}_p(q) = -\Psi^{(a)}_p(q)$ when the index $a$ is negative, and $\Psi^{(p+a)}_p(q) = -\Psi^{(p-a)}_p(q)$ when $a>p$. We also make use of the identity $\Psi^{(a)}_p(q^2)=\Psi^{(2a)}_{2p}(q)$. We may be tempted to write the above expression as a linear combination of the $q$-series $\Phi^{(a_i)}_{6p}(q)$, but this would not be correct. The correct definition is to collect the linear combination of false theta functions and then factor out $q^{-\Delta_{(p,j)}}$, corresponding to the rational power of $\Psi^{(a_1)}_{6p}(q)$ in the expression. The resulting expression is a $q$-series with integer coefficients and powers of $q$.
\\

\noindent{\bf Remarks:}
\begin{enumerate}
    \item The fact that each entry in $X_p$ is proportional to a sum of four different false theta functions mimics the fact that each component of the vector of integrals in \eqref{eq:ls}-\eqref{eq:l1s} is a linear combination of either four different $JS$ integrals or four different $JC$ integrals.
    \item The unary-side $q$-series and $\tilde{q}$-series identities \eqref{eq:mock_q2}-\eqref{eq:mock_minusq} are consistent with the modular properties of the vectors of integrals defined in \eqref{eq:eq-kl} on the non-unary side where $t>0$:
\begin{eqnarray}
K_p(t)&=&\left(\frac{\pi}{t}\right)^{3/2} 
M\,  K_p\left(\frac{\pi^2}{t}\right),
\label{eq:mock-k}
\\
L_p(t)&=&
\left(\frac{\pi}{t}\right)^{3/2} P_{\pi} N P^T_{\pi}\, L_p\left(\frac{\pi^2}{t}\right).
\label{eq:mock-l}
\end{eqnarray}
    \item The presence of the permutation matrices in the $L_p(t)$ integral relations is a choice we make so that the vector of $q$-series in the first term on the RHS is the same in both sets of relations \eqref{eq:mock_q2} and \eqref{eq:mock_minusq}. We note that this is different from the ordering given in \cite{GM12} for the order $5$ and order $7$ mock theta relations, but the proposed ordering of the $L_p(t)$ vector in \eqref{eq:eq-kl} is more natural for our goal, which is to find the dual $q$-series associated with all the entries  $X_p^{(j)}(q)$, for $j=1, \dots,(p-1)/2$.
\end{enumerate}

\subsection{Resurgence and the Property of Preservation of Relations}
\label{sec:preservation2}

The unary decomposition identities \eqref{eq:mock_q2}-\eqref{eq:mock_minusq} constitute a fundamental property of these particular linear combinations of the original Mordell-Borel integrals $JS_{(p, a)}(t)$ and $JC_{(p, a)}(t)$.

Therefore, the relations  \eqref{eq:mock_q2}-\eqref{eq:mock_minusq} must be preserved under unique continuation from $t<0$ back to the $t>0$ region in which the integrals were originally defined. Hence we are led to the following conjecture for the vectors of Mordell-Borel integral defined in \eqref{eq:mock-k}-\eqref{eq:mock-l}:
\\

\begin{conjecture}\label{conj:conjecture2}
For the linear combinations of Mordell-Borel integrals defined in \eqref{eq:ls}-\eqref{eq:eq-kl}, for all odd $p$ of the form $p=6k\pm 1$, $k\in \mathbb Z$, the Mordell-Borel integrals at $t>0$ decompose with the same algebraic form as in \eqref{eq:mock_q2}-\eqref{eq:mock_minusq}, with rational exponents given in \eqref{eq:delta-mock}:
\begin{eqnarray}
\sqrt{\frac{48p t}{\pi}}\, K_p^{(j)}(t) &=& q^{-2\Delta_{(p,j)}}\, X_p^{(j)}(q^2)^\vee + \sqrt{\frac{\pi}{t}} \sum_{k=1}^{\frac{(p-1)}{2}}  M_{jk} \, \tilde{q}^{\, -2\Delta_{(p,k)}}\, X_p^{(k)} \left(\tilde{q}^2\right)^\vee,
\label{eq:mock_q2_nonunary}
\\
\sqrt{\frac{24p t}{\pi}}\, L_p^{(j)}(t) &=& (-1)^j q^{-\Delta_{(p,j)}}\, X_p^{(j)}(-q)^\vee + \sqrt{\frac{\pi}{t}} \sum_{k=1}^{\frac{(p-1)}{2}}  (P_{\pi} N P^T_{\pi})_{jk}\, (-1)^k\tilde{q}^{\,- \Delta_{(p,k)}}\, 
X_p^{(k)} \left(-\tilde{q}\right)^\vee.
\nonumber\\
\label{eq:mock_minusq_nonunary}
\end{eqnarray}
Here $X^\vee$ is a $\frac{(p-1)}{2}$-component vector of $q$-series, with the $j^{th}$ entry multiplied by a rational power  $q^{-\Delta_{(p,j)}}$. The mixing matrices $M$ and $N$ (and the associated permutation matrix $P_\pi$) are the same as appearing in the decompositions \eqref{eq:mock_q2}-\eqref{eq:mock_minusq} on the unary side. These are also the same matrices appearing in the modular relations  \eqref{eq:mock-k}-\eqref{eq:mock-l} for the associated Mordell-Borel integrals.
\end{conjecture}

Therefore these $q$-series $X_k(q)^{\vee}$ serve as duals to the linear combinations of false theta functions in \eqref{eq:mock-x} under the operation $q \to \dfrac{1}{q}$:
\begin{eqnarray}
t \to -t\quad \left(q \to \dfrac{1}{q}\right) \quad \Rightarrow\quad
\Psi^{(a_1)+(a_2)+(a_3)+(a_4)}_{6p}(q) \longleftrightarrow q^{-\Delta_{(p,j)}}X^{(j)}_p(q)^{\vee},
\end{eqnarray}
where $\Psi$ and  $a_i$ are defined in \eqref{eq:mock-x}-\eqref{eq:ns}.

\subsection{Example: Order 5 Mock Theta Functions}
\label{sec:mock5}
For $p=5$, the integral vectors defined in \eqref{eq:eq-kl} have 2 components, 
\begin{eqnarray}
K_5(t)\!\! &=& \!\!
\begin{pmatrix}
    L^{(5,1)}(t)
    \\
    L^{(5,2)}(t)
\end{pmatrix}
=  \begin{pmatrix}
    \frac{1}{t}\int_0^\infty du\, e^{-60 u^2/t} \left[\frac{\sinh(2u)+\sinh(22 u)+\sinh(38 u)+\sinh(58 u)}{\sinh(60 u)}\right]
    \\
    \frac{1}{t}\int_0^\infty du\, e^{-60 u^2/t} \left[\frac{\sinh(14 u)+\sinh(26 u)+\sinh(34 u)+\sinh(46 u)}{\sinh(60 u)}\right]
\end{pmatrix},
\\ 
L_5(t)\!\! &=& \!\!
\begin{pmatrix}
    L_1^{(5,2)}(t) 
    \\
    L_1^{(5,1)}(t)
\end{pmatrix}
= 
\begin{pmatrix}
    \frac{1}{t}\int_0^\infty du\, e^{-30 u^2/t} \left[\frac{\cosh(u)+\cosh(11 u)+\cosh(19 u)-\cosh(29 u)}{\cosh(30 u)}\right]
    \\
    \frac{1}{t}\int_0^\infty du\, e^{-30 u^2/t} \left[\frac{\cosh(7 u)+\cosh(13 u)-\cosh(17 u)+\cosh(23 u)}{\cosh(30 u)}\right]
\end{pmatrix}.
    \label{eq:mock5-integrals}
\end{eqnarray}
The $q$-series decompositions \eqref{eq:mock_q2}-\eqref{eq:mock_minusq} on the unary side where $t<0$ ($q>1$) have the following $2$-component form:
\begin{eqnarray}
\hskip -1cm 
i\sqrt{\frac{240|t|}{\pi}} 
\begin{pmatrix}
L^{(5,1)}(t)
\\
L^{(5,2)}(t)
\end{pmatrix}
&=&
\begin{pmatrix}
q^{-\frac{1}{60}}X^{(1)}_5\left(\frac{1}{q^2}\right)
\\
q^{-\frac{49}{60}}X^{(2)}_5\left(\frac{1}{q^2}\right)
\end{pmatrix}
\nonumber
\\
&+& i\sqrt{\frac{\pi}{|t|}}  \frac{2}{\sqrt{5}} 
\begin{pmatrix}
\sin\left(\frac{\pi}{5}\right) & \sin\left(\frac{2\pi}{5}\right)
\\
\sin\left(\frac{2\pi}{5}\right) & - \sin\left(\frac{\pi}{5}\right) 
\end{pmatrix}
\begin{pmatrix}
\qt^{\, -\frac{1}{60}}X^{(1)}_5\left(\frac{1}{\qt^2}\right)
\\
\qt^{\, -\frac{49}{60}}X^{(2)}_5\left(\frac{1}{\qt^2}\right)
\end{pmatrix},
\label{eq:l5q}
\end{eqnarray}
\begin{eqnarray}
\hskip -1cm 
i\sqrt{\frac{120|t|}{\pi}} 
\begin{pmatrix}
L_1^{(5,2)}(t)
\\
L_1^{(5,1)}(t)
\end{pmatrix}
&=&
\begin{pmatrix}
-q^{-\frac{1}{120}}X^{(1)}_5\left(-\frac{1}{q}\right)
\\
q^{-\frac{49}{120}}X^{(2)}_5\left(-\frac{1}{q}\right)
\end{pmatrix}
\nonumber
\\
&+&i\sqrt{\frac{\pi}{|t|}} \frac{2}{\sqrt{5}} 
\begin{pmatrix}
-\sin\left(\frac{2\pi}{5}\right) & -\sin\left(\frac{\pi}{5}\right)
\\
-\sin\left(\frac{\pi}{5}\right) &  \sin\left(\frac{2\pi}{5}\right) 
\end{pmatrix}
\begin{pmatrix}
-\qt^{\, -\frac{1}{120}}X^{(1)}_5\left(-\frac{1}{\qt}\right)
\\
\qt^{\, -\frac{49}{120}}X^{(2)}_5\left(-\frac{1}{\qt}\right)
\end{pmatrix}.
\label{eq:l15q}
\end{eqnarray}
On the non-unary side ($t>0$; $q<1$), we can use the results in \cite{GM12} to write the identities for these integrals in $2\times 2$ matrix form:
\begin{eqnarray}
\hskip -1cm 
\sqrt{\frac{540t}{\pi}} 
\begin{pmatrix}
L^{(5,1)}(t)
\\
L^{(5,2)}(t)
\end{pmatrix}
&=&
\begin{pmatrix}
-q^{-\frac{1}{60}} (\chi_0(q^2)-2)
\\
-q^{-\frac{49}{60}} q^2\, \chi_1(q^2)
\end{pmatrix}
\nonumber
\\
&+&
\sqrt{\frac{\pi}{t}}  \frac{2}{\sqrt{5}} 
\begin{pmatrix}
\sin\left(\frac{\pi}{5}\right) & \sin\left(\frac{2\pi}{5}\right)
\\
\sin\left(\frac{2\pi}{5}\right) &  - \sin\left(\frac{\pi}{5}\right) 
\end{pmatrix}
\begin{pmatrix}
-\tilde{q}^{\, -\frac{1}{60}} (\chi_0(\tilde{q}^2)-2)
\\
-\tilde{q}^{\, -\frac{49}{60}} \tilde{q}^2 \, \chi_1(\tilde{q}^2)
\end{pmatrix},
\label{eq:mock5-chi02-chi12}
\end{eqnarray}
\begin{eqnarray}
\hskip -1cm 
\sqrt{\frac{270t}{\pi}} 
\begin{pmatrix}
L_1^{(5,2)}(t)
\\
L_1^{(5,1)}(t)
\end{pmatrix}
&=&
\begin{pmatrix}
-q^{-\frac{1}{120}} (\chi_0(-q)-2)
\\
-q^{-\frac{49}{120}} q\, \chi_1(-q)
\end{pmatrix}
\nonumber
\\
&+&
\sqrt{\frac{\pi}{t}} \frac{2}{\sqrt{5}} 
\begin{pmatrix}
\sin\left(\frac{\pi}{5}\right) &  \sin\left(\frac{2\pi}{5}\right)
\\
\sin\left(\frac{2\pi}{5}\right) &  - \sin\left(\frac{\pi}{5}\right) 
\end{pmatrix}
\begin{pmatrix}
-\tilde{q}^{\, -\frac{1}{120}} (\chi_0(-\tilde{q})-2)
\\
-\tilde{q}^{\, -\frac{49}{120}} \tilde{q} \, \chi_1(-\tilde{q})
\end{pmatrix}.
\label{eq:mock5-chi0m-chi1m}
\end{eqnarray}
Here $\chi_0(q)$ and $\chi_1(q)$ are the standard order 5 mock theta functions, with the following expressions and expansions \cite{GM12}:
\begin{eqnarray}
    (\chi_0(q)-2) &=&\sum_{n=0}^\infty \frac{q^n}{(q^{n+1}; q)_n} -2 \\
    &\sim & -1+q+q^2+2 q^3+q^4+3 q^5+2 q^6+ \dots \qquad,\quad q\to 0^+\\
    &\sim& -1-\frac{1}{q}-\left(\frac{1}{q}\right)^3-\left(\frac{1}{q}\right)^7+\left(\frac{1}{q}\right)^8+ \dots \qquad,\quad q\to \infty,
    \label{eq:mock5-chi0}
   \\
    \chi_1(q) &=&\sum_{n=0}^\infty \frac{q^n}{(q^{n+1}; q)_{n+1}}
    \\
    &\sim & 1+2 q+2 q^2+3 q^3+3 q^4+4 q^5+4 q^6+ \dots \qquad,\quad q\to 0^+\\
    &\sim& -\frac{1}{q}-\left(\frac{1}{q}\right)^2-\left(\frac{1}{q}\right)^3-\left(\frac{1}{q}\right)^5+\left(\frac{1}{q}\right)^{12}+ \dots \qquad,\quad q\to \infty.
   \label{eq:mock5-chi1}
\end{eqnarray}
Note that the large $q$ expansions, as occurs on the unary side where $q>1$, are unary series, while on the non-unary side where $q<1$, the small $q$ series are non-unary but with integer-valued coefficients. We also observe that both the unary-side and non-unary-side $q$-series decompositions are self-dual under $t\to \frac{\pi^2}{t}$, consistent with the modular properties of the integrals in \eqref{eq:mock-k}-\eqref{eq:mock-l}.

Comparing the unary-side relations \eqref{eq:l5q}-\eqref{eq:l15q} with the non-unary-side relations \eqref{eq:mock5-chi02-chi12}-\eqref{eq:mock5-chi0m-chi1m}, we see that the structure of Conjecture \ref{conj:conjecture2} is satisfied, with the identification of the duals under $q\to \frac{1}{q}$ as:
\begin{eqnarray}
    X^{(1)}_5(q)^{\vee} &=& -\frac{2}{3}(\chi_0(q)-2) =\frac{2}{3}\left(2-\sum_{n=0}^\infty \frac{q^n}{(q^{n+1}; q)_n}\right),
    \label{eq:mock51}
    \\
    X^{(2)}_5(q)^{\vee} &=& -\frac{2}{3}q\,\chi_1(q) = -\frac{2}{3}\sum_{n=0}^\infty \frac{q^n}{(q^{n+1}; q)_{n+1}}.
    \label{eq:mock52}
\end{eqnarray}

\subsection{The Importance of Uniqueness}
\label{sec:mock-uniqueness}

For the $p=5$ example discussed in the previous section, there are several relevant remarks to be made concerning uniqueness of the decomposition of Conjecture \ref{conj:conjecture2}, and the associated identifications in \eqref{eq:mock51}-\eqref{eq:mock52}.

For order 5 mock theta functions, on the non-unary side ($q<1$) there are two other, different, $2$-component vectors of $q$-series which satisfy the $K_5(t)$ vector identities \cite{GM12}:
\begin{eqnarray}
\sqrt{\frac{240t}{\pi}} \begin{pmatrix}
L^{(5,1)}(t)
\\
L^{(5,2)}(t)
\end{pmatrix}
\!\! = \!\!
\begin{pmatrix}
q^{-\frac{1}{60}} f_0(-q)
\\
-q^{\frac{11}{60}} f_1(-q)
\end{pmatrix}
\!+\! \sqrt{\frac{\pi}{t}} \frac{2}{\sqrt{5}} 
\!
\begin{pmatrix}
\sin\left(\frac{\pi}{5}\right) &\!\! \sin\left(\frac{2\pi}{5}\right)
\\
\sin\left(\frac{2\pi}{5}\right) &\!\! - \sin\left(\frac{\pi}{5}\right) 
\end{pmatrix}
\!\!
\begin{pmatrix}
\tilde{q}^{-\frac{1}{60}} f_0(-\tilde{q})
\\
-\tilde{q}^{\frac{11}{60}} f_1(-\tilde{q})
\end{pmatrix},
\label{eq:mock5-f0-f1}
\end{eqnarray}
\begin{eqnarray}
\sqrt{\frac{6t}{\pi}} \begin{pmatrix}
L^{(5,1)}(t)
\\
L^{(5,2)}(t)
\end{pmatrix}
\!\! = \!\!
\begin{pmatrix}
-q^{-\frac{1}{60}} \psi_0(-q)
\\
q^{\frac{11}{60}} \psi_1(-q)
\end{pmatrix}
\!\!+\!\!
\sqrt{\frac{\pi}{t}} \frac{2}{\sqrt{5}}
\!
\begin{pmatrix}
\sin\left(\frac{\pi}{5}\right) &\!\! \sin\left(\frac{2\pi}{5}\right)
\\
\sin\left(\frac{2\pi}{5}\right) &\!\! - \sin\left(\frac{\pi}{5}\right) 
\end{pmatrix}
\!\!\!
\begin{pmatrix}
-\tilde{q}^{-\frac{1}{60}} \psi_0(-\tilde{q})
\\
\tilde{q}^{\frac{11}{60}} \psi_1(-\tilde{q})
\end{pmatrix}.
\label{eq:mock5-psi0-psi1}
\end{eqnarray}
Here, $f_0(q)$, $f_1(q)$, $\psi_0(q)$ and $\psi_1(q)$ are other order 5 mock theta functions, defined as \cite{GM12}:
\begin{eqnarray}
    f_0(q):=\sum_{n=0}^\infty \frac{q^{n^2}}{(-q; q)_n} \qquad &;& \qquad  
    f_1(q):=\sum_{n=0}^\infty \frac{q^{n(n+1)}}{(-q; q)_n},
    \\
    \psi_0(q):=\sum_{n=0}^\infty q^{(n+1)(n+2)/2}\, (-q; q)_n \qquad &;&  \qquad  
    \psi_1(q):=\sum_{n=0}^\infty q^{n(n+1)/2}\, (-q; q)_n.
\end{eqnarray}
However, there are several features which preclude these identities from crossing the natural boundary from the $q<1$ region to the $q>1$ region:

\begin{enumerate}
    \item The large $q$ expansions of $f_0(q)$, $f_1(q)$, $\psi_0(q)$, and $\psi_1(q)$ are not unary, in contrast to the large $q$ expansions of $\chi_0(q)$ and $\chi_1(q)$ (recall \eqref{eq:mock5-chi0}-\eqref{eq:mock5-chi1}). So while $f_0(q)$, $f_1(q)$, $\psi_0(q)$, and $\psi_1(q)$ satisfy the correct $K_5(t)$ decomposition on the non-unary side, they cannot satisfy the corresponding decomposition of the integrals after continuation $t \to -t$, or $q\to 1/q$, where the decompositions are necessarily into unary series. In other words, the decompositions \eqref{eq:mock5-f0-f1}-\eqref{eq:mock5-psi0-psi1} do not cross the natural boundary, whereas the decompositions \eqref{eq:mock5-chi02-chi12}
    in terms of $\chi_0$ and $\chi_1$ do. 
    \item From the unary side, the $q$-series decompositions for $K_5(t)$ are in even powers of $1/q$ and $1/\qt$. However, on the non-unary side only the $\chi_0$ and $\chi_1$ expansions respect this parity structure.
    \item While $f_0(q)$, $f_1(q)$, $\psi_0(q)$, and $\psi_1(q)$ satisfy the decomposition identities for $K_5(t)$ on the non-unary side, $q<1$, they do not simultaneously satisfy also the $L_5(t)$ decomposition identities on the non-unary side. On the other hand, the $\chi_0$ and $\chi_1$ $q$-series  satisfy {\it both} sets of identities, as is the case on the unary side $q>1$.
    \item By construction, other $q$-series decompositions of both integral vectors $K_5(t)$ and $L_5(t)$, must differ from those above by modular forms. It can be proven rigorously that there is no such modular form that can be added to the decompositions in terms of $\chi_0$ and $\chi_1$ such that both decomposition identities \eqref{eq:mock5-chi02-chi12}-\eqref{eq:mock5-chi0m-chi1m} are satisfied \cite{CD25}. 
\end{enumerate}

\subsection{Example: Order 7 Mock Theta Functions}
\label{sec:mock7}
For $p=7$, the vector-valued integrals defined in \eqref{eq:eq-kl} have $\frac{(p-1)}{2}=3$ components, 
\begin{equation}
K_7(t)=
\begin{pmatrix}
    L^{(7,1)}(t)
    \\
    L^{(7,2)}(t)
    \\
    L^{(7,3)}(t)
\end{pmatrix}
=\begin{pmatrix}
    \frac{1}{t}\int_0^\infty du\, e^{-84 \frac{u^2}{t}}\left[\frac{-\sinh(2u)+\sinh(26 u)+\sinh(58 u)-\sinh(82 u)}{\sinh(84 u)}\right]
    \\
    \frac{1}{t}\int_0^\infty du\, e^{-84 \frac{u^2}{t}} \left[\frac{\sinh(10 u)+\sinh(38 u)+\sinh(46 u)+\sinh(74 u)}{\sinh(84 u)}\right]
    \\
    \frac{1}{t}\int_0^\infty du\, e^{-84 \frac{u^2}{t}} \left[\frac{\sinh(22 u)+\sinh(34 u)+\sinh(50 u)+\sinh(62 u)}{\sinh(84 u)}\right]
\end{pmatrix},
\label{eq:mock7-integralsL}
\end{equation}
\begin{equation}
L_7(t)=
\begin{pmatrix}
    L_1^{(7,3)}(t) 
    \\
    L_1^{(7,1)}(t)
    \\
    L_1^{(7,2)}(t)
\end{pmatrix}
= 
\begin{pmatrix}
    \frac{1}{t}\int_0^\infty du\, e^{-42 \frac{u^2}{t}} \left[\frac{\cosh(u)+\cosh(13 u)+\cosh(29 u)+\cosh(41 u)}{\cosh(42 u)}\right]
    \\
    \frac{1}{t}\int_0^\infty du\, e^{-42 \frac{u^2}{t}} \left[\frac{\cosh(5 u)- \cosh(19 u)+\cosh(23 u)+\cosh(37 u)}{\cosh(42 u)}\right]
    \\
    \frac{1}{t}\int_0^\infty du\, e^{-42 \frac{u^2}{t}} \left[\frac{\cosh(11 u)+\cosh(17 u)+\cosh(25 u)-\cosh(31 u)}{\cosh(42 u)}\right]
\end{pmatrix}.
\label{eq:mock7-integralsL1}
\end{equation}
After continuing to the unary side, $t<0$, these integrals uniquely split into real and imaginary parts, with  the following $3$-component structure (recall \eqref{eq:mock_q2}-\eqref{eq:pn}):
\begin{eqnarray}
\hskip -1cm 
i\sqrt{\frac{336|t|}{\pi}} 
\begin{pmatrix}
L^{(7,1)}(t)
\\
L^{(7,2)}(t)
\\
L^{(7,3)}(t)
\end{pmatrix}
&=&
\begin{pmatrix}
q^{-\frac{1}{84}}X^{(1)}_7\left(\frac{1}{q^2}\right)
\\
q^{-\frac{25}{84}}X^{(2)}_7\left(\frac{1}{q^2}\right)
\\
q^{-\frac{121}{84}}X^{(3)}_7\left(\frac{1}{q^2}\right)
\end{pmatrix} 
\nonumber
\\
 &\hskip -4cm + & \hskip -2cm i\sqrt{\frac{\pi}{|t|}}  \frac{2}{\sqrt{7}} 
\begin{pmatrix}
\sin\left(\frac{\pi}{7}\right) & \sin\left(\frac{2\pi}{7}\right) & \sin\left(\frac{3\pi}{7}\right)
\\
\sin\left(\frac{2\pi}{7}\right) &  -\sin\left(\frac{3\pi}{7}\right) & \sin\left(\frac{\pi}{7}\right)
\\
\sin\left(\frac{3\pi}{7}\right) &  \sin\left(\frac{\pi}{7}\right) & -\sin\left(\frac{2\pi}{7}\right)
\end{pmatrix}
\begin{pmatrix}
\qt^{\, -\frac{1}{168}}X^{(1)}_7\left(\frac{1}{\qt^2}\right)
\\
\qt^{\, -\frac{25}{168}}X^{(2)}_7\left(\frac{1}{\qt^2}\right)
\\
\qt^{\, -\frac{121}{168}}X^{(3)}_7\left(\frac{1}{\qt^2}\right)
\end{pmatrix},
\label{eq:l7}
\end{eqnarray}
\begin{eqnarray}
\hskip -1cm 
i\sqrt{\frac{168|t|}{\pi}}
\begin{pmatrix}
L_1^{(7,3)}(t)
\\
L_1^{(7,1)}(t)
\\
L_1^{(7,2)}(t)
\end{pmatrix}
&=&
\begin{pmatrix}
-q^{-\frac{1}{168}}X^{(1)}_7\left(-\frac{1}{q}\right)
\\
q^{-\frac{25}{168}}X^{(2)}_7\left(-\frac{1}{q}\right)
\\
-q^{-\frac{121}{168}}X^{(3)}_7\left(-\frac{1}{q}\right)
\end{pmatrix}
\nonumber\\
&\hskip -4cm + & \hskip -2cm i\sqrt{\frac{\pi}{|t|}}  \frac{2}{\sqrt{7}}
\begin{pmatrix}
\sin\left(\frac{3\pi}{7}\right) & \sin\left(\frac{\pi}{7}\right) & \sin\left(\frac{2\pi}{7}\right)
\\
\sin\left(\frac{\pi}{7}\right) &  \sin\left(\frac{2\pi}{7}\right) & -\sin\left(\frac{3\pi}{7}\right)
\\
\sin\left(\frac{2\pi}{7}\right) &  -\sin\left(\frac{3\pi}{7}\right) & -\sin\left(\frac{\pi}{7}\right)
\end{pmatrix}
\begin{pmatrix}
-\qt^{\, -\frac{1}{168}}X^{(1)}_7\left(-\frac{1}{\qt}\right)
\\
\qt^{\, -\frac{25}{168}}X^{(2)}_7\left(-\frac{1}{\qt}\right)
\\
-\qt^{\, -\frac{121}{168}}X^{(3)}_7\left(-\frac{1}{\qt}\right)
\end{pmatrix}.
\label{eq:l17}
\end{eqnarray}
Recall that the mixing matrices and the rational exponents follow uniquely from the Mordell-Borel integrals.

On the non-unary side ($t>0$; $q<1$), we can use the results in \cite{GM12} to write the identities for these integrals in $3\times 3$ matrix form:
\begin{eqnarray}
\hskip -1cm 
\sqrt{\frac{336 t}{\pi}} 
\begin{pmatrix}
L^{(7,1)}(t)
\\
L^{(7,2)}(t)
\\
L^{(7,3)}(t)
\end{pmatrix}
&=&
\begin{pmatrix}
-q^{-\frac{1}{84}}\mathcal F_0(q^2)
\\
-q^{-\frac{25}{84}}\mathcal F_1(q^2)
\\
-q^{-\frac{121}{84}} q^2 \mathcal F_1(q^2)
\end{pmatrix} 
\nonumber\\
&\hskip -4cm + & \hskip -2cm
\sqrt{\frac{\pi}{t}}  \frac{2}{\sqrt{7}} 
\begin{pmatrix}
\sin\left(\frac{\pi}{7}\right) & \sin\left(\frac{2\pi}{7}\right) & \sin\left(\frac{3\pi}{7}\right)
\\
\sin\left(\frac{2\pi}{7}\right) &  -\sin\left(\frac{3\pi}{7}\right) & \sin\left(\frac{\pi}{7}\right)
\\
\sin\left(\frac{3\pi}{7}\right) &  \sin\left(\frac{\pi}{7}\right) & -\sin\left(\frac{2\pi}{7}\right)
\end{pmatrix}
\begin{pmatrix}
-\qt^{\, -\frac{1}{84}}\mathcal F_0(\qt^2)
\\
-\qt^{\, -\frac{25}{84}}\mathcal F_1(\qt^2)
\\
-\qt^{\, -\frac{121}{84}} \qt^2 \mathcal F_1(\qt^2)
\end{pmatrix},
\label{eq:l-7}
\end{eqnarray}
\begin{eqnarray}
\hskip -1cm 
\sqrt{\frac{168 t}{\pi}}
\begin{pmatrix}
L_1^{(7,3)}(t)
\\
L_1^{(7,1)}(t)
\\
L_1^{(7,2)}(t)
\end{pmatrix}
&=&
\begin{pmatrix}
q^{-\frac{1}{168}}\mathcal F_0(-q)
\\
-q^{-\frac{25}{168}}\mathcal F_1(-q)
\\
-q^{-\frac{121}{168}} q\,\, \mathcal F_2(-q)
\end{pmatrix}
\nonumber\\
&\hskip -4cm + & \hskip -2cm \sqrt{\frac{\pi}{t}}  \frac{2}{\sqrt{7}}
\begin{pmatrix}
\sin\left(\frac{3\pi}{7}\right) & \sin\left(\frac{\pi}{7}\right) & \sin\left(\frac{2\pi}{7}\right)
\\
\sin\left(\frac{\pi}{7}\right) &  \sin\left(\frac{2\pi}{7}\right) & -\sin\left(\frac{3\pi}{7}\right)
\\
\sin\left(\frac{2\pi}{7}\right) &  -\sin\left(\frac{3\pi}{7}\right) & -\sin\left(\frac{\pi}{7}\right)
\end{pmatrix}
\begin{pmatrix}
\qt^{\, -\frac{1}{168}}\mathcal F_0(-\qt)
\\
-\qt^{\, -\frac{25}{168}}\mathcal F_1(-\qt)
\\
-\qt^{\, -\frac{121}{168}}\qt \, \mathcal F_2(-\qt)
\end{pmatrix}.
\label{eq:l1-7}
\end{eqnarray}
Comparing the unary-side relations \eqref{eq:l7}-\eqref{eq:l17} with the non-unary-side relations \eqref{eq:l-7}-\eqref{eq:l1-7}, we see that the structure of Conjecture \ref{conj:conjecture2} is satisfied, with the identification of the duals under $q\to \frac{1}{q}$ as:
\begin{eqnarray}
    X_1^{(7)}(q)^\vee &=& -\mathcal{F}_0(q) =-\sum_{n=0}^\infty \frac{q^{n^2}}{(q^{n+1}; q)_n} 
    \label{eq:mock7-ids-1}, \\
    X_2^{(7)}(q)^\vee &=& -\mathcal{F}_1(q) =-\sum_{n=0}^\infty \frac{q^{(n+1)^2}}{(q^{n+1}; q)_{n+1}} ,
    \label{eq:mock7-ids-2} \\
    X_3^{(7)}(q)^\vee &=& -q \,\mathcal{F}_2(q) =-q\sum_{n=0}^\infty \frac{q^{n(n+1)}}{(q^{n+1}; q)_{n+1}}.
  \label{eq:mock7-ids-3}  
\end{eqnarray}
\noindent{\bf Remarks:} 
\begin{enumerate}
    \item 
    The self-dual $q$-series decompositions \eqref{eq:l-7}-\eqref{eq:l1-7} are consistent with the modular properties under $t\to \frac{\pi^2}{t}$ of the integral vectors defined in \eqref{eq:eq-kl}. This self-duality structure is an important part of our numerical algorithm for finding the dual $q$-series.

    \item As in the order 5 mock example in Section \ref{sec:mock5}, the large $q$ expansions of $\mathcal F_0(q)$, $\mathcal F_1(q)$ and $\mathcal F_2(q)$ are manifestly unary, as they must be in order to agree with the decompositions of the analytically continued integrals under $t\to -t$.
    \item

  The non-unary-side relations \eqref{eq:l-7}-\eqref{eq:l1-7} identities involve a small, but significant, rearrangement of the way they are expressed in \cite{GM12}. As written there, the relevant identities on page 119 of \cite{GM12} are not self-dual under $t\to \frac{\pi^2}{t}$. However, this can be repaired by a simple redefinition of one of the integrals. First, in the first set of identities we rescale $t\to 2t$. But the asymmetry arises in the expressions involving $\mathcal F_0(q)$ and $\mathcal F_0(-q)$, where subtractions $-2$ appear on the left-hand-side, but not on the right-hand-side. However, this can be fixed by absorbing these subtraction terms into a simple redefinition of the integrals on the right-hand-side. This then corresponds to our definitions in \eqref{eq:eq-kl}. Fundamentally, this is then a more natural way to express these relations, ensuring that the Borel transform functions have a well-defined Fourier transform, which therefore implies the relations between the small $t$ and large $t$ expansions. Recall Section \ref{sec:mordell-borel}.
 
    This ensures that the self-dual relations under $t \rightarrow \pi^2/t$ will hold with a similar structure to the order $5$ examples. Our definitions of the integrals $L^{(p,n)}(t)$ and $L_1^{(p,n)}(t)$ in \eqref{eq:ls}-\eqref{eq:l1s} automatically take care of these shifts and are manifestly self-dual for all $p$.
\item While our redefinition of the integrals makes  manifest the self-duality of the $q$-series relations, we note that the integrals that require this subtraction are in 1-1 correspondence with the complex $SL(2,\mathbb{C})$ connections on $\overline{\Sigma(2,3,p)}$. This is evident by the form of the integrals appearing in the $K_p(t)$ vector (\ref{eq:ls}), where minus signs appear in the first two integrals when $p>6n$, i.e. depending on the sign of $6n-p$. The presence of these minus signs leads to the vanishing of the transseries parameters $n_{\alpha_n}$: see equation (3.70) in \cite{GMP}. In Table \ref{tab:subtractions} we indicate which integrals for a given value of $p$ require these subtractions. 
\begin{table}[h!]
\begin{tabular}{ | m{2em} | m{5em}| m{5em} |} 
  \hline
  $p$ & $n$ & $|6n-p|$\\ 
  \hline
  $5$ & none & none \\
  \hline
  $7$ & $1$ & $1$ \\
  \hline
  $11$ & $1$ & $5$ \\
  \hline
  $13$ & $1$, $2$ & $7$, $1$\\
  \hline
  $17$ & $1$, $2$ & $11$, $5$ \\
  \hline
  $19$ & $1$, $2$, $3$ & $13$, $7$, $1$ \\
  \hline
\end{tabular}
\caption{Integrals in the vector $K_p(t)$ (\ref{eq:eq-kl}) which correspond to complex $SL(2,\mathbb{C})$ connections on $\overline{\Sigma(2,3,p)}$. The second column denotes the entry in the vector, and the third column gives the corresponding value of $|6n-p|$, related to the location of the poles in the Borel plane corresponding to the complex connection, see Table 2 in \cite{Wu20}.} 
\label{tab:subtractions}
\end{table}
\end{enumerate}

\section{Numerical Fitting Algorithm on the Non-Unary Side}
\label{sec:algorithm}

Based on the algebraic and functional relations underlying the two resurgence orbits of vectors of Mordell-Borel integrals defined above in Sections \ref{sec:class1} and \ref{sec:class2}, concerning their unique decompositions into unary $q$-series and $\qt$-series, we invoke the property of {\it preservation of relations} of resurgent functions to argue that after analytic continuation back to the non-unary side ($t>0$; $q<1$) this algebraic and functional structure should be preserved. This leads to \textbf{\conjref{conj:conjecture1}} for the first resurgence orbit involving the duals of the false theta functions, and \textbf{\conjref{conj:conjecture2}} for the second resurgence orbit involving the duals associated with the orientation reversed Brieskorn spheres $\overline{\Sigma(2, 3, 6k\pm 1)}$. We have shown that this structure gives a consistent crossing of the boundary for the first two of the false theta duals (i.e., the order 3 and order 10 mock theta functions) and for the first two of the Brieskorn sphere duals (i.e., the order 5 and order 7 mock theta functions). 

To proceed beyond these known examples, we now introduce a numerical procedure that finds these identities numerically, and which can be extended to generate further examples for each of the two resurgence orbits.

The numerical task becomes that of finding a unique decomposition of the relevant vectors of Mordell-Borel integrals into the conjectured algebraic form, on the non-unary side. Recall that the decomposition into unary series is uniquely fixed on the Stokes line where $t<0$, where the decomposition is simply the unique decomposition into real and imaginary parts. 

The basic idea is simple. We take advantage of the duality structure, under $t \to \frac{\pi^2}{t}$, of the decomposition identities \eqref{eq:aq-dual}-\eqref{eq:dq-dual} in \textbf{\conjref{conj:conjecture1}} and \eqref{eq:mock_q2_nonunary}-\eqref{eq:mock_minusq_nonunary} in \textbf{\conjref{conj:conjecture2}}, to fit the $q$-series and $\qt$-series to the same functional form, in the most sensitive region, which is near the self-dual point $t=\pi$ where $q=\qt$:
\begin{eqnarray}
    t=\pi \qquad \longleftrightarrow \qquad q=\qt.
    \label{eq:sdt}
\end{eqnarray}
Note that at this point, $q=\qt$ is small:
\begin{eqnarray}
    q_c=\qt_c=e^{-\pi} \approx 0.0432.
    \label{eq:qc}
\end{eqnarray}
This basic idea applies to both classes of algebraic structures, but there are some minor strategic differences because of the different algebraic structures.

We do not yet have a rigorous theory of convergence for the numerical procedure described above, which generates the \( q \), \( \tilde{q} \) decomposition of Mordell integrals. For an empirical analysis, see \cite{CDGG}. 

We also observe, at an empirical level, that the eigenvectors corresponding to the smallest eigenvalue of the matrices arising from the matching problem, approach solutions of the associated homogeneous problem in the limit of large matrix size. Properly adapted, this provides a practical method for generating the \( q \)-series of various Hauptmoduls and modular forms. 
This observation also suggests that convergence is intimately linked to the uniqueness of the \( q \)–\( \tilde{q} \) decompositions, for which we have obtained rigorous results in certain special cases \cite{CD25}.

\subsection{Fitting Strategy for the First Resurgence Orbit: Duals of the False Thetas}
\label{sec:algorihm1}

For the first resurgence orbit, comprised of duals of the false theta functions, discussed in Section \ref{sec:class1}, we are seeking $q$-series expansions that match the structure in \textbf{\conjref{conj:conjecture1}} \eqref{eq:aq-dual}-\eqref{eq:dq-dual}. Recall that this algebraic structure involves two decompositions that are manifestly self-dual under $t\to \frac{\pi^2}{t}$, namely the A and B sets in\eqref{eq:aq-dual}-\eqref{eq:bq-dual}, and two decompositions that mix under $t\to \frac{\pi^2}{t}$, namely the C and D sets in \eqref{eq:cq-dual}-\eqref{eq:dq-dual}. 

Superficially, it looks like the most efficient strategy would be to solve \eqref{eq:aq-dual} to find the $\Phi_p^{(2j)}$ with {\bf even} $a$ indices, and \eqref{eq:bq-dual} to find the $\Phi_p^{(2j-1)}$ with {\bf odd} $a$ indices. However, recall that already for the known examples, $p=3$ and $p=5$ discussed in Sections \ref{sec:p3} and \ref{sec:p5}, uniqueness of the solutions required also satisfying the mixed identities \eqref{eq:cq-dual}-\eqref{eq:dq-dual}. 
This suggests that the most efficient strategy to find a {\bf unique} solution would be to solve one of the self-dual decomposition identities, and then use the resulting $q$-series in one of the mixed identities to generate the other set. Empirically we find that the fitting procedure is more numerically stable for set A, \eqref{eq:aq-dual}, which generates the $\Phi_p^{(2j)}$ with {\bf even} $a$ indices. We can then feed these into the mixed identity for set C, \eqref{eq:cq-dual}, which generates the $\Phi_p^{(2j-1)}$ with {\bf odd} $a$ indices. 
We then check by plotting that the remaining identities, \eqref{eq:bq-dual} and \eqref{eq:dq-dual}, are satisfied. Note that \eqref{eq:dq-dual} is in fact equivalent to \eqref{eq:cq-dual}, so we are really just checking consistency with \eqref{eq:bq-dual}.

\begin{itemize}
    \item

The first step of the numerical algorithm is to generate expansions for set A of the integrals on the LHS of the relevant decomposition identity \eqref{eq:aq-dual}, by expanding in $s=1-\frac{t}{\pi}$, and collecting terms in powers of $s$. This can be done to very high order in $s$, and with very high numerical precision.

\item
The second step of the numerical algorithm is based on the fact that the decomposition on the RHS of \eqref{eq:aq-dual} involves {\bf the same} series $\Phi_p^{(2j)}(q)$ and $\Phi_p^{(2j)}(\qt)$. So we make the ans\"atze
\begin{eqnarray}
    \Phi_p^{(2j)}(q)^\vee =\sum_{n=0}^O c_n^{(2j)}\, q^{n} 
    \qquad, \qquad
    \Phi_p^{(2j)}(\qt)^\vee =\sum_{n=0}^O c_n^{(2j)}\, 
    \qt^{\,n},
    \label{eq:a-ansatz}
\end{eqnarray}
with the same to-be-determined coefficients $c_n^{(2j)}$ appearing in both $\Phi_p^{(2j)}(q)^\vee$ and $\Phi_p^{(2j)}(\qt)^\vee$. These expansions are made to some chosen order $n=O$. 
Since the same coefficients $c_n^{(2j)}$ appear in the expansions of $\Phi_p^{(2j)}(q)^\vee$ and $\Phi_p^{(2j)}(\qt)^\vee$, we then have to solve a linear algebra problem to determine the coefficients $c_n^{(2j)}$.

\item The third step is to use this consistent set of coefficients $c_n^{(2j)}$, with {\bf even} $a$ indices, in the mixed decomposition identity for the C set, \eqref{eq:cq-dual}, in order to determine the coefficients $c_n^{(2j-1)}$ for {\bf odd} $a$ indices:
\begin{eqnarray}
    \Phi_p^{(2j-1)}(q)^\vee =\sum_{n=0}^O c_n^{(2j-1)}\, q^{2n}. 
    \label{eq:c-ansatz}
\end{eqnarray}
We have encoded the requirement that these be expansions in $q^2$, not just in $q$. This structure is inherited from the unary side, and is a fundamental property of the integrals that we are trying to decompose.
The $\qt$ part of the RHS of the decomposition \eqref{eq:cq-dual} is expressed in terms of the $\Phi_p^{(2j)}(q)$ with {\bf even} $a$ indices, and is therefore now known. So we can expand the integrals on the LHS also near the self-dual point $t=\pi$, producing a linear algebra problem to determine the coefficients $c_n^{(2j-1)}$ of the $q$ part of the RHS. 

This then means that both the self-dual A set and the mixed C set of decomposition identities, \eqref{eq:aq-dual} and \eqref{eq:cq-dual} respectively, are satisfied. The D set decomposition \eqref{eq:dq-dual} is automatically satisfied because it is equivalent to the C set under $q\leftrightarrow \qt$.

\item 
Finally, we can make plots to confirm that the B set decomposition \eqref{eq:bq-dual} is satisfied.
\end{itemize}
\bigskip 

\noindent{\bf Remarks:}
\begin{enumerate}

\item 
As a check, we have confirmed that our numerical algorithm generates the known results for $p=3$ and $p=5$, which were discussed in Sections \ref{sec:p3} and \ref{sec:p5}.

\item 
Our numerical algorithm finds new results for the duals of false theta functions, $\Phi_p^{(a)}(q)^\vee$ for $a=1, 2, \dots, (p-1)$, for higher odd values of $p$. These are listed and discussed in Section \ref{sec:results-p-class} for $p=7$ and $p=9$, and in the Appendix \ref{sec:appendix-results1} for $p=11$, $p=13$, $p=15$, $p=17$, and $p=19$. All of these are new, beyond the known results for $p=3$ and $p=5$ discussed already in Sections \ref{sec:p3} and \ref{sec:p5}.  
    
\item We do not impose the condition that the coefficients $c_n^{(a)}$, in the ans\"atze \eqref{eq:a-ansatz}-\eqref{eq:c-ansatz}, should be integers, but we find that they are integers to high precision, up to a common overall multiplicative rational number which for this class is a power of $1/2$. See the results listed in Section \ref{sec:results-p-class} and in the Appendix \ref{sec:appendix-results1}. 
    This overall $(1/2)^c$ factor is consistent with Conjecture 2.1 of \cite{GPPV}:
\begin{equation}
   \widehat{Z}_b(q) \in 2^{-c} q^{\Delta_b} \mathbb{Z}[[q]] \quad \Delta_b \in \mathbb{Q}, \quad c \in \mathbb{Z}_{+} .
   \label{eq:2c}
\end{equation}
    
    \item We can extend the number of coefficients by a bootstrapping procedure, as follows. We first solve for a small value of the order $O$, for example $O=10$. After finding integer coefficients to a desired precision, we can restart the fitting procedure taking these first coefficients to be exactly integers, and then add another 10 unknown coefficients. With this procedure running on a simple laptop we can easily find $60$ integer coefficients for low values of $p$, and up to $30$ integer coefficients for higher values such as $p=17$ and $p=19$.
    
    \item The algebraic complexity of this procedure clearly increases as $p$ increases, because we are solving for a vector of $\frac{(p-1)}{2}$ integrals and $(p-1)$ different $q$-series, all of which are coupled. However, we have been able to extend this method up to $p=19$ on a simple laptop. With more serious programming it should be possible to probe even higher values of $p$.

    \item As a further consistency check, in \cite{ceff2} we analyze the large order growth of the integer coefficients, and show that these match the analytic asymptotic  form that can be derived from the original Mordell-Borel integrals.
\end{enumerate}

\subsection{Fitting Strategy for the Second Resurgence Orbit: Duals for Orientation Reversed Brieskorn Spheres}
\label{sec:algorihm2}

For the second resurgence orbit of dual $q$-series, related to Chern-Simons theory on the orientation reversed Brieskorn spheres $\overline{\Sigma(2, 3, 6k\pm 1)}$, the algebraic structure is different, given by \textbf{\conjref{conj:conjecture2}} in Section \ref{sec:preservation2}: see \eqref{eq:mock_q2_nonunary}-\eqref{eq:mock_minusq_nonunary}. Since the algebraic structure is different, the form of the ansatz that we use is correspondingly different. However, the basic strategy is the same: we expand the relevant Mordell-Borel integrals near $t=\pi$ and match to ansatz expansions satisfying \eqref{eq:mock_q2_nonunary}-\eqref{eq:mock_minusq_nonunary}. Now there is only one vector of unknown dual $q$-series $X^{(j)}_p(q)^{\vee}$, but these must satisfy the decompositions of {\bf two separate vectors} of Mordell-Borel integrals. For this class, there are no mixed identities, so the procedure is somewhat simpler. 

We find that it is more numerically stable to use the self-dual identities \eqref{eq:mock_q2_nonunary} involving the $K^{(j)}_p(t)$ integrals to solve for the coefficients of $X^{(j)}_p(q^2)^{\vee}$. We then check that the $L^{(j)}_p(t)$ identities are also satisfied. 
\\

\noindent{\bf Remarks:}
\begin{enumerate}

\item 
As a check, we have confirmed that our numerical algorithm generates the known results for the orientation reversed Brieskorn spheres $\overline{\Sigma(2, 3, 5)}$ and $\overline{\Sigma(2, 3, 7)}$, (i.e., for $p=5$ and $p=7$), in terms of the order 5 and order 7 mock theta functions, respectively, as discussed in Sections \ref{sec:mock5} and \ref{sec:mock7}.

\item 
Our numerical algorithm finds new results for the dual $q$-series for the orientation reversal of the  Brieskorn spheres $\overline{\Sigma(2, 3, 11)}$ and $\overline{\Sigma(2, 3, 13)}$, discussed in Section \ref{sec:results-mock-class}, and for $\overline{\Sigma(2, 3, 17)}$, $\overline{\Sigma(2, 3, 19)}$, $\overline{\Sigma(2, 3, 25)}$, and $\overline{\Sigma(2, 3, 35)}$ in Appendix \ref{sec:appendix-results2}.

\item 
In addition, we present new results for other Brieskorn sphere manifolds $\overline{\Sigma(2, 5, 7)}$, $\overline{\Sigma(2, 5, 11)}$, $\overline{\Sigma(3,4, 5)}$ and $\overline{\Sigma(3, 5, 7)}$ in the Appendix \ref{sec:other-brieskorn}.

    \item As a further consistency check, in \cite{ceff2} we analyze the large order growth of the integer coefficients, and show that these match the analytic form that can be derived from the original Mordell-Borel integrals.
    \item As with the first class, we find that the coefficients of the $q$-series are integers up to an overall rational factor for each value of $p$. However, we do not find the universal factor of $1/2^c$ as in the first class for all of our examples, see Table \ref{tab:mock35secondclass}.
\end{enumerate}

\section{Results: Duals for False Theta Functions}
\label{sec:results-p-class}

In this section we display plots and tables with the results of our numerical algorithm for the known case $p=5$, and the new results for $p=7$ and $p=9$.\footnote{In accompanying supplementary files we list all the expansions of these new $q$-series.} For each value of $p$ we list the first terms of the relevant $q$-series on both sides of the natural boundary: i.e., the $\Psi_p^{(a)}$ appearing in the decompositions \eqref{eq:aq} and \eqref{eq:cq}, and the $\Psi_p^{(a)\,^\vee}$ in \eqref{eq:aq-dual}-\eqref{eq:cq-dual}. We also present plots to illustrate the smooth crossing of the natural boundary. For brevity's sake, we only include plots for the set $A$ and set $C$ relations, as these cover all of the dual false thetas. In Appendix \ref{sec:appendix-results1} we list the analogous results for $p=11, 13, 15, 17, 19$. When plotting truncated $q$-series, we use diagonal Pad\'e approximants to improve the convergence near the boundary. 

\subsection{Warm up: $p=5$. Order $10$ Mock Theta Functions}
Here we give results from our algorithm for the known case of $p=5$. Recall Section \ref{sec:p5}.
The results of our numerical fitting procedure for $p=5$, showing the false theta function $\Psi^{(a)}_5(q)$ and its numerically found dual $\tilde\Psi^{(a)}_5(q)^\vee$, are displayed in Table \ref{tab:p5}.\\ 
\begin{center}
\begin{table}[H]
\begin{tabular}{ | m{1em} | m{7.6cm}| m{7.6cm}| } 
  \hline
  $a$ & Unary false theta $\Psi^{(a)}_5(q)$ & 
  Non-unary dual $q$-series $\Psi^{(a)}_5(q)^{\vee}$ \\ 
  \hline
  $1$ & $q^{1/20}\left(1-q^4+q^6-q^{18}+q^{22}-q^{42}+\dots\right)$ 
  & $q^{-1/20}\left(1-q^2+q^4+q^8-2q^{10}+\dots\right)$ \\
  \hline
  $2$ & $q^{1/5}\left(1 - q^3 + q^7 - q^{16} + q^{24} - q^{39}+\dots\right)$ & $q^{-1/5}\left(q - q^2 + 2 q^3 - 2 q^4 + 2 q^5+\dots\right)$ \\
  \hline
  $3$ & $q^{9/20}\left(1 - q^2 + q^8 - q^{14} + q^{26} - q^{36}+\dots\right)$ & $q^{-9/20}\left(q^2 - q^4 + q^6 - 2 q^8 + 2  q^{10}+\dots\right)$ \\
  \hline
  $4$ & $q^{4/5}\left(1 - q + q^9 - q^{12} + q^{28} - q^{33}+\dots\right)$ & $q^{-4/5}\left(q-2  q^2 + 2 q^3 - 3 q^4 + 4 q^5+\dots\right)$ \\
  \hline
\end{tabular}
\caption{Small $q$ expansions of the $p=5$ false theta functions $\Psi^{(a)}_5(q)$ and their duals $\Psi^{(a)}_5(q)^\vee$, for $1\leq a \leq 4$.}
\label{tab:p5}
\end{table}
\end{center} 
The results in Table \ref{tab:p5} should be compared with the known exact identifications in \eqref{eq:p5-ids-x-chi}-\eqref{eq:p5-ids-psi-phi} and \eqref{eq:p5-x-chi}-\eqref{eq:p5-psi-phi}, from which the small $q$ and large $q$ expansions are:
\begin{eqnarray}
    X(q^2)&\sim& 1-q^2+q^4+q^8-2 q^{10}+\dots \\
    &\sim & 1-\frac{1}{q^4}+\frac{1}{q^6}-\frac{1}{q^{18}}+\frac{1}{q^{22}}-\frac{1}{q^{42}}
   +\dots, \\
   \chi(q^2) &\sim& q^2-q^4+q^6-2 q^8+2 q^{10}+ \dots \\
    &\sim & 1-\frac{1}{q^2}+\frac{1}{q^8}-\frac{1}{q^{14}}+\frac{1}{q^{26}}-\frac{1}{q^{36}}+ \dots, \\
    -\psi(-q) &\sim& q-q^2+2 q^3-2 q^4+2 q^5 +\dots \\
    &\sim& 1-\frac{1}{q^3}+\frac{1}{q^7}-\frac{1}{q^{16}}+\frac{1}{q^{24}}-\frac{1}{q^{39}}+ \dots, \\
    q\, \phi(-q) &\sim & q-2 q^2+2 q^3-3 q^4+4 q^5-4 q^6+ \dots \\
    &\sim& 1-\frac{1}{q}+\frac{1}{q^9}-\frac{1}{q^{12}}+\frac{1}{q^{28}}-\frac{1}{q^{33}}+ \dots .
    \label{eq:p5-table-match}
\end{eqnarray}
After including the prefactor powers $q^{-a^2/(4p)}$, we see that the false theta functions $\Psi^{(a)}_5(q)$ in Table \ref{tab:p5} match the unary large $q$ expansions, and the dual theta functions $\Psi^{(a)}_5(q)^\vee$ in Table \ref{tab:p5} match the non-unary small $q$ expansions. This shows that our numerical algorithm produces results that agree with the identifications listed in \eqref{eq:p5-ids-x-chi}-\eqref{eq:p5-ids-psi-phi} of the duals in terms of order 10 mock theta functions.
\graphicspath{ {./paper plots/} }
\begin{figure}[h!]
\begin{subfigure}[t]{.5\textwidth}
  \centering
  \includegraphics[width=.8\linewidth]{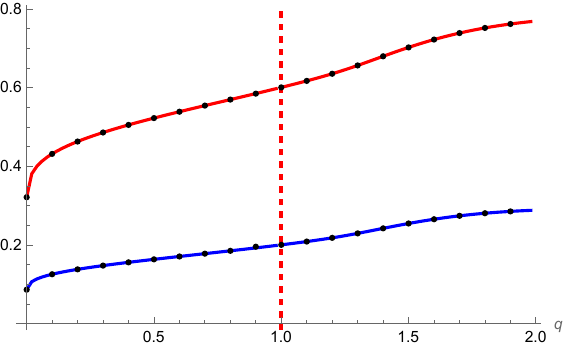}
\end{subfigure}
~
\begin{subfigure}[t]{.5\textwidth}
  \centering
  \includegraphics[width=.8\linewidth]{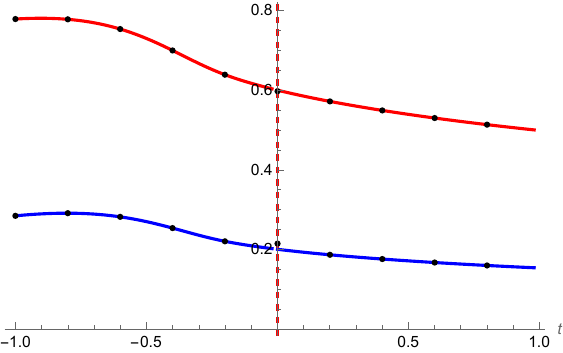}
\end{subfigure}
\\
\begin{subfigure}[t]{.5\textwidth}
  \centering
  \includegraphics[width=.8\linewidth]{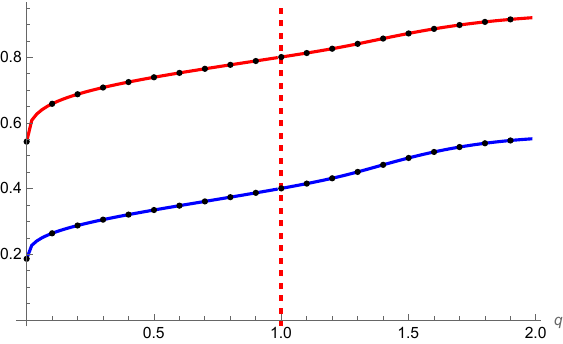}
\end{subfigure}
~
\begin{subfigure}[t]{.5\textwidth}
  \centering
  \includegraphics[width=.8\linewidth]{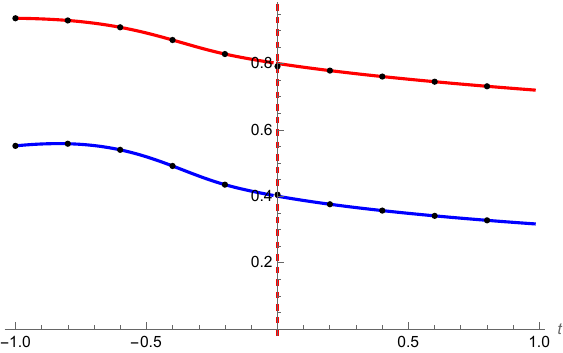}
\end{subfigure}
  \caption{ 
Plots of the real parts of the set A [top], and C [bottom] relations \eqref{eq:ja5qunary} and \eqref{eq:jc5qunary}, and \eqref{eq:ja5q} and \eqref{eq:jc5q}, for $p=5$ on both sides of the natural boundary, in terms of both $q=e^{-t}$ (left) and $t$ (right). On the unary side ($q>1$, $t<0$), we plot the upper/lower entry of the LHS of \eqref{eq:ja5qunary} and \eqref{eq:jc5qunary} as a solid red/blue  line, and the real part of the RHS as black points. Similarly, on the non-unary side ($q<1$ and $t>0$) we plot the upper/lower entry of the LHS of \eqref{eq:ja5q} and \eqref{eq:jc5q} as a solid red/blue line, and the RHS as black points.
}
  \label{fig:p5}
\end{figure}
In Figure \ref{fig:p5} we show how these numerical solutions cross smoothly the natural boundary ($t=0$ or $q=1$), for the algebraic sets $A$ [top] and $C$ [bottom]. 
The plots in Figure \ref{fig:p5} show with solid lines the real parts of the integral combinations in the respective decomposition identities, \eqref{eq:ja5qunary}-\eqref{eq:jc5qunary} and \eqref{eq:ja5q}-\eqref{eq:jc5q}, both inside and outside the natural boundary. These figures illustrate the smooth crossing of the natural boundary.
\subsection{New territory: $p=7$ and $p=9$. Duals for False Theta Functions 
$\Psi_7^{(a)}(q)$ and $\Psi_9^{(a)}(q)$}
\label{sec:p7-results}
\begin{figure}[h!]
\begin{subfigure}[t]{.5\textwidth}
  \centering
  \includegraphics[width=.8\linewidth]{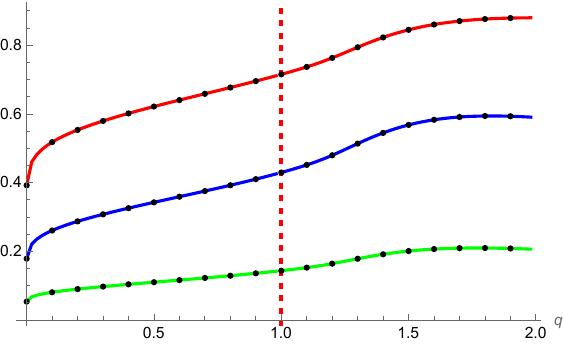}
\end{subfigure}
~
\begin{subfigure}[t]{.5\textwidth}
  \centering
  \includegraphics[width=.8\linewidth]{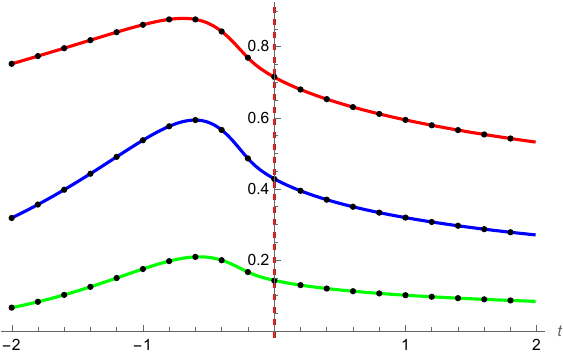}
\end{subfigure}
\centering
\begin{subfigure}[t]{.5\textwidth}
  \centering
  \includegraphics[width=.8\linewidth]{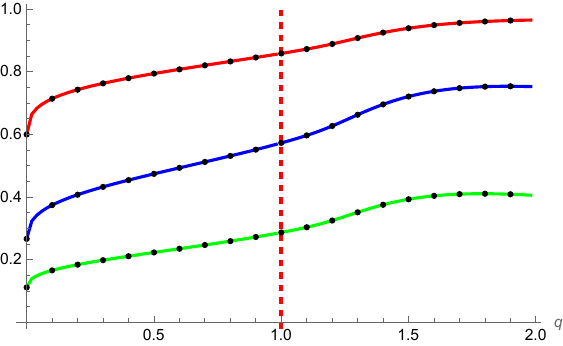}
\end{subfigure}
~
\begin{subfigure}[t]{.5\textwidth}
  \centering
  \includegraphics[width=.8\linewidth]{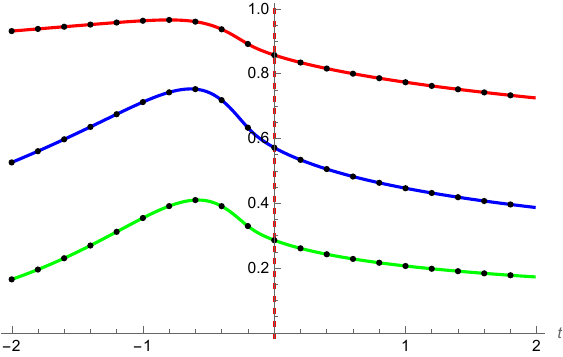}
\end{subfigure}
  \caption{Plots of the real parts of the set A [top] and C [bottom] relations \eqref{eq:aq}-\eqref{eq:cq} and \eqref{eq:aq-dual} and \eqref{eq:cq-dual} for $p=7$ on both sides of the natural boundary, in terms of both $q=e^{-t}$ (left) and $t$ (right). On the unary side ($q>1$, $t<0$), we plot the upper/middle/lower entry of the LHS of \eqref{eq:aq} and \eqref{eq:cq} as a solid red/blue/green line, and the real part of the RHS as black points. Similarly, on the non-unary side ($q<1$ and $t>0$) we plot the upper/middle/lower entry of the LHS of \eqref{eq:aq-dual} and \eqref{eq:cq-dual} as a solid red/blue/green line, and the RHS as black points.  }
  \label{fig:p7}
\end{figure}

\begin{figure}[H]
\centering
\begin{subfigure}[t]{.5\textwidth}
  \centering
  \includegraphics[width=.8\linewidth]{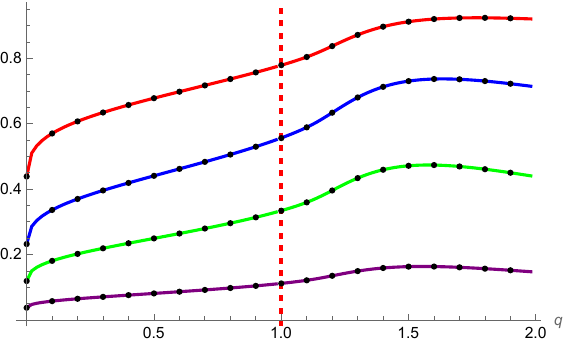}
\end{subfigure}
~
\begin{subfigure}[t]{.5\textwidth}
  \centering
  \includegraphics[width=.8\linewidth]{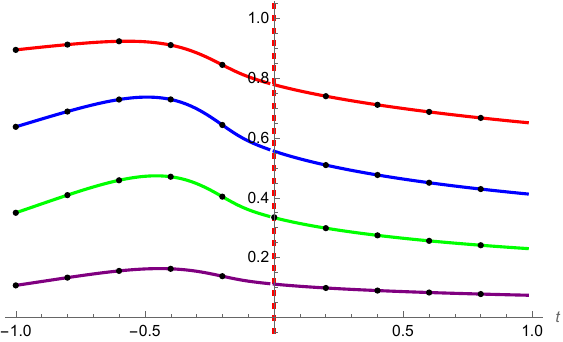}
\end{subfigure}
\\
\begin{subfigure}[t]{.5\textwidth}
  \centering
  \includegraphics[width=.8\linewidth]{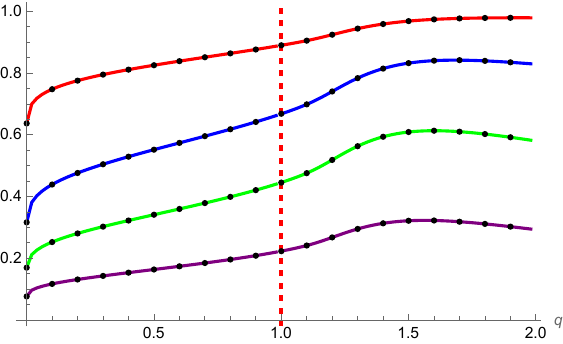}
\end{subfigure}
~
\begin{subfigure}[t]{.5\textwidth}
  \centering
  \includegraphics[width=.8\linewidth]{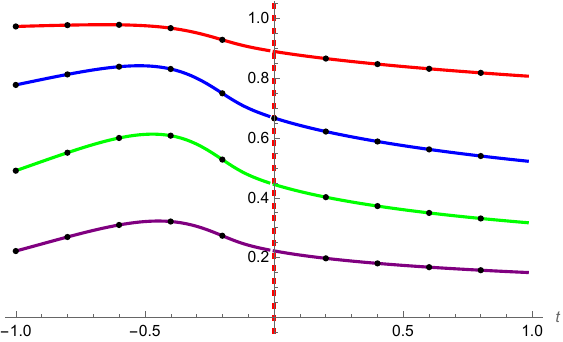}
\end{subfigure}
  \caption{Plots of the real parts of the set A [top] and C [bottom] relations \eqref{eq:aq} and \eqref{eq:cq}, and \eqref{eq:aq-dual} and \eqref{eq:cq-dual}, for $p=9$ on both sides of the natural boundary, in terms of both $q=e^{-t}$ (left) and $t$ (right). On the unary side ($q>1$, $t<0$), we plot the  LHS of \eqref{eq:aq} and \eqref{eq:cq} as a solid line, and the real part of the RHS as black points. Similarly, on the non-unary side ($q<1$ and $t>0$) we plot the  LHS of \eqref{eq:aq-dual} and \eqref{eq:cq-dual} as a solid  line, and the RHS as black points.}
  \label{fig:p9}
\end{figure}

The results for $p=7$, showing the false theta function $\Psi^{(a)}_7(q)$ and its numerically found dual $\tilde\Psi^{(a)}_7(q)^\vee$, are shown in Table \ref{tab:p7}. The analogous results for $p=9$, showing the false theta function $\Psi^{(a)}_9(q)$ and its numerically found dual $\tilde\Psi^{(a)}_9(q)^\vee$, are shown in Table \ref{tab:p9}. 
For $p=7$ and $p=9$ we have no mock theta functions with which we can compare. But in Figures \ref{fig:p7} and \ref{fig:p9} we show the behavior of the decomposition identities on both sides of the natural boundary, in terms of $t$, and in terms of $q$.
In figures \ref{fig:p7} and \ref{fig:p9} we show the behavior of the decomposition identities on both sides of the natural boundary, in terms of $t$, and in terms of $q$. 
The behavior is qualitatively the same as the $p=5$ case (order 10 mock and false theta functions), shown in Figure \ref{fig:p5}.
\begin{center}
\begin{table}[h!]
\begin{tabular}{ | m{1em} | m{7.6cm}| m{7.6cm}| } 
  \hline
  $a$ & Unary false theta $\Psi^{(a)}_7(q)$ & Non-unary dual $q$-series $\Psi^{(a)}_7(q)^{\vee}$ \\ 
  \hline
  $1$ & $q^{1/28}\left(1-q^6+q^8-q^{26}+q^{30}-q^{60}+\dots\right)$ 
  & $\frac{1}{4}q^{-1/28}\left(3+3 q^4-7 q^6+9 q^8+\dots\right)$ \\
  \hline
  $2$ & $q^{1/7}\left(1-q^5+q^9-q^{24}+q^{32}-q^{57}+\dots\right)$ & $q^{-1/7}\left(3 q - 9 q^2 + 23 q^3 - 52 q^4 +\dots\right)$ \\
  \hline
  $3$ & $q^{9/28}\left(1-q^4+q^{10}-q^{22}+q^{34}-q^{54}+\dots\right)$ & $\ff{1}{4}q^{-9/28}\left(1+3 q^2-6 q^4+14 q^6+\dots\right)$ \\
  \hline
  $4$ & $q^{4/7}\left(1-q^3+q^{11}-q^{20}+q^{36}-q^{51}+\dots\right)$ & $q^{-4/7}\left(q-3 q^2+8 q^3-17 q^4+36 q^5+\dots\right)$ \\
  \hline
  $5$ & $q^{25/28}\left(1-q^2+q^{12}-q^{18}+q^{38}-q^{48}+\dots\right)$ & $\ff{1}{4}q^{-25/28}\left(3 q^2-7 q^4+14 q^6-21 q^8+\dots\right)$ \\
  \hline
  $6$ & $q^{9/7}\left(1-q+q^{13}-q^{16}+q^{40}-q^{45}+\dots\right)$ & $q^{-9/7}\left(q^2-5 q^3+15 q^4-35 q^5+\dots\right)$ \\
  \hline
\end{tabular}
\caption{Small $q$ expansions of the $p=7$ false theta functions $\Psi^{(a)}_7(q)$ and their duals $\Psi^{(a)}_7(q)^\vee$, for $1\leq a \leq 6$.}
\label{tab:p7}
\end{table}
\end{center}
\graphicspath{ {./paper plots/} }
\begin{table}[H]
\begin{tabular}{ | m{1em} | m{7.5cm}| m{7.5cm}| } 
  \hline
  $a$ & Unary false theta $\Psi^{(a)}_9(q)$ & Non-unary dual $q$-series $\Psi^{(a)}_9(q)^{\vee}$ \\ 
  \hline
  $1$ & $q^{1/36}\left(1-q^8+q^{10}-q^{34}+q^{38}-q^{78}+\dots\right)$ 
  & $q^{-1/36}\left(2+2 q^4+2 q^8-2 q^{10}+2 q^{16}+\dots\right)$ \\
  \hline
  $2$ & $q^{1/9}\left(1-q^7+q^{11}-q^{32}+q^{40}-q^{75}+\dots\right)$ & $q^{-1/9}\left(-2 q^2+2 q^5-2 q^6-2 q^8+2 q^9+\dots\right)$ \\
  \hline
  $3$ & $q^{9/36}\left(1-q^6+q^{12}-q^{30}+q^{42}-q^{72}+\dots\right)$ & $q^{-9/36}\left(2 q^2-2 q^4+2 q^{10}-2 q^{12}+\dots\right)$ \\
  \hline
  $4$ & $q^{4/9}\left(1-q^5+q^{13}-q^{28}+q^{44}-q^{69}+\dots\right)$ & $q^{-4/9}\left(2 q+2 q^3-2 q^4+2 q^5-2 q^6+\dots\right)$ \\
  \hline
  $5$ & $q^{25/36}\left(1-q^4+q^{14}-q^{26}+q^{46}-q^{66}+\dots\right)$ & $q^{-25/36}\left(-2 q^2-2 q^{10}-2 q^{18}+2 q^{20}+\dots\right)$ \\
  \hline
  $6$ & $q\left(1-q^3+q^{15}-q^{24}+q^{48}-q^{63}+\dots\right)$ & $q^{-1}\left(q-2 q^2+2 q^3-2 q^4+2 q^5-2 q^6+\dots\right)$ \\
  \hline
   $7$ & $q^{49/36}\left(1-q^2+q^{16}-q^{22}+q^{50}-q^{60}+\dots\right)$ & $q^{-49/36}\left(2 q^2+2 q^6-2 q^8+2 q^{10}+\dots\right)$ \\
  \hline
   $8$ & $q^{16/9}\left(1-q+q^{17}-q^{20}+q^{52}-q^{57}+\dots\right)$ & $q^{-16/9}\left(-2 q^2+2 q^3-2 q^4+2 q^{5}+\dots\right)$ \\
  \hline
\end{tabular}
\caption{Small $q$ expansions of the $p=9$ false theta functions $\Psi^{(a)}_{9}(q)$ and their duals $\Psi^{(a)}_{9}(q)^\vee$, for $1\leq a \leq 8$.} 
\label{tab:p9}
\end{table}
\graphicspath{ {./paper plots/} }
\subsection{Comparison with other results for duals of false theta functions}
\label{sec:comparison1}
In this section we discuss some comparisons with other proposals in the literature for duals of individual false theta functions.
\begin{itemize}
    \item In \cite{CDGG} a plot-fitting approach to extract numerically the first few coefficients of $\Psi^{(1)}_7(q)^{\vee}$ was proposed, but it does not match the results for $\Psi^{(1)}_7(q)^{\vee}$ in Table \ref{tab:p7}. This is because in \cite{CDGG} the full vector-valued algebraic structure was not imposed. In the current paper we have defined {\it resurgence orbits} by imposing the full orbit of the Mordell-Borel integrals under the modular transformations which probe the neighborhood of the natural boundary.  
    \item 
    In their work on optimal mock Jacobi functions \cite{CD20c}, Cheng and Duncan give a finite list of such `optimal' functions in Tables 3-18. Several of our results for duals of individual false theta functions appear in these tables. 
    \begin{enumerate}
    \item 
   The first row of Table 3 in \cite{CD20c} gives the coefficients of $-2f(q)$, the order 3 mock theta function, while the coefficients of $q\,\omega(q)$ do not seem to appear in the tables. Both these order 3 mock theta function cases are found by our numerical procedure. 
   \item 
   For $p=5$ (order $10$ mock theta functions), the coefficients of $-2X(q)$ and $-2\chi(q)$ appear in Table 4 in \cite{CD20c}, in the $r=1,3$ rows respectively, while the coefficients of $2\psi(q)$ and $2q\,\phi(q)$ appear in the rows $r=4,8$ of Table 9 in \cite{CD20c}.  

   \item 
   The previous two items correspond to known mock theta functions. However, our new results (see Table \ref{tab:p9} in Section \ref{sec:p7-results} of this paper) for the $p=9$ duals of false theta functions also appear in Tables in \cite{CD20c}. The coefficients of the series $-\Phi^{(a)}_9(\sqrt{q})^\vee$ for odd $a$ appear in rows $r=1,3,5,7$ in Table 6 in \cite{CD20c}, and the coefficients of the series $-\Phi^{(a)}_9(-q)^\vee$ for even $a$ appear in rows $r=4,8,12,16$ of Table 14 in \cite{CD20c}.
    \end{enumerate}
   This overlap of results is interesting, given that the optimality condition in \cite{CD20c} (a condition on the growth rate of the coefficients of the $q$-series) is not the same as our resurgence condition of preservation of relations when crossing the natural boundary. Nevertheless, there appears to be a connection. The correspondence between our results and optimal mock Jacobi functions is discussed further in \cite{ceff2}.
\end{itemize}

\subsection{Higher $p$ Duals for False Theta Functions $\Psi_p^{(a)}(q)$}
In the previous subsections, we gave the results of our numerical method for $p=5$, to compare with the known results in Section \ref{sec:p5}, and for the first new results: $p=7$ and $p=9$. In the Appendix \ref{sec:results-p-class} we present the  corresponding tables for $p=11$, $p=13$, $p=15$, $p=17$ and $p=19$. For brevity's sake, we only include plots for the set $A$ and set $C$ relations, as these cover all of the dual false thetas.

\section{Results: Duals for the Orientation Reversal of Brieskorn Spheres}
\label{sec:results-mock-class}

In this section we list the results for the dual $q$-series associated with the orientation reversal of Brieskorn spheres $\Sigma(2,3,7)$, $\Sigma(2,3,11)$ and $\Sigma(2,3,13)$. The first terms of the relevant 
$q$-series are shown in Tables \ref{tab:mock7-table-match}, \ref{tab:mock11-table-match} and \ref{tab:mock13-table-match}, respectively. The results for $\overline{\Sigma(2,3,7)}$ agree with the known order 7 false and  mock theta functions, while the results for $\overline{\Sigma(2,3,11)}$ and $\overline{\Sigma(2,3,13)}$, shown in Tables \ref{tab:mock11-table-match} and \ref{tab:mock13-table-match}, are new. In accompanying supplementary files we list all the expansions of these new $q$-series.

\subsection{Warm up: the orientation reversed Brieskorn sphere $\overline{\Sigma(2, 3, 7)}$ and order 7 mock theta functions} 
\label{sec:237}

Here we give the results of our numerical algorithm applied to the Brieskorn sphere resurgence orbit of Mordell-Borel integrals for $p=7$, which match the expansions of the order $7$ mock theta functions $-\mathcal{F}_0(q)$, $-\mathcal{F}_1(q)$, and $-q \, \mathcal{F}_2(q)$. 
\begin{table}[h!] \centering  \begin{tabular}{ | m{1em} | m{7.7cm} | m{8.2cm} | } 
\hline $j$ & Unary $q$-series $X^{(j)}_7(q)$ & 
Non-unary dual $q$-series $X^{(j)}_7(q)^{\vee}$ \\ 
\hline $1$ & $q^{1/168}\left(-1 + q + q^5 - q^{10} + q^{11} - q^{18}+\dots\right)$ & $q^{-1/168}\left(-1-q - q^3 - q^4 - q^5 - 2 q^7+\dots\right)$ \\ 
\hline $2$ & $q^{25/168}\left(1 + q^2 + q^3 + q^8 - q^{13} - q^{22}+\dots\right)$ & $q^{-25/168}\left(-q - q^2 - q^3 - 2 q^4 - q^5 - 2 q^6+\dots\right)$ \\ 
\hline $3$ & $q^{121/168}\left(1 + q + q^3 + q^5 - q^{16} - q^{20}+\dots\right)$ & $q^{-121/168}\left(-q - q^2 - 2 q^3 - q^4 - 2 q^5+\dots\right)$ \\ \hline 
\end{tabular}
\caption{Small $q$ expansions of the $p=7$ $q$- series $X^{(j)}_7(q)$ and their duals $X^{(j)}_7(q)^\vee$ for $1\leq j \leq 3$, for the Brieskorn sphere $\Sigma(2, 3, 7)$.}
\label{tab:mock7-table-match}
\end{table}
The results for $p=7$,  are shown in Table \ref{tab:mock7-table-match}. These should be compared with the identifications in 
\eqref{eq:mock7-ids-1}-\eqref{eq:mock7-ids-3}, for which the small and large $q$ expansions are:
\begin{eqnarray}
    -\mathcal F_0 (q)&\sim& -1-q-q^3-q^4-q^5-2 q^7- \dots \\
    &\sim & -1+\frac{1}{q}+ \frac{1}{q^5}- \frac{1}{q^{10}}+ \frac{1}{q^{11}}-\frac{1}{q^{18}}+ \dots ,\\
   -\mathcal F_1 (q) &\sim& -q-q^2-q^3-2 q^4-q^5-2 q^6- \dots \\
    &\sim & 1+\frac{1}{q^2}+ \frac{1}{q^3}+ \frac{1}{q^8}- \frac{1}{q^{13}} - \frac{1}{q^{22}} - \dots ,\\
   -q\, \mathcal F_2 (q) &\sim& -q-q^2-2 q^3-q^4-2 q^5- \dots \\
    &\sim& 1+\frac{1}{q}+\frac{1}{q^3}+\frac{1}{q^5} - \frac{1}{q^{16}} - \frac{1}{q^{20}}- \dots .
    \label{eq:mock7-table-match}
\end{eqnarray}
We observe that our numerical method has found the known order 7 mock theta functions which smoothly cross the boundary. The small $q$ expansions in \eqref{eq:mock7-table-match} match the expansions of $X_7(q)^\vee$, while the large $q$ expansions match the unary expressions for $X_7(1/q)$.

\subsection{New territory: The orientation reversed Brieskorn spheres $\overline{\Sigma(2, 3, 11)}$ and $\overline{\Sigma(2, 3, 13)}$ and associated $q$-series}
\label{sec:2311}
Tables \ref{tab:mock11-table-match} and \ref{tab:mock13-table-match} show the results of our algebraic procedure for the unary $q$-series $X_{p}^{(j)}(q)$ and their dual $q$-series $X_{p}^{(j)}(q)^\vee$, for $p=11$ and $p=13$, respectively, for $j=1,\dots, (p-1)/2$,

\FloatBarrier
\begin{table}[H]
\centering
\begin{tabular}{ | m{1em} | m{7.7cm}| m{8.2cm} | } 
  \hline
  $j$ & Unary $q$-series $X^{(j)}_{11}(q)$ & 
  Non-unary dual $q$-series $X^{(j)}_{11}(q)^{\vee}$  \\ 
  \hline
  $1$ & $q^{25/264}\left(-1+q+q^9-q^{14}+q^{19}-q^{26}+\dots\right)$ &$\frac{1}{6}q^{-25/264}\left(-1+15 q+65 q^2+175 q^3+\dots\right)$ 
  \\
  \hline
  $2$ & $q^{1/264}\left(1+q^2+q^7+q^{16}-q^{17}-q^{30}+\dots\right)$ & $\frac{1}{6}q^{-1/264}\left(5+5 q+15 q^2+60 q^3+125 q^4+\dots\right)$ 
  \\
  \hline
  $3$ & $q^{49/264}\left(1+q^3+q^5+q^{13}-q^{20}-q^{34}+\dots\right)$ &$\frac{1}{6}q^{-49/264}\left(-16 q-55 q^2-155 q^3-385 q^4+\dots\right)$ 
  \\
  \hline
  $4$ & $q^{169/264}\left(1+q^3+q^4+q^{10}-q^{23}-q^{35}+\dots\right)$ & $\frac{1}{6}q^{-169/264}\left(-5 q-22 q^2-60 q^3-155 q^4+\dots\right)$ 
  \\
  \hline
  $5$ & $q^{361/264}\left(1+q+q^5+q^7-q^{26}-q^{30}+\dots\right)$  & $\frac{1}{6}q^{-361/264}\left(5 q^2+33 q^3+99 q^4+268 q^5+\dots\right)$ 
  \\
  \hline
\end{tabular}
\caption{Small $q$ expansions of the $p=11$ $q$-series $X^{(j)}_{11}(q)$ and their duals $X^{(j)}_{11}(q)^\vee$ for $1\leq j \leq 5$, for $\Sigma(2, 3, 11)$.} 
\label{tab:mock11-table-match}
\end{table}
\begin{table}[h!]
\begin{center}
\begin{tabular}{ | m{1em} | m{7.5cm}| m{8.0cm} | } 
  \hline
  $j$ & Unary $q$-series $X^{(j)}_{13}(q)$ & 
  Non-unary dual $q$-series $X^{(j)}_{13}(q)^{\vee}$  \\ 
  \hline
  $1$ & $q^{49/312}\left(-1+q+q^{11}-q^{16}+q^{23}+\dots\right)$ & $q^{-49/312}\left(-2 q-2 q^7-2 q^9-2 q^{13}+\dots\right)$ 
  \\
  \hline
  $2$ & $q^{1/312}\left(-1+q^2+q^9-q^{19}+q^{20}+\dots\right)$ & $q^{-1/312}\left(-2-2 q^3-2 q^4-2 q^7+\dots\right)$ 
  \\
  \hline
  $3$ & $q^{25/312}\left(1+q^3+q^7+q^{17}-q^{22}+\dots\right)$ & $q^{-25/312}\left(-2 q-2 q^3-2 q^5-2 q^7+\dots\right)$ 
  \\
  \hline
  $4$ & $q^{121/312}\left(1+q^4+q^5+q^{14}-q^{25}+\dots\right)$ & $q^{-121/312}\left(-2 q-2 q^2-2 q^4-2 q^6+\dots\right)$ 
  \\
  \hline
  $5$ & $q^{289/312}\left(1+q^3+q^5+q^{11}-q^{28}+\dots\right)$ & $q^{-289/312}\left(-2 q-2 q^3-2 q^4-2 q^5+\dots\right)$ 
  \\
  \hline
  $6$ & $q^{529/312}\left(1+q+q^6+q^8-q^{31}+\dots\right)$ & $q^{-529/312}\left(-2 q^2-2 q^3-2 q^4-2 q^6+\dots\right)$ 
  \\
  \hline
\end{tabular}
\end{center}
\caption{Small $q$ expansions of the $p=13$ $q$-series $X^{(j)}_{13}(q)$ and their duals $X^{(j)}_{13}(q)^\vee$ for $1\leq j \leq 6$, for $\Sigma(2, 3, 13)$.}
\label{tab:mock13-table-match}
\end{table}
In Appendix \ref{sec:appendix-results2} we present the tables of $q$-series coefficients, and associated rational prefactor exponents, for higher members of the $\overline{\Sigma(2,3,6k\pm 1)}$ class:  $\overline{\Sigma(2,3,17)}$, $\overline{\Sigma(2,3,19)}$, $\overline{\Sigma(2,3,25)}$ and $\overline{\Sigma(2,3,35)}$. These results are all new.

\subsection{Comparison with other results for Brieskorn Spheres}
\label{sec:comparison}

In this section we discuss some comparisons with other proposals in the literature for dual $q$-series for certain Brieskorn spheres.

\begin{itemize}
    \item In \cite{Zag09}, Zagier proposes a criterion very similar to ours, namely that the full $SL(2, \mathbb Z)$ action should be imposed. For the example of order 5 mock theta functions, he shows that this criterion selects the pair $(\chi_0(q), \chi_1(q))$ over the other 4 possible pairs of order 5 mock theta functions.\footnote{Zagier's expression for the order 5 mock theta function case also includes the unusual $\frac{2}{3}$ normalization factor in \eqref{eq:mock51}-\eqref{eq:mock52}, which is also found by our numerical procedure.} 
    Zagier also gives an analogous result for order 7 mock theta functions:
    \begin{eqnarray}
        M_{7,j}(\tau) = \frac{1}{\eta(\tau)^3}\,\sum\limits_{\substack{m>2|n|/9 \\ n \equiv j \,(\text{mod }7)}} \left(\frac{-4}{m}\right)\left(\frac{12}{n}\right) \left( m \,\text{sgn}(n) - \frac{3n}{14} \right) \, q^{m^2/8 - n^2/168},
        \label{eq:zagier7}
    \end{eqnarray}
    again in agreement with our reasoning and our numerical results in Table \ref{tab:mock7-table-match}. Zagier then argues that a similar construction exists for a vector-valued mock modular form $M_p(q)$, with $\frac{(p-1)}{2}$ elements, for any prime $p$. An explicit expression (on page 15 of \cite{Zag09}) for $p=11$ reads:
    \begin{eqnarray}
        M_{11,j}(\tau) = \frac{1}{\eta(\tau)^3}\,\sum\limits_{\substack{m>2|n|/11 \\ n \equiv j \,(\text{mod } 11)}} \left(\frac{-4}{m}\right)\left(\frac{12}{n}\right) \left( m \,\text{sgn}(n) - \frac{n}{6} \right) \, q^{m^2/8 - n^2/264}.
        \label{eq:zagier11}
    \end{eqnarray}
    Interestingly, this result matches precisely the $q$-series produced by our numerical procedure, in Section \ref{sec:2311} (see Table \ref{tab:mock11-table-match}), just with a different indexing order $j=1, ..., 5$.  This agreement is surprising, given that the methods used in \cite{Zag09} and here are so very different, and is suggestive of something deeper.
    
    \item However, in discussions with Prof. Zagier, we have learned that for higher prime $p$ values his generalized formulas do not match the dual $q$-series produced by our numerical procedure described in Section \ref{sec:algorihm2}. The reason for this distinction is that our resurgence-based approach, based on preservation of relations under unique continuation, is different from Zagier's approach based on indefinite theta series, holomorphic projection and Rankin-Cohen brackets \cite{Zag09}. This distinction deserves to be investigated further. 
    \item 
    On the other hand, our $p=13$ numerical results for $X_{13}(q)^\vee$ in Table \ref{tab:mock13-table-match} match the {\it optimal} mock Jacobi theta function expansions of Cheng and Duncan in \cite{CD20c} for $C^{(78+6,26,39)}(r^2-312n,r)$. This case was identified with $\overline{\Sigma(2, 3, 13)}$ in \cite{CCFGH}). The first 15 coefficients for this case appear in Table 18 on page 55 of \cite{CD20c}. Those coefficients match our numerical results, for which we have more than $60$ terms, the first few of which are shown in Table \ref{tab:mock13-table-match}, up to a simple relabelling of the list of $q$-series. Further coefficients can be found in the accompanying supplementary files. This example is particularly interesting because it has extremely slow growth of the expansion coefficients, as discussed further in \cite{ceff2}.

    \item In a more recent paper \cite{CCKPG},
    in section (5.5.2) the authors present results for the Chern-Simons $\widehat{Z}$-invariants of $\overline{\Sigma(2,3,13)}$ along with the invariants with defects, based on Weil representations of indefinite theta series. The authors note the interesting fact that these results do not agree with the $p=13$ optimal mock Jacobi theta function results in \cite{CD20c} (and therefore also not with our results for $p=13$ presented here in Table \ref{tab:mock13-table-match}). The same comment applies to higher values of $p=6k+1$, where the expression in equation (3.40) in Conjecture 5 of \cite{CCKPG} does not match the numerical results for $p=19$ and $p=25$, listed here in Appendix \ref{sec:appendix-results2}.

    \item In Appendix \ref{sec:other-brieskorn}, we outline a procedure to extend the analysis of $\overline{\Sigma(2,3,6k\pm1)}$ to more general orientation reversed Brieskorn spheres $\overline{\Sigma(p_1,p_2,p_3)}$, with $p_1$, $p_2$, and $p_3$ pairwise coprime, and give results for four examples. See the Tables \ref{tab:sigma257}, \ref{tab:sigma2511}, \ref{tab:sigma345}, and \ref{tab:sigma357}, in the Appendix \ref{sec:other-brieskorn}. We note that the coefficients of the $q$-series associated to $\overline{\Sigma(3,4,5)}$ and $\overline{\Sigma(2,5,7)}$ match the coefficients given in Tables 15 and 17 of \cite{CD20c}. The coefficients of these $q$-series are optimal in the sense of \cite{CD20c}. For the other two examples, $\overline{\Sigma(2,5,11)}$ and $\overline{\Sigma(3,5,7)}$, the coefficients are not optimal. Nevertheless, the associated dual $q$-series Tables \ref{tab:sigma257}, \ref{tab:sigma2511}, \ref{tab:sigma345}, and \ref{tab:sigma357},  satisfy all the required conditions for crossing the natural boundary. See further discussion in \cite{ceff2}.

\end{itemize}

\section{Conclusions}

We have shown that the property of preservation of relations for resurgent functions yields a rigid algebraic structure for the decomposition of Mordell-Borel integrals into $q$-series and $\qt$-series parts. The unique decomposition of these Mordell-Borel integrals on the Stokes line into real and imaginary parts, expressed in terms of unary false theta functions, provides a template for the decomposition on the other (non-unary) side of the $|q|=1$ natural boundary. The resulting continuation 
can be implemented numerically, producing dual $q$-series with coefficients that are, to high precision, integer-valued, and which match known examples where unique decomposition identities exist. Recall that the appearance of integer-valued coefficients matches physical expectations because these series arise in the analysis of complex Chern-Simons theory on non-trivial 3 dimensional manifolds under the operation of orientation reversal, where the integer coefficients are interpreted as counting BPS states. We have confirmed agreement with known cases and provided many new results within two {\it resurgence orbits}: (i) duals $\Psi_p^{(a)}(q)^\vee$ of individual false theta functions $\Psi_p^{(a)}(q)$ for $p$ an odd integer, and $0<a<p$ also integer; (ii) the duals of special combinations of four false theta functions which arise in the analysis of Chern-Simons theory on Brieskorn sphere manifolds. Further examples and results will be presented in a forthcoming publication.

The results for higher values of $p$, for both resurgence orbits, show that the rate of growth of the integer coefficients of the dual $q$-series depends sensitively on $p$. Contrast, for example, the growth of the coefficients of $\Phi_5^{(a)}(q)^\vee$ in Table \ref{tab:p5} with the much more rapid growth of the coefficients of $\Phi_7^{(a)}(q)^\vee$ in Table \ref{tab:p7}, and with the slower again growth of the coefficients of $\Phi_9^{(a)}(q)^\vee$ in Table \ref{tab:p9}. This dependence on $p$ (and less sensitively on $a$) is explained analytically in \cite{ceff2}, where we show that the algebraic structure imposed by the unique decomposition into real and imaginary parts on the Stokes line places strong constraints on the form of the dual $q$-series. When combined with one piece of numerical data, the leading term in each dual $q$-series, we obtain an analytic prediction for the growth rate of the integer coefficients of the dual $q$-series. In \cite{ceff2} this analytic result is derived and shown to agree with known growth-rate results, and also to match our numerical results. In \cite{CD25} it is proved that the decomposition of the Mordell-Borel integrals into $q$-series and $\qt$-series with this structure is unique, for the first case of each of the two resurgence orbits of dual series analyzed in this paper.

Modular symmetry is a well-studied aspect of complex Chern-Simons theory.\footnote{Note, while the origin of the modular group --- or, depending on rank and topology, its congruence subgroup --- is clear from the string theory realization of non-perturbative complex Chern-Simons, it is less clear in the existing mathematical approaches. Thus, $SL(2,\mathbb{Z})$ is manifest as a symmetry of the physical setup \cite{Gopakumar:1998ii,Gopakumar:1998ki,Gopakumar:1998jq,Ooguri:1999bv,deHaro:2004id,deHaro:2005rz} on which many of the recent works \cite{GMP,CCFGH,CCKPG} are based. It would be helpful to find a similar realization, e.g. as a manifest symmetry of the category of representations of a quantum group or other elements of complex Chern-Simons theory that appear in potentially relevant work \cite{Garoufalidis:2021lcp,Fantini:2024ihf}.}
However, modular symmetry is not essential for the argument of preserving algebraic structures under unique continuation of resurgent functions, but it plays a role in our numerical method because we take advantage of sub-structure with respect to the modular operation $t\to\pi^2/t$, which relates expansions near $q=0$ to expansions near $|q|\to 1^-$, as well as expansions near $q\to\infty$ to expansions near $|q|\to 1^+$. By reducing the algebraic structure to orbits that are self-dual under this operation we can simplify the decomposition algorithm. In this respect the role of modularity is that it includes the natural operations of probing the radial and transverse directions in the neighborhood of the natural boundary, so we are effectively rephrasing the problem in terms of orbits of the Mordell-Borel integrals under operations that probe the natural boundary.

Our resurgent approach is quite different from other approaches \cite{CCFGH,CFG19,CCKPG,Zag09,CD20c,GM,HM14,Mort14}
to the problem of crossing the natural boundary with false and mock theta functions, such as the indefinite theta series, Appell-Lerch sums, and representation theory of logarithmic vertex operator algebras.  Understanding the intricacies of the relations between these diverse methods is an important and interesting open problem, which we expect to yield both physical and mathematical insights. 

\section*{Acknowledgements}

Thanks to John Chae, Miranda C. N. Cheng, Daniele Dorigoni, Jean \'Ecalle, Angus Gruen, Mrunmay Jagadale, Albrecht Klemm, and Don Zagier for discussions, ideas, help, advice, support and inspiration that have greatly benefited this project.
The work of OC is supported in part by the U.S. National Science Foundation, Division of Mathematical Sciences, Award NSF DMS-2206241.
The work of GD and GA is supported in part by the U.S. Department of Energy, Office of High Energy Physics, Award DE-SC0010339.
The work of SG was supported in part by a Simons Collaboration Grant on New Structures in Low-Dimensional Topology, by the NSF grant DMS-2245099, and by the U.S. Department of Energy, Office of Science, Office of High Energy Physics, under Award No. DE-SC0011632.
The authors OC, GD and SG  thank the Galileo Galilei Institute for Theoretical Physics for hospitality, and the INFN for partial support, during the workshop “Resurgence and Modularity in QFT and String Theory”, Spring 2024. GD thanks the Max Planck Institute for Mathematics, Bonn, for support during the program "Combinatorics, Resurgence and Algebraic Geometry in Quantum Field Theory", August 2024. GA and O\"O thank L'\'Ecole de Physique des Houches for support during the Les Houches School "Quantum Geometry", Summer 2024.

\pagebreak

\appendix

\numberwithin{equation}{section}

\section{Results for Duals of False Thetas $\Psi_p^{(a)}(q)$}
\label{sec:appendix-results1}
In this appendix we display tables containing the small $q$ expansions of the unary false thetas $\Psi^{(a)}_p(q)$ for $p=11,13,15,17,19$, and their corresponding duals $\Psi^{(a)}_p(q)^\vee$.
\begin{table}[h!]
\begin{tabular}{ | m{1em} | m{7.6cm}| m{7.6cm}| } 
  \hline
  $a$ & Unary false theta $\Psi^{(a)}_{11}(q)$ & Non-unary dual $q$-series $\Psi^{(a)}_{11}(q)^{\vee}$ \\ 
  \hline
  $1$ & $q^{1/44}\left(1-q^{10}+q^{12}-q^{42}+q^{46}-q^{96}+\dots\right)$ 
  & $\frac{1}{2}q^{-1/44}\left(2-q^2+q^4-q^6+4 q^8+\dots\right)$ \\
  \hline
  $2$ & $q^{1/11}\left(1-q^9+q^{13}-q^{40}+q^{48}-q^{93}+\dots\right)$ 
  & $q^{-1/11}\left(2 q-4 q^2+9 q^3-16 q^4+28 q^5+\dots\right)$ \\
  \hline
  $3$ & $q^{9/44}\left(1-q^8+q^{14}-q^{38}+q^{50}-q^{90}+\dots\right)$ 
  & $\frac{1}{2}q^{-9/44}\left(1+q^2-3 q^4+6 q^6-7 q^8+\dots\right)$ \\
  \hline
  $4$ & $q^{4/11}\left(1-q^7+q^{15}-q^{36}+q^{52}-q^{87}+\dots\right)$ 
  & $q^{-4/11}\left(2 q-5 q^2+10 q^3-18 q^4+\dots\right)$ \\
  \hline
  $5$ & $q^{25/44}\left(1-q^6+q^{16}-q^{34}+q^{54}-q^{84}+\dots\right)$ 
  & $\frac{1}{2}q^{-25/44}\left(2 q^2-q^4+q^6-4 q^8+5 q^{10}+\dots\right)$ \\
  \hline
  $6$ & $q^{9/11}\left(1-q^5+q^{17}-q^{32}+q^{56}-q^{81}+\dots\right)$ 
  & $q^{-9/11}\left(q-2 q^2+4 q^3-8 q^4+14 q^5+\dots\right)$ \\
  \hline
  $7$ & $q^{49/44}\left(1-q^4+q^{18}-q^{30}+q^{58}-q^{78}+\dots\right)$ 
  & $\frac{1}{2}q^{-49/44}\left(q^2-q^4+3 q^6-6 q^8+7 q^{10}+\dots\right)$ \\
  \hline
  $8$ & $q^{16/11}\left(1-q^3+q^{19}-q^{28}+q^{60}-q^{75}+\dots\right)$ 
  & $q^{-16/11}\left(2 q^4-4 q^5+7 q^6-14 q^7+\dots\right)$ \\
  \hline
  $9$ & $q^{81/44}\left(1-q^2+q^{20}-q^{26}+q^{62}-q^{72}+\dots\right)$ 
  & $\frac{1}{2}q^{-81/44}\left(q^2-3 q^4+5 q^6-6 q^8+9 q^{10}+\dots\right)$ \\
  \hline
  $10$ & $q^{25/11}\left(1-q+q^{21}-q^{24}+q^{64}-q^{69}+\dots\right)$ 
  & $q^{-25/11}\left(-q^3+4 q^4-8 q^5+16 q^6+\dots\right)$ \\
  \hline
\end{tabular}
\caption{Small $q$ expansions of $\Psi^{(a)}_{11}(q)$ and their duals $\Psi^{(a)}_{11}(q)^\vee$ for $1\leq a \leq 10$.} 
\label{tab:p=11firstclass}
\end{table}
\graphicspath{ {./paper plots/} }
\begin{figure}[h!]
\centering
\begin{subfigure}[t]{.5\textwidth}
  \centering
  \includegraphics[width=.8\linewidth]{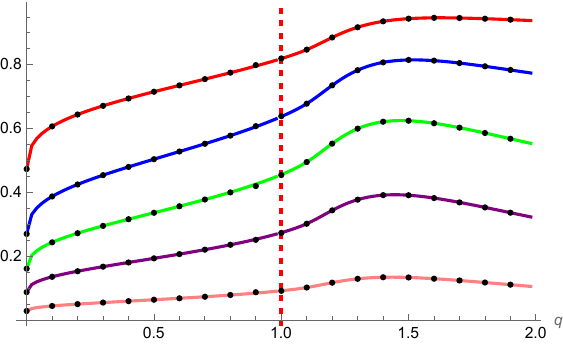}
\end{subfigure}
~
\begin{subfigure}[t]{.5\textwidth}
  \centering
  \includegraphics[width=.8\linewidth]{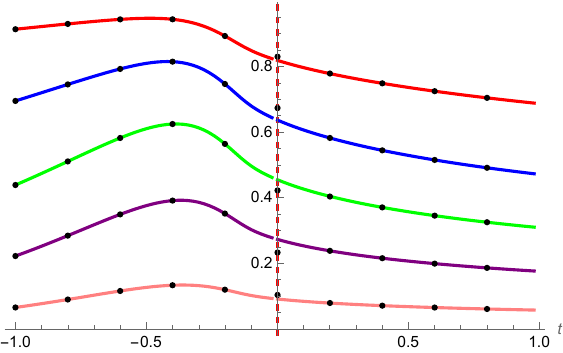}
\end{subfigure}
\\
\begin{subfigure}[t]{.5\textwidth}
  \centering
  \includegraphics[width=.8\linewidth]{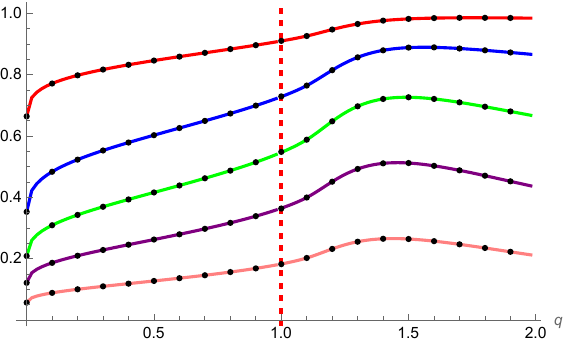}
\end{subfigure}
~
\begin{subfigure}[t]{.5\textwidth}
  \centering
  \includegraphics[width=.8\linewidth]{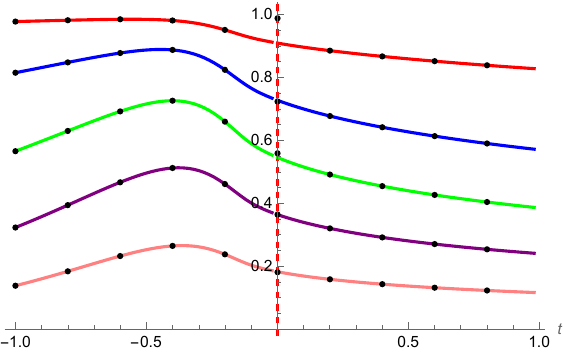}
\end{subfigure}
  \caption{Plots of the real parts of the set A [top] and C [bottom] relations \eqref{eq:aq} and \eqref{eq:cq}, and \eqref{eq:aq-dual} and \eqref{eq:cq-dual}, for $p=11$ on both sides of the natural boundary, in terms of both $q=e^{-t}$ (left) and $t$ (right). On the unary side ($q>1$, $t<0$), we plot the  LHS of \eqref{eq:aq} and \eqref{eq:cq} as a solid line, and the real part of the RHS as black points. Similarly, on the non-unary side ($q<1$ and $t>0$) we plot the  LHS of \eqref{eq:aq-dual} and \eqref{eq:cq-dual} as a solid line, and the RHS as black points.}
  \label{fig:p11}
  \end{figure}
\begin{table}[H]
\begin{tabular}{ | m{1em} | m{7.6cm}| m{7.6cm}| } 
\hline
$a$ & Unary false theta $\Psi^{(a)}_{13}(q)$ & Non-unary dual $q$-series $\Psi^{(a)}_{13}(q)^{\vee}$ \\ 
\hline
$1$ & $q^{1/52}\left(1-q^{12}+q^{14}-q^{50}+q^{54}-q^{114}+\dots\right)$ 
& $q^{-1/52}\left(2 q^2-2 q^4+3 q^6-5 q^8+7 q^{10}+\dots\right)$ \\
\hline
$2$ & $q^{1/13}\left(1-q^{11}+q^{15}-q^{48}+q^{56}-q^{111}+\dots\right)$ 
& $q^{-1/13}\left(2 q-5 q^2+10 q^3-16 q^4+\dots\right)$ \\
\hline
$3$ & $q^{9/52}\left(1-q^{10}+q^{16}-q^{46}+q^{58}-q^{108}+\dots\right)$ 
& $q^{-9/52}\left(1-q^2+q^4+q^8-4 q^{10}+\dots\right)$ \\
\hline
$4$ & $q^{4/13}\left(1-q^9+q^{17}-q^{44}+q^{60}-q^{105}+\dots\right)$ 
& $q^{-4/13}\left(3 q-6 q^2+12 q^3-22 q^4+\dots\right)$ \\
  \hline
  $5$ & $q^{25/52}\left(1-q^8+q^{18}-q^{42}+q^{62}-q^{102}+\dots\right)$ 
  & $q^{-25/52}\left(2 q^2-3 q^4+3 q^6-5 q^8+8 q^{10}+\dots\right)$ \\
  \hline
  $6$ & $q^{9/13}\left(1-q^7+q^{19}-q^{40}+q^{64}-q^{99}+\dots\right)$ 
  & $q^{-9/13}\left(q-4 q^2+8 q^3-14 q^4+26 q^5+\dots\right)$ \\
  \hline
  $7$ & $q^{49/52}\left(1-q^6+q^{20}-q^{38}+q^{66}-q^{96}+\dots\right)$ 
  & $q^{-49/52}\left(2 q^6-2 q^8+3 q^{10}-5 q^{12}+\dots\right)$ \\
  \hline
  $8$ & $q^{16/13}\left(1-q^5+q^{21}-q^{36}+q^{68}-q^{93}+\dots\right)$ 
  & $q^{-16/13}\left(2 q^3-3 q^4+5 q^5-10 q^6+\dots\right)$ \\
  \hline
  $9$ & $q^{81/52}\left(1-q^4+q^{22}-q^{34}+q^{70}-q^{90}+\dots\right)$ 
  & $q^{-81/52}\left(q^2-q^4+q^6-4 q^8+5 q^{10}+\dots\right)$ \\
  \hline
  $10$ & $q^{25/13}\left(1-q^3+q^{23}-q^{32}+q^{72}-q^{87}+\dots\right)$ 
  & $q^{-25/13}\left(q^2-2 q^3+4 q^4-8 q^5+12 q^6+\dots\right)$ \\
  \hline
  $11$ & $q^{121/52}\left(1-q^2+q^{24}-q^{30}+q^{74}-q^{84}+\dots\right)$ 
  & $q^{-121/52}\left(-q^4+3 q^6-3 q^8+4 q^{10}+\dots\right)$ \\
  \hline
  $12$ & $q^{36/13}\left(1-q+q^{25}-q^{28}+q^{76}-q^{81}+\dots\right)$ 
  & $q^{-36/13}\left(-2 q^3+5 q^4-8 q^5+16 q^6+\dots\right)$ \\
  \hline
\end{tabular}
\caption{Small $q$ expansions of $\Psi^{(a)}_{13}(q)$ and their duals $\Psi^{(a)}_{13}(q)^\vee$ for $1\leq a \leq 12$.}  
\label{tab:p=13firstclass}
\end{table}
\graphicspath{ {./paper plots/} }
\begin{figure}[H]
\centering
\begin{subfigure}[t]{.5\textwidth}
  \centering
  \includegraphics[width=.8\linewidth]{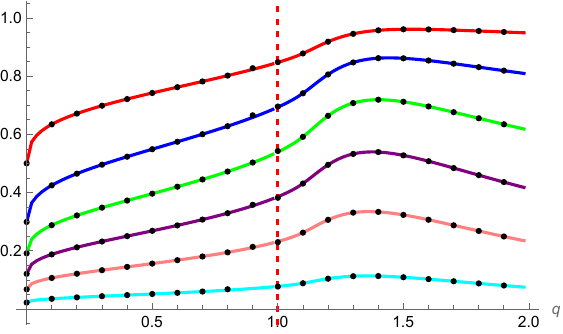}
\end{subfigure}
~
\begin{subfigure}[t]{.5\textwidth}
  \centering
  \includegraphics[width=.8\linewidth]{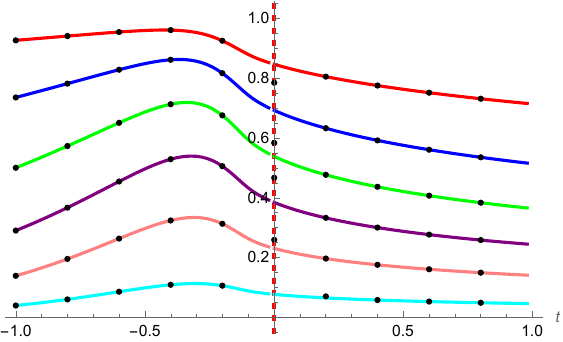}
\end{subfigure}
\\
\begin{subfigure}[t]{.5\textwidth}
  \centering
  \includegraphics[width=.8\linewidth]{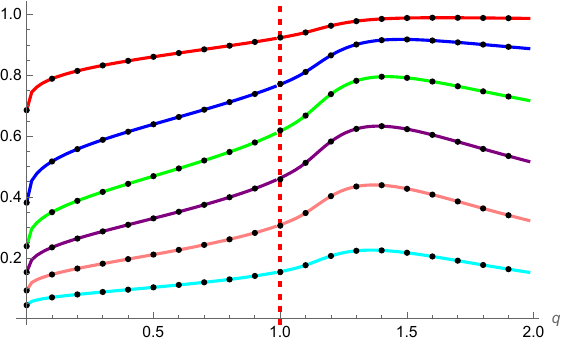}
\end{subfigure}
~
\begin{subfigure}[t]{.5\textwidth}
  \centering
  \includegraphics[width=.8\linewidth]{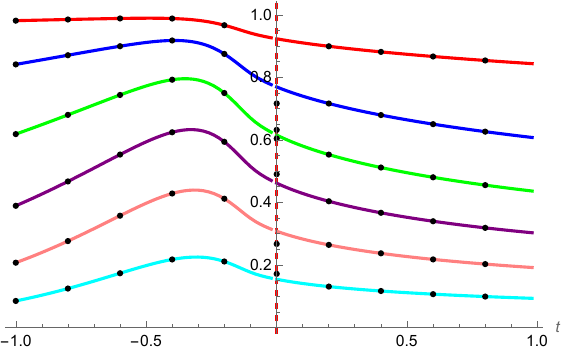}
\end{subfigure}
  \caption{Plots of the real parts of the set A [top] and C [bottom] relations \eqref{eq:aq} and \eqref{eq:cq}, and \eqref{eq:aq-dual} and \eqref{eq:cq-dual}, for $p=13$ on both sides of the natural boundary, in terms of both $q=e^{-t}$ (left) and $t$ (right). On the unary side ($q>1$, $t<0$), we plot the LHS of \eqref{eq:aq} and \eqref{eq:cq} as a solid line, and the real part of the RHS as black points. Similarly, on the non-unary side ($q<1$ and $t>0$) we plot the LHS of \eqref{eq:aq-dual} and \eqref{eq:cq-dual} as a solid line, and the RHS as black points.}
  \label{fig:p13}
\end{figure}
\begin{table}[H]
\begin{tabular}{ | m{1em} | m{8cm}| m{8.5cm}| } 
  \hline
  $a$ & Unary false theta $\Psi^{(a)}_{15}(q)$ & Non-unary dual $q$-series $\Psi^{(a)}_{15}(q)^{\vee}$ \\ 
  \hline
  $1$ & $q^{1/60}\left(1-q^{14}+q^{16}-q^{58}+q^{62}+\dots\right)$ 
  & $\frac{1}{4}q^{-1/60}\left(3+3 q^2-8 q^4+16 q^6-29 q^8+\dots\right)$ \\
  \hline
  $2$ & $q^{1/15}\left(1-q^{13}+q^{17}-q^{56}+q^{64}+\dots\right)$ 
  & $q^{-1/15}\left(4 q-13 q^2+37 q^3-93 q^4+213 q^5+\dots\right)$ \\
  \hline
  $3$ & $q^{9/60}\left(1-q^{12}+q^{18}-q^{54}+q^{66}-q^{126}\dots\right)$ 
  & $\frac{1}{2}q^{-9/60}\left(1+q^2- q^4+2 q^6-4 q^8+5 q^{10}+\dots\right)$ \\
  \hline
  $4$ & $q^{4/15}\left(1-q^{11}+q^{19}-q^{52}+q^{68}+\dots\right)$ 
  & $q^{-4/15}\left(3 q-11 q^2+32 q^3-80 q^4+186 q^5+\dots\right)$ \\
  \hline
  $5$ & $q^{25/60}\left(1-q^{10}+q^{20}-q^{50}+q^{70}-q^{120}\dots\right)$ 
  & $\frac{1}{4}q^{-25/60}\left(1+3 q^2-6 q^4+17 q^6-34 q^8+\dots\right)$ \\
  \hline
  $6$ & $q^{9/15}\left(1-q^9+q^{21}-q^{48}+q^{72}-q^{117}+\dots\right)$ 
  & $q^{-9/15}\left(q-2 q^2+4 q^3-6 q^4+10 q^5+\dots\right)$ \\
  \hline
  $7$ & $q^{49/60}\left(1-q^8+q^{22}-q^{46}+q^{74}-q^{114}\dots\right)$ 
  & $\frac{1}{4}q^{-49/60}\left(5 q^2-10 q^4+19 q^6-33 q^8+\dots\right)$ \\
  \hline
  $8$ & $q^{16/15}\left(1-q^7+q^{23}-q^{44}+q^{76}-q^{111}+\dots\right)$ 
  & $q^{-16/15}\left(2 q^2-9 q^3+31 q^4-83 q^5+197 q^6+\dots\right)$ \\
  \hline
  $9$ & $q^{81/60}\left(1-q^6+q^{24}-q^{42}+q^{78}-q^{108}\dots\right)$ 
  & $\frac{1}{2}q^{-81/60}\left(q^2- q^4+2 q^6-3 q^8+4 q^{10}+\dots\right)$ \\
  \hline
  $10$ & $q^{25/15}\left(1-q^5+q^{25}-q^{40}+q^{80}-q^{105}+\dots\right)$ 
  & $q^{-25/15}\left(q^2-5 q^3+18 q^4-50 q^5+124 q^6+\dots\right)$ \\
  \hline
  $11$ & $q^{121/60}\left(1-q^4+q^{26}-q^{38}+q^{82}-q^{102}\dots\right)$ 
  & $\frac{1}{4}q^{-121/60}\left(q^2-3 q^4+12 q^6-24 q^8+37 q^{10}+\dots\right)$ \\
  \hline
  $12$ & $q^{36/15}\left(1-q^3+q^{27}-q^{36}+q^{84}-q^{99}+\dots\right)$ 
  & $q^{-36/15}\left(q^4-2 q^5+3 q^6-6 q^7+10 q^8+\dots\right)$ \\
  \hline
  $13$ & $q^{169/60}\left(1-q^2+q^{28}-q^{34}+q^{86}-q^{96}\dots\right)$ 
  & $\frac{1}{4}q^{-169/60}\left(q^4-2 q^6+7 q^8-21 q^{10}+43 q^{12}+\dots\right)$ \\
  \hline
  $14$ & $q^{49/15}\left(1-q+q^{29}-q^{32}+q^{88}-q^{93}+\dots\right)$ 
  & $q^{-49/15}\left(-q^4+7 q^5-24 q^6+66 q^7-164 q^8+\dots\right)$ \\
  \hline
\end{tabular}
\caption{Small $q$ expansions of $\Psi^{(a)}_{15}(q)$ and their duals $\Psi^{(a)}_{15}(q)^\vee$ for $1\leq a \leq 14$. We note the growth of the coefficients in the non-unary $q$-series is much slower for $a$ such that $\gcd(15,a)\neq1$.} 
\label{tab:p=15firstclass}
\end{table}
\graphicspath{ {./paper plots/} }
\begin{figure}[H]
\centering
\begin{subfigure}[t]{.5\textwidth}
  \centering
  \includegraphics[width=.8\linewidth]{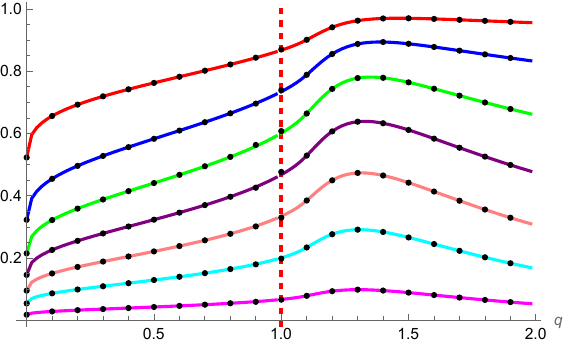}
\end{subfigure}
~
\begin{subfigure}[t]{.5\textwidth}
  \centering
  \includegraphics[width=.8\linewidth]{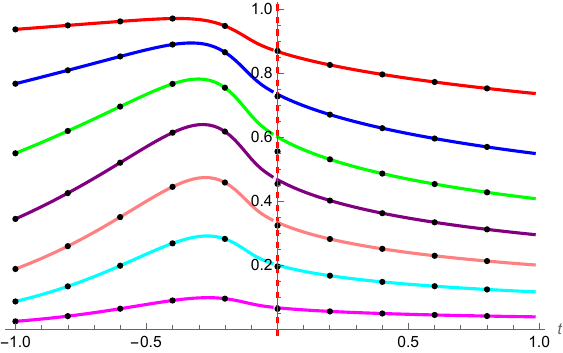}
\end{subfigure}
\\
\begin{subfigure}[t]{.5\textwidth}
  \centering
  \includegraphics[width=.8\linewidth]{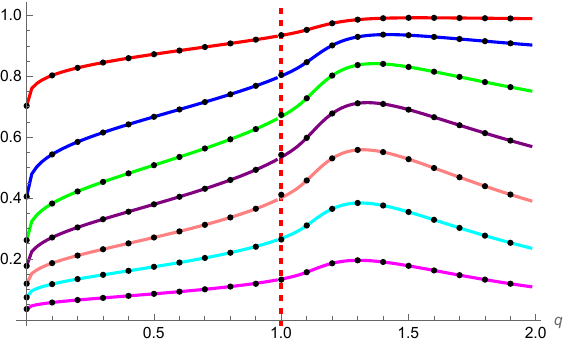}
\end{subfigure}
~
\begin{subfigure}[t]{.5\textwidth}
  \centering
  \includegraphics[width=.8\linewidth]{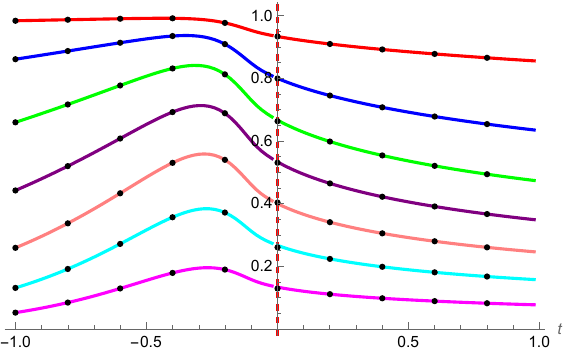}
\end{subfigure}
  \caption{Plots of the real parts of the set A [top] and C [bottom] relations \eqref{eq:aq} and \eqref{eq:cq}, and \eqref{eq:aq-dual} and \eqref{eq:cq-dual}, for $p=15$ on both sides of the natural boundary, in terms of both $q=e^{-t}$ (left) and $t$ (right). On the unary side ($q>1$, $t<0$), we plot the  LHS of \eqref{eq:aq} and \eqref{eq:cq} as a solid  line, and the real part of the RHS as black points. Similarly, on the non-unary side ($q<1$ and $t>0$) we plot the LHS of \eqref{eq:aq-dual} and \eqref{eq:cq-dual} as a solid  line, and the RHS as black points.}
  \label{fig:p15}
\end{figure}
\begin{table}[H]
\begin{tabular}{ | m{1em} | m{7.0cm}| m{7.7cm}| } 
  \hline
  $a$ & Unary false theta $\Psi^{(a)}_{17}(q)$ & Non-unary dual $q$-series  
  $\Psi^{(a)}_{17}(q)^{\vee}$ \\ 
  \hline
  $1$ & $q^{1/68}\left(1-q^{16}+q^{18}-q^{66}+q^{70}+\dots\right)$ 
  & $q^{-1/68}\left(1+q^2-2 q^4+q^6-2 q^8+\dots\right)$ \\
  \hline
  $2$ & $q^{1/17}\left(1-q^{15}+q^{19}-q^{64}+q^{72}+\dots\right)$ 
  & $q^{-1/17}\left(q-2 q^2+4 q^3-6 q^4+10 q^5+\dots\right)$ \\
  \hline
  $3$ & $q^{9/68}\left(1-q^{14}+q^{20}-q^{62}+q^{74}+\dots\right)$ 
  & $q^{-9/68}\left(1-q^6+q^8-q^{10}+q^{12}+\dots\right)$ \\
  \hline
  $4$ & $q^{4/17}\left(1-q^{13}+q^{21}-q^{60}+q^{76}+\dots\right)$ 
  & $q^{-4/17}\left(2 q-4 q^2+6 q^3-9 q^4+15 q^5+\dots\right)$ \\
  \hline
  $5$ & $q^{25/68}\left(1-q^{12}+q^{22}-q^{58}+q^{78}+\dots\right)$ 
  & $q^{-25/68}\left(q^2-q^4+2 q^6-3 q^8+4 q^{10}+\dots\right)$ \\
  \hline
  $6$ & $q^{9/17}\left(1-q^{11}+q^{23}-q^{56}+q^{80}+\dots\right)$ 
  & $q^{-9/17}\left(2 q-3 q^2+6 q^3-10 q^4+14 q^5+\dots\right)$ \\
  \hline
  $7$ & $q^{49/68}\left(1-q^{10}+q^{24}-q^{54}+q^{82}+\dots\right)$ 
  & $q^{-49/68}\left(2 q^6-q^8-q^{12}+2 q^{14}-2 q^{16}+\dots\right)$ \\
  \hline
  $8$ & $q^{16/17}\left(1-q^9+q^{25}-q^{52}+q^{84}+\dots\right)$ 
  & $q^{-16/17}\left(q-2 q^2+4 q^3-6 q^4+10 q^5+\dots\right)$ \\
  \hline
  $9$ & $q^{81/68}\left(1-q^8+q^{26}-q^{50}+q^{86}+\dots\right)$ 
  & $q^{-81/68}\left(q^2-q^4+q^6-2 q^8+3 q^{10}+\dots\right)$ \\
  \hline
  $10$ & $q^{25/17}\left(1-q^7+q^{27}-q^{48}+q^{88}+\dots\right)$ 
  & $q^{-25/17}\left(-q^2+2 q^3-2 q^4+4 q^5-6 q^6+\dots\right)$ \\
  \hline
  $11$ & $q^{121/68}\left(1-q^6+q^{28}-q^{46}+q^{90}+\dots\right)$ 
  & $q^{-121/68}\left(q^2-q^4+q^6-2 q^8+2 q^{10}+\dots\right)$ \\
  \hline
  $12$ & $q^{36/17}\left(1-q^5+q^{29}-q^{44}+q^{92}+\dots\right)$ 
  & $q^{-36/17}\left(-q^5+2 q^6-2 q^7+5 q^8-6 q^9+\dots\right)$ \\
  \hline
  $13$ & $q^{169/68}\left(1-q^4+q^{30}-q^{42}+q^{94}+\dots\right)$ 
  & $q^{-169/68}\left(q^6-2 q^8+2 q^{10}-q^{12}+2 q^{14}+\dots\right)$ \\
  \hline
  $14$ & $q^{49/17}\left(1-q^3+q^{31}-q^{40}+q^{96}+\dots\right)$ 
  & $q^{-49/17}\left(2 q^4-3 q^5+4 q^6-8 q^7+12 q^8+\dots\right)$ \\
  \hline
  $15$ & $q^{225/68}\left(1-q^2+q^{32}-q^{38}+q^{98}+\dots\right)$ 
  & $q^{-225/68}\left(-q^4+2 q^6-2 q^8+3 q^{10}+\dots\right)$ \\
  \hline
  $16$ & $q^{64/17}\left(1-q+q^{33}-q^{36}+q^{100}+\dots\right)$ 
  & $q^{-64/17}\left(q^4-2 q^5+4 q^6-8 q^7+12 q^8+\dots\right)$ \\
  \hline
\end{tabular}
\caption{Small $q$ expansions of $\Psi^{(a)}_{17}(q)$ and their duals $\Psi^{(a)}_{17}(q)^\vee$ for $1\leq a \leq 16$.} 
\label{tab:p=17firstclass}
\end{table}
\graphicspath{ {./paper plots/} }
\begin{figure}[H]
\centering
\begin{subfigure}[t]{.5\textwidth}
  \centering
  \includegraphics[width=.8\linewidth]{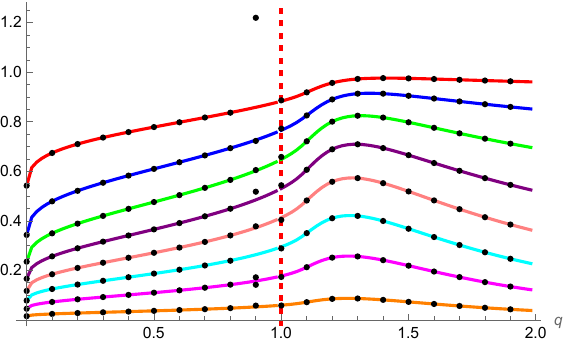}
\end{subfigure}
~
\begin{subfigure}[t]{.5\textwidth}
  \centering
  \includegraphics[width=.8\linewidth]{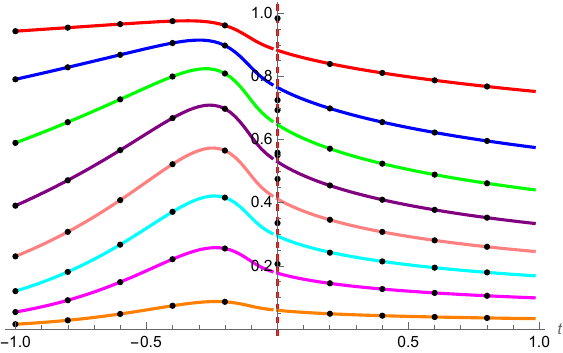}
\end{subfigure}
\\
\begin{subfigure}[t]{.5\textwidth}
  \centering
  \includegraphics[width=.8\linewidth]{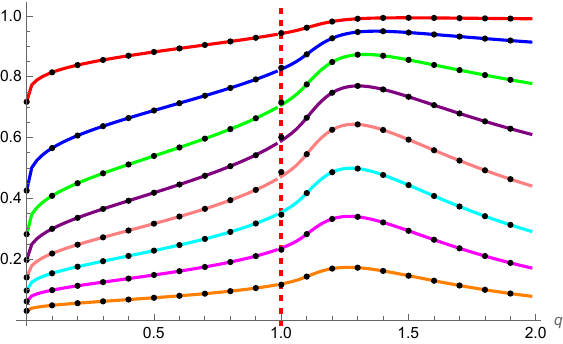}
\end{subfigure}
~
\begin{subfigure}[t]{.5\textwidth}
  \centering
  \includegraphics[width=.8\linewidth]{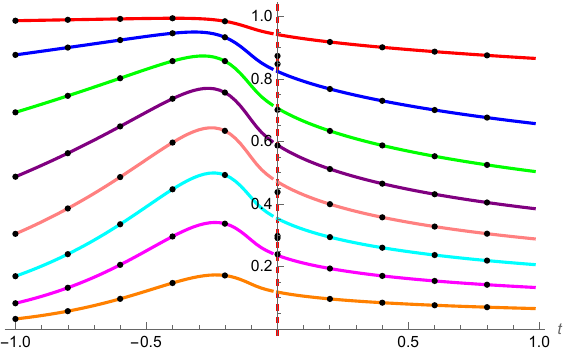}
\end{subfigure}
 \caption{Plots of the real parts of the set A [top] and C [bottom] relations \eqref{eq:aq} and \eqref{eq:cq}, and \eqref{eq:aq-dual} and \eqref{eq:cq-dual}, for $p=17$ on both sides of the natural boundary, in terms of both $q=e^{-t}$ (left) and $t$ (right). On the unary side ($q>1$, $t<0$), we plot the  LHS of \eqref{eq:aq} and \eqref{eq:cq} as a solid line, and the real part of the RHS as black points. Similarly, on the non-unary side ($q<1$ and $t>0$) we plot the LHS of \eqref{eq:aq-dual} and \eqref{eq:cq-dual} as a solid  line, and the RHS as black points.}
  \label{fig:p17}
\end{figure}
\begin{table}[H]
\begin{center}
\begin{tabular}{ | m{1em} | m{7.1cm}| m{7.9cm}| } 
  \hline
  $a$ & Unary false theta  $\Psi^{(a)}_{19}(q)$ & Non-unary dual $q$-series 
  $\Psi^{(a)}_{19}(q)^{\vee}$ \\ 
  \hline
  $1$ & $q^{1/76}\left(1-q^{18}+q^{20}-q^{74}+q^{78}+\dots\right)$ 
  & $\frac{1}{8}q^{-1/76}\left(10-7 q^2+6 q^4-10 q^6+18 q^8+\dots\right)$ \\
  \hline
  $2$ & $q^{1/19}\left(1-q^{17}+q^{21}-q^{72}+q^{80}+\dots\right)$ 
  & $\frac{1}{2}q^{-1/19}\left(6 q-14 q^2+31 q^3-61 q^4+\dots\right)$ \\
  \hline
  $3$ & $q^{9/76}\left(1-q^{16}+q^{22}-q^{70}+q^{82}+\dots\right)$ 
  & $\frac{1}{8}q^{-9/76}\left(7-2 q^2-3 q^4+2 q^6+13 q^8+\dots\right)$ \\
  \hline
  $4$ & $q^{4/19}\left(1-q^{15}+q^{23}-q^{68}+q^{84}+\dots\right)$ 
  & $\frac{1}{2}q^{-4/19}\left(2 q-3 q^2+9 q^3-18 q^4+39 q^5+\dots\right)$ \\
  \hline
  $5$ & $q^{25/76}\left(1-q^{14}+q^{24}-q^{66}+q^{86}+\dots\right)$ 
  & $\frac{1}{8}q^{-25/76}\left(1+8 q^2-6 q^4+13 q^6-25 q^8+\dots\right)$ \\
  \hline
  $6$ & $q^{9/19}\left(1-q^{13}+q^{25}-q^{64}+q^{88}+\dots\right)$ 
  & $\frac{1}{2}q^{-9/19}\left(4 q-9 q^2+17 q^3-33 q^4+\dots\right)$ \\
  \hline
  $7$ & $q^{49/76}\left(1-q^{12}+q^{26}-q^{62}+q^{90}+\dots\right)$ 
  & $\frac{1}{8}q^{-49/76}\left(8 q^2-3 q^4+8 q^6-28 q^8+\dots\right)$ \\
  \hline
  $8$ & $q^{16/19}\left(1-q^{11}+q^{27}-q^{60}+q^{92}+\dots\right)$ 
  & $\frac{1}{2}q^{-16/19}\left(3 q-6 q^2+12 q^3-22 q^4+\dots\right)$ \\
  \hline
  $9$ & $q^{81/76}\left(1-q^{10}+q^{28}-q^{58}+q^{94}+\dots\right)$ 
  & $\frac{1}{8}q^{-81/76}\left(7 q^2-10 q^4+20 q^6-30 q^8+\dots\right)$ \\
  \hline
  $10$ & $q^{25/19}\left(1-q^9+q^{29}-q^{56}+q^{96}+\dots\right)$ 
  & $\frac{1}{2}q^{-25/19}\left(q^2-5 q^3+19 q^4-42 q^5+\dots\right)$ \\
  \hline
  $11$ & $q^{121/76}\left(1-q^8+q^{30}-q^{54}+q^{98}+\dots\right)$ 
  & $\frac{1}{8}q^{-121/76}\left(5 q^2-13 q^4+28 q^6-38 q^8+\dots\right)$ \\
  \hline
  $12$ & $q^{36/19}\left(1-q^7+q^{31}-q^{52}+q^{100}+\dots\right)$ 
  & $\frac{1}{2}q^{-36/19}\left(-q^2+q^3+4 q^4-9 q^5+25 q^6+\dots\right)$ \\
  \hline
  $13$ & $q^{169/76}\left(1-q^6+q^{32}-q^{50}+q^{102}+\dots\right)$ 
  & $\frac{1}{8}q^{-169/76}\left(3 q^2-2 q^4-4 q^6+8 q^8+\dots\right)$ \\
  \hline
  $14$ & $q^{49/19}\left(1-q^5+q^{33}-q^{48}+q^{104}+\dots\right)$ 
  & $\frac{1}{2}q^{-49/19}\left(q^3-q^4+q^5-5 q^7+14 q^8+\dots\right)$ \\
  \hline
  $15$ & $q^{225/76}\left(1-q^4+q^{34}-q^{46}+q^{106}+\dots\right)$ 
  & $\frac{1}{8}q^{-225/76}\left(2 q^4+2 q^6-5 q^8-q^{10}+4 q^{12}+\dots\right)$ \\
  \hline
  $16$ & $q^{64/19}\left(1-q^3+q^{35}-q^{44}+q^{108}+\dots\right)$ 
  & $\frac{1}{2}q^{-64/19}\left(3 q^4-8 q^5+13 q^6-25 q^7+\dots\right)$ \\
  \hline
  $17$ & $q^{289/76}\left(1-q^2+q^{36}-q^{42}+q^{110}+\dots\right)$ 
  & $\frac{1}{8}q^{-289/76}\left(-5 q^4+16 q^6-12 q^8+6 q^{10}+\dots\right)$ \\
  \hline
  $18$ & $q^{81/19}\left(1-q+q^{37}-q^{40}+q^{112}+\dots\right)$ 
  & $\frac{1}{2}q^{-81/19}\left(q^5-4 q^6+11 q^7-30 q^8+\dots\right)$ \\
  \hline
\end{tabular}
\end{center}
\caption{Small $q$ expansions of $\Psi^{(a)}_{19}(q)$ and their duals $\Psi^{(a)}_{19}(q)^\vee$ for $1\leq a \leq 18$.} 
\label{tab:p=19firstclass}
\end{table}
\graphicspath{ {./paper plots/} }
\begin{figure}[H]
\centering
\begin{subfigure}[t]{.5\textwidth}
  \centering
  \includegraphics[width=.8\linewidth]{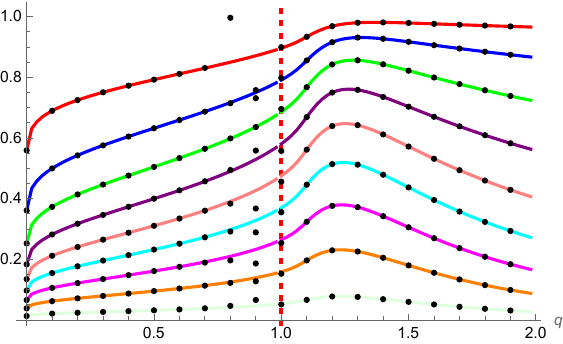}
\end{subfigure}
~
\begin{subfigure}[t]{.5\textwidth}
  \centering
  \includegraphics[width=.8\linewidth]{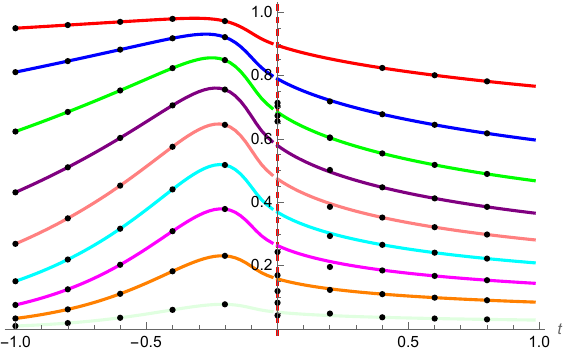}
\end{subfigure}
\\
\begin{subfigure}[t]{.5\textwidth}
  \centering
  \includegraphics[width=.8\linewidth]{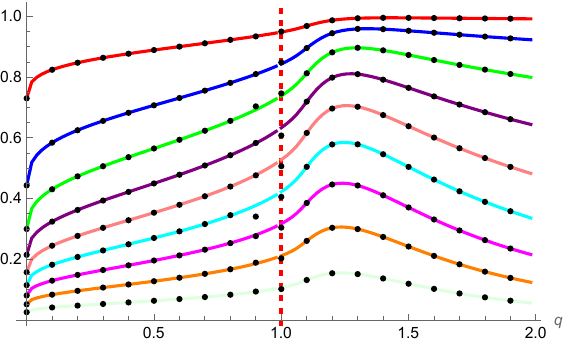}
\end{subfigure}
~
\begin{subfigure}[t]{.5\textwidth}
  \centering
  \includegraphics[width=.8\linewidth]{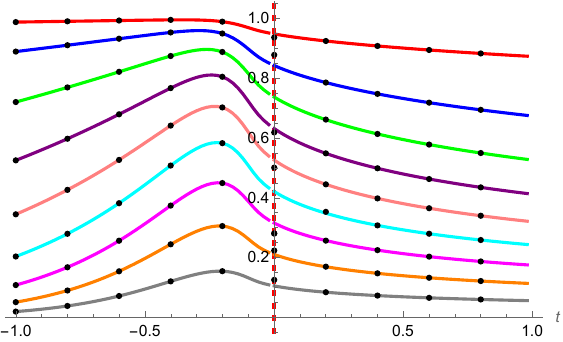}
\end{subfigure}
  \caption{Plots of the real parts of the set A [top] and C [bottom] relations \eqref{eq:aq} and \eqref{eq:cq}, and \eqref{eq:aq-dual} and \eqref{eq:cq-dual}, for $p=19$ on both sides of the natural boundary, in terms of both $q=e^{-t}$ (left) and $t$ (right). On the unary side ($q>1$, $t<0$), we plot the LHS of \eqref{eq:aq} and \eqref{eq:cq} as a solid line, and the real part of the RHS as black points. Similarly, on the non-unary side ($q<1$ and $t>0$) we plot the LHS of \eqref{eq:aq-dual} and \eqref{eq:cq-dual} as a solid  line, and the RHS as black points.}
  \label{fig:p19}
\end{figure}

\section{Results for Duals for the Orientation Reversal of Brieskorn Spheres $\overline{\Sigma(2, 3, 6k\pm 1)}$ and a strategy for general $\overline{\Sigma(p_1,p_2,p_3)}$}
\label{sec:appendix-results2}

In this appendix we give the tables of $q$-series expansions of the duals $X^{(j)}_p(q)^\vee$ for $j=1,2,\dots,(p-1)/2$, omitting the unary expansions.
Here we tabulate the results for $p=17, 19, 25, 35$. The latter two cases are  the first instances of $p=6k\pm1$ not being prime. We see that a new substructure  emerges -- the $q$-series coefficients grow more slowly than the others with $j=5,10$ for $p=25$, or with  $j=5,10, 14, 15, 21$ for $p=35$.
We also outline a strategy for extending our methods past the class $\overline{\Sigma(2,3,6k\pm1)}$, and display some results.
\begin{table}[H]
\begin{tabular}{ | m{1em} | m{11cm}| } 
  \hline
  $j$ & dual $q$-series $X^{(j)}_{17}(q)^{\vee}$  \\ 
  \hline
  $1$ & $\frac{1}{3}q^{-121/408}\left(3 q+13 q^2+26 q^3+59 q^4+103 q^5+197 q^6+\dots\right)$ 
  \\
  \hline
  $2$ & $\frac{1}{3}q^{-25/408}\left(-1+7 q+16 q^2+41 q^3+76 q^4+153 q^5+\dots\right)$ 
  \\
  \hline
  $3$ & $\frac{1}{3}q^{-1/408}\left(3+q+8 q^2+14 q^3+31 q^4+56 q^5+102 q^6+\dots\right)$ 
  \\
  \hline
  $4$ & $\frac{1}{3}q^{-49/408}\left(-3 q-12 q^2-21 q^3-45 q^4-76 q^5-146 q^6+\dots\right)$ 
  \\
  \hline
  $5$ & $\frac{1}{3}q^{-169/408}\left(-6 q-15 q^2-31 q^3-68 q^4-125 q^5-229 q^6+\dots\right)$ 
  \\
  \hline
  $6$ & $\frac{1}{3}q^{-361/408}\left(-2 q-7 q^2-16 q^3-35 q^4-61 q^5-120 q^6+\dots\right)$ 
  \\
  \hline
  $7$ & $\frac{1}{3}q^{-625/408}\left(-q^2+q^3+4 q^4+11 q^5+16 q^6+42 q^7+61 q^8+\dots\right)$ 
  \\
  \hline
  $8$ & $\frac{1}{3}q^{-961/408}\left(3 q^3+12 q^4+27 q^5+61 q^6+113 q^7+209 q^8+\dots\right)$ 
  \\
  \hline
\end{tabular}
\caption{Small $q$ expansions of the dual $q$-series $X^{(j)}_{17}(q)^\vee$ for $1\leq j \leq 8$.}  
\label{tab:mock17secondclass}
\end{table}
\begin{table}[H]
\begin{tabular}{ | m{1em} | m{11cm}| } 
  \hline
  $j$ & dual $q$-series $X^{(j)}_{19}(q)^{\vee}$  \\ 
  \hline
  $1$ & $\frac{1}{2}q^{-169/456}\left(3 q+8 q^2+16 q^3+37 q^4+59 q^5+111 q^6+\dots\right)$ 
  \\
  \hline
  $2$ & $\frac{1}{2}q^{-49/456}\left(6 q+12 q^2+30 q^3+53 q^4+99 q^5+166 q^6+\dots\right)$ 
  \\
  \hline
  $3$ & $\frac{1}{2}q^{-1/456}\left(-1+4 q+9 q^2+18 q^3+37 q^4+62 q^5+110 q^6+\dots\right)$ 
  \\
  \hline
  $4$ & $\frac{1}{2}q^{-25/456}\left(1-4 q^2-3 q^3-11 q^4-16 q^5-32 q^6-41 q^7+\dots\right)$ 
  \\
  \hline
  $5$ & $\frac{1}{2}q^{-121/456}\left(-3 q-11 q^2-21 q^3-42 q^4-74 q^5-133 q^6+\dots\right)$ 
  \\
   \hline
  $6$ & $\frac{1}{2}q^{-289/456}\left(-2 q-9 q^2-19 q^3-38 q^4-70 q^5-127 q^6+\dots\right)$ 
  \\
  \hline
  $7$ & $\frac{1}{2}q^{-529/456}\left(-3 q^2-5 q^3-15 q^4-24 q^5-48 q^6-75 q^7+\dots\right)$ 
  \\
  \hline
  $8$ & $\frac{1}{2}q^{-841/456}\left(q^2+2 q^3+6 q^4+10 q^5+20 q^6+37 q^7+64 q^8+\dots\right)$ 
  \\
  \hline
  $9$ & $\frac{1}{2}q^{-1225/456}\left(3 q^3+7 q^4+18 q^5+32 q^6+67 q^7+111 q^8+\dots\right)$ 
  \\
  \hline
\end{tabular}
\caption{Small $q$ expansions of the dual $q$-series $X^{(j)}_{19}(q)^\vee$ for $1\leq j \leq 9$.}  
\label{tab:mock19secondclass}
\end{table}
\begin{table}[H]
\begin{tabular}{ | m{1em} | m{11cm}| } 
  \hline
  $j$ & dual $q$-series $X^{(j)}_{25}(q)^{\vee}$  \\ 
  \hline
  $1$ & $q^{-361/600}\left(-2 q-2 q^2-6 q^3-8 q^4-18 q^5-22 q^6-40 q^7+\dots\right)$ 
  \\
    \hline
  $2$ & $q^{-169/600}\left(2 q+6 q^2+8 q^3+18 q^4+30 q^5+46 q^6+70 q^7+\dots\right)$ 
  \\
    \hline
  $3$ & $q^{-49/600}\left(2 q+6 q^2+12 q^3+18 q^4+32 q^5+50 q^6+78 q^7+\dots\right)$ 
  \\
    \hline
  $4$ & $q^{-1/600}\left(-2-2 q-6 q^2-6 q^3-14 q^4-20 q^5-34 q^6+\dots\right)$ 
  \\
    \hline
  $5$ & $\frac{2}{5}q^{-1/24}\left(1-4 q+2 q^2-2 q^3-3 q^5-4 q^6+2 q^8+\dots\right)$ 
  \\
    \hline
  $6$ & $q^{-121/600}\left(2 q+2 q^2+6 q^3+10 q^4+20 q^5+28 q^6+48 q^7+\dots\right)$ 
  \\
    \hline
  $7$ & $q^{-289/600}\left(-4 q-4 q^2-10 q^3-16 q^4-28 q^5-42 q^6+\dots\right)$ 
  \\
    \hline
  $8$ & $q^{-529/600}\left(-2 q-4 q^2-8 q^3-14 q^4-22 q^5-34 q^6+\dots\right)$ 
  \\
    \hline
  $9$ & $q^{-841/600}\left(2 q^3+6 q^4+10 q^5+14 q^6+28 q^7+\dots\right)$ 
  \\
    \hline
  $10$ & $\frac{2}{5}q^{-49/24}\left(-2 q^2-2 q^3-4 q^4-q^5-4 q^7-2 q^8+\dots\right)$ 
  \\
    \hline
  $11$ & $q^{-1681/600}\left(-2 q^3-4 q^4-4 q^5-10 q^6-14 q^7-24 q^8+\dots\right)$ 
  \\
    \hline
  $12$ & $q^{-2209/600}\left(4 q^5+6 q^6+14 q^7+22 q^8+40 q^9+58 q^{10}+\dots\right)$ 
  \\
  \hline
\end{tabular}
\caption{Small $q$ expansions of the dual $q$-series $X^{(j)}_{25}(q)^\vee$ for $1\leq j \leq 12$.} 
\label{tab:mock25secondclass}
\end{table}
\begin{table}[H]
\begin{tabular}{ | m{1em} | m{11cm}| } 
  \hline
  $j$ & dual $q$-series $X^{(j)}_{35}(q)^{\vee}$  \\ 
  \hline
  $1$ & $\frac{1}{6}q^{-841/840}\left(q+9 q^2+17 q^3+43 q^4+65 q^5+137 q^6+\dots\right)$ 
  \\
    \hline
  $2$ & $\frac{1}{6}q^{-529/840}\left(7 q+20 q^2+42 q^3+81 q^4+149 q^5+262 q^6+\dots\right)$ 
  \\
    \hline
  $3$ & $\frac{1}{6}q^{-289/840}\left(9 q+27 q^2+50 q^3+106 q^4+177 q^5+315 q^6+\dots\right)$ 
  \\
    \hline
  $4$ & $\frac{1}{6}q^{-121/840}\left(10 q+23 q^2+44 q^3+82 q^4+135 q^5+240 q^67+\dots\right)$ 
  \\
    \hline
  $5$ & $\frac{1}{2}q^{-5/168}\left(-1+q+3 q^2+4 q^3+7 q^4+10 q^5+16 q^6+\dots\right)$ 
  \\
    \hline
  $6$ & $\frac{1}{6}q^{-1/840}\left(5-3 q-11 q^2-25 q^3-47 q^4-95 q^5-169 q^6+\dots\right)$ 
  \\
    \hline
  $7$ & $\frac{1}{3}q^{-7/120}\left(1-4 q-12 q^2-23 q^3-49 q^4-87 q^5-153 q^6+\dots\right)$ 
  \\
    \hline
  $8$ & $\frac{1}{6}q^{-169/840}\left(-11 q-24 q^2-52 q^3-99 q^4-177 q^5-307 q^6+\dots\right)$ 
  \\
    \hline
  $9$ & $\frac{1}{6}q^{-361/840}\left(-5 q-18 q^2-34 q^3-66 q^4-108 q^5-202 q^6+\dots\right)$ 
  \\
    \hline
  $10$ & $\frac{1}{2}q^{-125/168}\left(-q-3 q^2-4 q^3-7 q^4-10 q^5-16 q^6-20 q^7+\dots\right)$ 
  \\
    \hline
  $11$ & $\frac{1}{6}q^{-961/840}\left(-2 q^2+5 q^3+8 q^4+34 q^5+57 q^6+132 q^7+\dots\right)$ 
  \\
    \hline
  $12$ & $\frac{1}{6}q^{-1369/840}\left(q^2+8 q^3+18 q^4+51 q^5+95 q^6+190 q^7+\dots\right)$ 
  \\
    \hline
  $13$ & $\frac{1}{6}q^{-1849/840}\left(5 q^3+12 q^4+40 q^5+69 q^6+147 q^7+\dots\right)$ 
  \\
    \hline
  $14$ & $\frac{1}{3}q^{-343/120}\left(q^3+3 q^4+9 q^5+15 q^6+35 q^7+58 q^8+\dots\right)$ 
  \\
    \hline
  $15$ & $\frac{1}{2}q^{-605/168}\left(q^5+q^6+3 q^7+4 q^8+7 q^9+8 q^{10}+\dots\right)$ 
  \\
    \hline
  $16$ & $\frac{1}{6}q^{-3721/840}\left(q^5-6 q^6-10 q^7-30 q^8-60 q^9-118 q^{10}+\dots\right)$ 
  \\
      \hline
  $17$ & $\frac{1}{6}q^{-4489/840}\left(-3 q^6-9 q^7-26 q^8-58 q^9-111 q^{10}+\dots\right)$ 
  \\
  \hline
\end{tabular}
\caption{Small $q$ expansions of the dual $q$-series $X^{(j)}_{35}(q)^\vee$ for $1\leq j \leq 17$.} 
\label{tab:mock35secondclass}
\end{table}

\pagebreak

\subsection{Other Brieskorn Spheres}
\label{sec:other-brieskorn}
The previous examples involving Brieskorn spheres were limited to the class $\overline{\Sigma(2,3,6k\pm1)}$, due to the fact they have been studied extensively in previous work. For $k=1$, the  structure that continues from the unary side matches the order $5$ and order $7$ mock theta relations. While in this work we focused on extending this to $k>1$, in principle there is nothing preventing us also from studying the more general $\overline{\Sigma(p_1,p_2,p_3)}$ using our machinery.\\
Here, we sketch a procedure for extracting the $q$-series associated to orientation reversed Brieskorn spheres outside of the class considered in section \ref{sec:class2}. Given the $\widehat{Z}$-invariant for $\overline{\Sigma(p_1,p_2,p_3)}$, with $p_1,p_2,p_3$ pairwise coprime, in terms of false theta functions, given in \cite{GM}, we can form the Mordell-Borel integral whose real part for $t<0$ is equal to the $\widehat{Z}$-invariant. The imaginary part of this integral will contain all of the information necessary to construct the entire vector of Mordell-Borel integrals of length $N=\frac{1}{4}(p_1-1)(p_2-1)(p_3-1)$. Let $p=p_1p_2p_3$ and
\begin{eqnarray*}
    a_{11} &=& p_1 p_2 p_3 - p_1 p_2-p_1 p_3-p_2p_3, \\
    a_{12} &=& p_1 p_2p_3 + p_1 p_2-p_1 p_3-p_2p_3, \\
    a_{13} &=& p_1 p_2 p_3 - p_1 p_2+p_1 p_3-p_2p_3, \\
    a_{14} &=& p_1 p_2 p_3 + p_1 p_2+p_1 p_3-p_2p_3,
\end{eqnarray*}
and define
\begin{eqnarray}
    \widehat{JS}_{(p,\vb{A}_1)}(t)=\sum\limits_{i=1}^{4} \text{sign}(a_i) JS_{(p,a_i)}(t) 
\end{eqnarray}
with $\vb{A}_1=(a_{11})+(a_{12})+(a_{13})+(a_{14})$. Then, we have
\begin{eqnarray}
    \operatorname{Im}\left[\sqrt{\frac{4pt}{\pi}}\widehat{JS}_{(p,\vb{A}_1)}(2t)\right] &=& \sqrt{\frac{\pi}{t}}\sum\limits_{j=1}^{N} M_{1j} \Psi^{(\vb{A}_j)}_p\left(\frac{1}{\qt^2}\right) \\
    &=&  \sqrt{\frac{\pi}{t}}\sum\limits_{j=1}^{N} M_{1j} \operatorname{Re}\left[\sqrt{\frac{4pt}{\pi}}\widehat{JS}_{(p,\vb{A}_j)}\left(\frac{2\pi^2}{t}\right)\right]
\end{eqnarray}
where $(\vb{A}_j) = (a_{j1})+(a_{j2})+(a_{j3})+(a_{j4})$ is a list of integers indexing the $a$ values appearing in the false theta functions and Mordell-Borel integrals, similar to the notation used in (\ref{eq:ns}). From the list of indices $(\vb{A}_j)$, we can construct the vector of integrals and take their respective imaginary decompositions to find the mixing matrices:
\begin{eqnarray}
    \operatorname{Im}\left[\sqrt{\frac{4pt}{\pi}}\widehat{JS}_{(p,\vb{A}_i)}(2t)\right] &=& \sqrt{\frac{\pi}{t}}\sum\limits_{j=1}^{N} M_{ij} \Psi^{(\vb{A}_j)}_p\left(\frac{1}{\qt^2}\right) \\
    &=&  \sqrt{\frac{\pi}{t}}\sum\limits_{j=1}^{N} M_{ij} \operatorname{Re}\left[\sqrt{\frac{4pt}{\pi}}\widehat{JS}_{(p,\vb{A}_j)}\left(\frac{2\pi^2}{t}\right)\right]
\end{eqnarray}
With the mixing matrix and the vector of Mordell-Borel integrals that satisfy the modular relation, we can form the expected relations when the integrals are rotated back to $t>0$, similar in structure to (\ref{eq:mock_q2_nonunary}), and then implement our numerical algorithm to find the coefficients of the $q$-series. We present results for $\overline{\Sigma(2,5,7)}$, $\overline{\Sigma(2,5,11)}$, and $\overline{\Sigma(3,5,7)}$ below. We note that the coefficients of the $q$-series for $\overline{\Sigma(3,4,5)}$ and $\overline{\Sigma(2,5,7)}$ match those presented in Tables 15 and 17 of \cite{CD20c}, as mentioned in Section {\ref{sec:comparison}}.
\label{sec:257}
\begin{table}[h!]
\begin{tabular}{ | m{1em} | m{13cm}| } 
  \hline
  $j$ & dual $q$-series $X^{(j)}_{(2,5,7)}(q)^{\vee}$  \\
  \hline
  $1$ & $q^{-1/280}(-2 - 2 q^2 - 2 q^3 - 2 q^6 - 2 q^8 - 2 q^9 - 2 q^{10} - 2 q^{11} - 2 q^{12} - 2 q^{14} + \dots)$ \\ 
  \hline
    $2$ & $q^{-9/280}(-2 q^2 - 2 q^4 - 2 q^7 - 2 q^8 - 2 q^{10} - 2 q^{11} - 2 q^{12} - 2 q^{14}  + \dots) $ \\ 
  \hline
    $3$ & $q^{-81/280}(-2 q - 2 q^3 - 2 q^5 - 2 q^6 - 2 q^7 - 2 q^9 - 2 q^{10} - 2 q^{11} - 2 q^{12} - 2 q^{13} + \dots)$ \\ 
  \hline
  $4$ & $q^{-121/280}(-2 q - 2 q^5 - 2 q^7 - 2 q^{11} - 2 q^{13} - 2 q^{15} - 2 q^{16} - 2 q^{17}  + \dots)$ \\ 
  \hline
  $5$ & $q^{-169/280}(-2 q - 2 q^2 - 2 q^4 - 2 q^5 - 2 q^6 - 2 q^7 - 2 q^8 - 2 q^9 - 2 q^{10} - 2 q^{11} + \dots)$ \\ 
  \hline
  $6$ & $q^{-529/280}(-2 q^2 - 2 q^3 - 2 q^4 - 2 q^5 - 2 q^6 - 2 q^7 - 2 q^8 - 2 q^9 - 2 q^{10} - 2 q^{11} + \dots)$ \\ 
  \hline
\end{tabular}
\caption{Small $q$ expansions of the dual $q$-series for  $\overline{\Sigma(2,5,7)}$.} 
\label{tab:sigma257}
\end{table}
\label{sec:2511}
\begin{table}[h!]
\begin{tabular}{ | m{1em} | m{14cm}| } 
  \hline
  $j$ & dual $q$-series $X^{(j)}_{(2,5,11)}(q)^{\vee}$  \\
  \hline
    $1$ & $\frac{2}{3}q^{-1/440}\left(-2 - q - 2 q^2 - 3 q^3 - 3 q^4 - 5 q^5 - 5 q^6 - 9 q^7 - 9 q^8 - 13 q^9 + \dots\right)$ \\
  \hline
    $2$ & $\frac{2}{3}q^{-9/440}\left(-1 + 2 q^2 + q^3 + 4 q^4 + 4 q^5 + 6 q^6 + 7 q^7 + 11 q^8 + 14 q^9 + 18 q^{10}+ \dots\right)$ \\
  \hline
    $3$ & $\frac{2}{3}q^{-49/440}\left(-q - q^2 - 3 q^3 - 2 q^4 - 3 q^5 - 4 q^6 - 5 q^7 - 6 q^8 - 9 q^9 - 9 q^{10}+ \dots\right)$ \\
  \hline
    $4$ & $\frac{2}{3}q^{-81/440}\left(-2 q - 3 q^2 - 4 q^3 - 5 q^4 - 5 q^5 - 9 q^6 - 13 q^7 - 15 q^8 - 20 q^9 + \dots\right)$ \\
  \hline
    $5$ & $\frac{2}{3}q^{-169/440}\left(q + 2 q^2 + 4 q^3 + 6 q^4 + 7 q^5 + 10 q^6 + 14 q^7 + 18 q^8 + 23 q^9 + \dots\right)$ \\
  \hline
    $6$ & $\frac{2}{3}q^{-289/440}\left(-2 q - 3 q^2 - 4 q^3 - 5 q^4 - 8 q^5 - 10 q^6 - 12 q^7 - 17 q^8 - 22 q^9+ \dots\right)$ \\
  \hline
    $7$ & $\frac{2}{3}q^{-361/440}\left(-q - 2 q^2 - 2 q^3 - 3 q^4 - 3 q^5 - 5 q^6 - 5 q^7 - 6 q^8 - 9 q^9 - 10 q^{10}+ \dots\right)$ \\
  \hline
  $8$ & $\frac{2}{3}q^{-529/440}\left(2 q^2 + q^3 + 4 q^4 + 4 q^5 + 6 q^6 + 8 q^7 + 13 q^8 + 13 q^9 + 20 q^{10} + \dots \right)$ \\
  \hline
    $9$ & $\frac{2}{3}q^{-841/440}\left(-q^2 - q^4 - q^6 + 2 q^7 + 3 q^9 + 2 q^{10}+ \dots\right)$ \\
  \hline
    $10$ & $\frac{2}{3}q^{-1521/440}\left(-q^4 + q^5 + q^6 + 2 q^7 + 3 q^8 + 6 q^9 + 5 q^{10}+ \dots\right)$ \\
  \hline
\end{tabular}
\caption{Small $q$ expansions of the dual $q$-series for  $\overline{\Sigma(2,5,11)}$. } 
\label{tab:sigma2511}
\end{table}
\label{sec:345}
\begin{table}[h!]
\begin{tabular}{ | m{1em} | m{8cm}| } 
  \hline
  $j$ & dual $q$-series $X^{(j)}_{(3,4,5)}(q)^{\vee}$  \\
  \hline
  $1$ & $q^{-1/240}(-2 - 2 q - 2 q^2 - 2 q^4 - 2 q^5 - 2 q^6 + \dots)$ \\ 
  \hline
    $2$ & $q^{-4/240}(-2 q - 2 q^3 - 2 q^4 - 2 q^6 - 2 q^7 - 2 q^8  + \dots) $ \\ 
  \hline
    $3$ & $q^{-49/240}(-2 q^2 - 2 q^5 - 2 q^6 - 2 q^8 - 2 q^{10} + \dots)$ \\ 
  \hline
  $4$ & $q^{-121/240}(-2 q - 2 q^3 - 2 q^4 - 2 q^5 - 2 q^7 + \dots)$ \\ 
  \hline
  $5$ & $q^{-169/240}(-2 q - 2 q^3 - 2 q^5 - 2 q^7 - 2 q^9 + \dots)$ \\ 
  \hline
  $6$ & $q^{-196/240}(-2 q - 2 q^2 - 2 q^3 - 2 q^4 - 2 q^5 - 2 q^6+  \dots)$ \\ 
  \hline
\end{tabular}
\caption{Small $q$ expansions of the dual $q$-series for  $\overline{\Sigma(3,4,5)}$.} 
\label{tab:sigma345}
\end{table}
\label{sec:357}
\begin{table}[h!]
\begin{tabular}{ | m{1em} | m{13cm}| } 
  \hline
 $j$ & dual $q$-series $X^{(j)}_{(3,5,7)}(q)^{\vee}$  \\ 
  \hline
    $1$ & $q^{-1/420}(-2 q - 2 q^2 - 2 q^3 - 4 q^4 - 2 q^5 - 4 q^6 - 6 q^7 - 6 q^8 - 6 q^9 - 8 q^{10}+ \dots)$ \\ 
  \hline
    $2$ & $q^{-4/420}(2 + 2 q + 2 q^2 + 2 q^3 + 2 q^4 + 2 q^5 + 4 q^6 + 4 q^7 + 4 q^8 + 6 q^9 + 6 q^{10}+ \dots)$ \\ 
  \hline
    $3$ & $q^{-16/420}(2 q + 2 q^3 + 2 q^4 + 2 q^5 + 2 q^6 + 2 q^7 + 4 q^8 + 4 q^9 + 4 q^{10}+ \dots)$ \\ 
  \hline
    $4$ & $q^{-64/420}(-2 q^2 - 2 q^3 - 2 q^5 - 4 q^6 - 2 q^7 - 4 q^8 - 4 q^9 - 4 q^{10}+ \dots)$ \\ 
  \hline
    $5$ & $q^{-121/420}(-2 q^2 - 2 q^4 - 2 q^5 - 2 q^6 - 2 q^7 - 4 q^8 - 2 q^9 - 6 q^{10}+ \dots)$ \\ 
  \hline
    $6$ & $q^{-169/420}(2 q + 2 q^3 + 2 q^4 + 2 q^5 + 2 q^6 + 4 q^7 + 2 q^8 + 4 q^9 + 4 q^{10}+ \dots)$ \\ 
  \hline
    $7$ & $q^{-256/420}(-2 q - 2 q^2 - 2 q^3 - 4 q^4 - 4 q^5 - 4 q^6 - 6 q^7 - 6 q^8 - 8 q^9 - 8 q^{10}+ \dots)$ \\ 
  \hline
    $8$ & $q^{-289/420}(2 q + 2 q^3 + 2 q^5 + 4 q^7 + 2 q^9 + 2 q^{10}+ \dots)$ \\ 
  \hline
    $9$ & $q^{-361/420}(2 q + 2 q^2 + 4 q^3 + 2 q^4 + 4 q^5 + 4 q^6 + 6 q^7 + 6 q^8 + 8 q^9 + 8 q^{10}+ \dots)$ \\ 
  \hline
  $10$ & $q^{-529/420}(-2 q^2 - 2 q^4 - 2 q^5 - 4 q^6 - 2 q^7 - 4 q^8 - 2 q^9 - 6 q^{10}+ \dots)$ \\ 
  \hline
    $11$ & $q^{-1024/420}(-2 q^5 - 2 q^8 - 2 q^9 - 2 q^{10}+ \dots)$ \\ 
  \hline
  $12$ & $q^{-1156/420}(2 q^3 + 2 q^4 + 2 q^5 + 2 q^6 + 4 q^7 + 4 q^8 + 4 q^9 + 4 q^{10}+ \dots)$ \\ 
  \hline
\end{tabular}
\caption{Small $q$ expansions of the dual $q$-series for  $\overline{\Sigma(3,5,7)}$.} 
\label{tab:sigma357}
\end{table}


\begin{thebibliography}{1234}

\bibitem{baxter}
Baxter, R.J.,  Rogers-Ramanujan identities in the hard hexagon model. J Stat Phys 26, 427–452 (1981).

\bibitem{andrews}
George E. Andrews, The hard-hexagon model and Rogers—Ramanujan type identities,
PNAS 78 (9) 5290-5292 (1981).

\bibitem{nickel}
Orrick, W.P., Nickel, B., Guttmann, A.J. et al. The Susceptibility of the Square Lattice Ising Model: New Developments. Journal of Statistical Physics 102, 795–841 (2001). 

\bibitem{perk}
Orrick, W. P., Nickel, B. G., Guttmann, A. J. and Perk, J. H. H., Critical Behavior of the Two-Dimensional Ising Susceptibility, Phys. Rev. Lett. {\bf 86}, 4120-4123 (2001).

\bibitem{mccoy}
B.M. McCoy, M. Assis, S. Boukraa, S. Hassani, J-M Maillard, W.P. Orrick, and N. Zenine, The saga of the Ising susceptibility, arXiv:1003.0751. 

\bibitem
{BMO95}
Alexander Berkovich, Barry M. McCoy, William P. Orrick 
\newblock {Polynomial Identities, Indices, and Duality for the N=1 Superconformal Model SM(2,4v)}, 1995.
\newblock {\em Journal of Statistical Physics, Vol. 83. Nos. 5/6, 1996}, 
\newblock arXiv:hep-th/9507072

\bibitem
{GMP}
Sergei Gukov, Marcos Marino, and Pavel Putrov.
\newblock {Resurgence in complex Chern-Simons theory}, 2016.
\newblock arXiv:1605.07615.

\bibitem
{CCFGH}
Miranda C.~N. Cheng, Sungbong Chun, Francesca Ferrari, Sergei Gukov, and
  Sarah~M. Harrison.
\newblock 3d modularity.
\newblock {\em JHEP}, 10:010, 2019.
\newblock arXiv:1809.10148.

\bibitem
{CFG19}
Miranda C.~N. Cheng, Francesca Ferrari, and Gabriele Sgroi.
\newblock {Three-Manifold Quantum Invariants and Mock Theta Functions}.
\newblock {\em Phil. Trans. Roy. Soc. Lond.}, 378(2163):20180439, 2019.

\bibitem
{CCKPG}
Miranda C. N. Cheng, Ioana Coman, Piotr Kucharski, Davide Passaro, Gabriele Sgroi.
\newblock 3d Modularity Revisited.
\newblock arXiv:2403.14920.

\bibitem
{CDGG}
Ovidiu Costin, Gerald V. Dunne, Angus Gruen, and Sergei Gukov.
\newblock {Going to the Other Side via the Resurgent Bridge}.
\newblock arXiv:2310.12317.

\bibitem{Ram}
Srinivasa Ramanujan,
\newblock The Lost Notebook and Other Unpublished Papers.
\newblock Narosa, New Delhi, 1988.

\bibitem
{Watson}
G.~N. Watson.
\newblock The final problem: An account of the mock theta functions.
\newblock {\em J., London Math. Soc.}, 11(2):55--80, 1936.

\bibitem
{GM12}
B.~Gordon and R.~J. McIntosh.
\newblock A survey of classical mock theta functions.
\newblock In {\em Partitions, q-Series and Modular Forms}, volume~23 of {\em
  Developments in Mathematics}, pages 95--143. Springer, 2012.

  
  \bibitem{Co08}
O.~Costin.
\newblock {\em {\it Asymptotics and Borel summability}}.
\newblock Chapman and Hall/CRC, 2008.

\bibitem{Marino:2012zq}
M.~Mari\~no,
``Lectures on non-perturbative effects in large $N$ gauge theories, matrix models and strings,''
Fortsch. Phys. \textbf{62}, 455-540 (2014),
[arXiv:1206.6272 [hep-th]].

\bibitem{Dorigoni:2014hea}
D.~Dorigoni,
``An Introduction to Resurgence, Trans-Series and Alien Calculus,''
Annals Phys. \textbf{409}, 167914 (2019),
[arXiv:1411.3585 [hep-th]].

\bibitem{ABS19}
I.~Aniceto, G.~Basar, and R.~Schiappa.
\newblock {A Primer on Resurgent Transseries and Their Asymptotics}.
\newblock {\em {Physics Reports}}, 809:1--135, 2019.


\bibitem{CD20b}
Ovidiu Costin and Gerald~V. Dunne.
\newblock {Uniformization and Constructive Analytic Continuation of Taylor Series}.
\newblock {\em Commun. Math. Phys.}, 392:863--906, 2022.
\newblock arXiv:2009.01962.

\bibitem{sauzin}
David Sauzin, {\it Resurgent functions and splitting problems}, RIMS Kokyuroku, (2006), 1493, pp.48-117.

\bibitem
{Zwe08}
Sander Zwegers.
\newblock {\em {Mock Theta Functions}}.
\newblock Phd thesis, 7 2008.


\bibitem
{Zag09} Don Zagier.
\newblock {\em Ramanujan's mock theta functions and their applications [d'après Zwegers and Bringmann-Ono]}, 
\newblock Séminaire Bourbaki, 60ème année, 2007-2008, $n^{\circ} 986$, Astérisque 326 (2009), Soc. Math. de France, 143-164.

\bibitem{rudin}
W. Rudin, Real and Complex Analysis, pp. 377-383 (1987).

\bibitem
{GJ}
Sergei Gukov and Mrunmay Jagadale.
\newblock $c_{\text {eff }}$ for 3d 
$\mathcal{N}=2$ theories.
\newblock 
arXiv:2308.05360.

\bibitem
{ceff2}
Griffen Adams, Ovidiu Costin, Gerald V. Dunne, Sergei Gukov and O\u{g}uz \"Oner.
\newblock {$c_{\rm eff}$ from Resurgence at the Stokes Line}.
\newblock preprint, May, 2025.

\bibitem{CD25}
Ovidiu Costin and Gerald~V. Dunne,
\newblock {Resurgence, Natural Boundaries and the Uniqueness of Solutions of Mock Modular Identities},
\newblock to appear.

\bibitem
{GPV}
Sergei Gukov, Pavel Putrov, and Cumrun Vafa.
\newblock Fivebranes and 3-manifold homology.
\newblock {\em JHEP}, 07:071, 2017.
\newblock arXiv:1602.05302.

\bibitem
{Bring15}
Kathrin Bringmann and Antun Milas.
\newblock {W-Algebras, False Theta Functions and Quantum Modular Forms, I}.
\newblock {\em International Mathematics Research Notices},
  2015(21):11351--11387, 02 2015.

\bibitem{HLSS22}
Li~Han, Yong Li, David Sauzin, and Shanzhong Sun.
\newblock Resurgence and partial theta series, 2022.

\bibitem
{Bringmann}
R.~C.~Rhoades K.~Bringmann, A.~Folsom.
\newblock Partial theta functions and mock modular forms as q-hypergeometric series.
\newblock {\em Ramanujan J.}, 29:295--310, 2012.

\bibitem
{GPPV}
Sergei Gukov, Du~Pei, Pavel Putrov, and Cumrun Vafa.
\newblock {BPS} spectra and 3-manifold invariants, 2017.
\newblock arXiv:1701.06567.

\bibitem{Chun17}
Sungbong Chun.
\newblock {A resurgence analysis of the $SU(2)$ Chern-Simons partition functions on a Brieskorn homology sphere $\Sigma(2,5,7)$}.
\newblock 1 2017.

\bibitem{Wu20}
David~H. Wu.
\newblock {Resurgent analysis of SU(2) Chern-Simons partition function on Brieskorn spheres $\Sigma(2,3,6n+5)$}.
\newblock {\em JHEP}, 21:008, 2020.

\bibitem{Chung20}
Hee-Joong Chung.
\newblock {Resurgent Analysis for Some 3-manifold Invariants}.
\newblock {\em JHEP}, 05:106, 2021.

\bibitem{AM22}
J\o rgen~Ellegaard Andersen and William Elb\ae k Misteg\aa rd.
\newblock Resurgence analysis of quantum invariants of {S}eifert fibered
  homology spheres.
\newblock {\em J. Lond. Math. Soc. (2)}, 105(2):709--764, 2022.

\bibitem
{CD20c}
Miranda C. Cheng and  John F.R. Duncan.
\newblock {Optimal mock Jacobi theta functions}.
\newblock {\em Advances in Mathematics, 372, 107284, 2020}.
\newblock arXiv:1605.04480.

\bibitem{Gopakumar:1998ii}
R.~Gopakumar and C.~Vafa,
``M theory and topological strings. 1.,''
[arXiv:hep-th/9809187 [hep-th]].

\bibitem{Gopakumar:1998ki}
R.~Gopakumar and C.~Vafa,
``On the gauge theory / geometry correspondence,''
Adv. Theor. Math. Phys. \textbf{3}, 1415-1443 (1999)
[arXiv:hep-th/9811131 [hep-th]].

\bibitem{Gopakumar:1998jq}
R.~Gopakumar and C.~Vafa,
``M theory and topological strings. 2.,''
[arXiv:hep-th/9812127 [hep-th]].

\bibitem{Ooguri:1999bv}
H.~Ooguri and C.~Vafa,
``Knot invariants and topological strings,''
Nucl. Phys. B \textbf{577}, 419-438 (2000)
[arXiv:hep-th/9912123 [hep-th]].

\bibitem{deHaro:2004id}
S.~de Haro and M.~Tierz,
``Brownian motion, Chern-Simons theory, and 2-D Yang-Mills,''
Phys. Lett. B \textbf{601}, 201-208 (2004)
[arXiv:hep-th/0406093 [hep-th]].

\bibitem{deHaro:2005rz}
S.~de Haro and M.~Tierz,
``Discrete and oscillatory matrix models in Chern-Simons theory,''
Nucl. Phys. B \textbf{731}, 225-241 (2005)
[arXiv:hep-th/0501123 [hep-th]].

\bibitem{Garoufalidis:2021lcp}
S.~Garoufalidis and D.~Zagier,
``Knots, Perturbative Series and Quantum Modularity,''
SIGMA \textbf{20}, 055 (2024),
[arXiv:2111.06645 [math.GT]].

\bibitem{Fantini:2024ihf}
V.~Fantini and C.~Rella,
``Modular resurgent structures,''
[arXiv:2404.11550 [math.NT]].

\bibitem{GM}
Sergei Gukov and Ciprian Manolescu.
\newblock A two-variable series for knot complements.
\newblock {\em Quantum Topol.}, 12(1):1--109, 2021.


\bibitem
{HM14}
D.~Hickerson and E.~Mortenson.
\newblock Hecke-type double sums, Appell–Lerch sums, and mock theta functions.
\newblock {\em Proc. London Math. Soc.}, 109(3):382--422, 2014.

\bibitem
{Mort14}
E.~T. Mortenson.
\newblock On the dual nature of partial theta functions and Appell–Lerch
  sums.
\newblock {\em Advances in Mathematics}, 264:236--260, 2014.




\end{thebibliography}
\end{document}